\documentclass[reprint, amsmath,amssymb, prb]{revtex4-2}

\usepackage{graphicx}
\usepackage{dcolumn}
\usepackage{bm}
\usepackage[dvipsnames]{xcolor}
\usepackage{hyperref}
\usepackage[inline]{enumitem}

\begin{document}

\title{Planar Josephson junctions in germanium: Effect of cubic spin-orbit interaction}

\author{Melina Luethi}
\email{melina.luethi@unibas.ch}
\author{Katharina Laubscher}
\author{Stefano Bosco}
\author{Daniel Loss}
\author{Jelena Klinovaja}
\affiliation{
	Department of Physics, University of Basel, Klingelbergstrasse 82, CH-4056 Basel, Switzerland}

\date{\today}

\begin{abstract}
Planar Josephson junctions comprising semiconductors with strong spin-orbit interaction (SOI) are promising platforms to host Majorana bound states (MBSs). 
Previous works on MBSs in planar Josephson junctions have focused on electron gases, where SOI is linear in momentum. 
In contrast, a two-dimensional hole gas in planar germanium (Ge) exhibits SOI that is cubic in momentum. Nevertheless, we show here that due to the particularly large SOI, Ge is a favorable material. 
Using a discretized model, we numerically simulate a Ge planar Josephson junction and demonstrate that also cubic SOI can lead to the emergence of MBSs.
Interestingly, we find that the cubic SOI yields an asymmetric phase diagram as a function of the superconducting phase difference across the junction. We also find that trivial Andreev bound states can imitate the signatures of MBSs in a Ge planar Josephson junction, therefore making the experimental detection of MBSs difficult. We use experimentally realistic parameters to assess if the topological phase is accessible within experimental limitations. Our analysis shows that two-dimensional Ge is an auspicious candidate for topological phases.
\end{abstract}

\maketitle

\section{\label{sec:introduction}Introduction}
\begin{figure}
	\centering
	\includegraphics[width=\columnwidth]{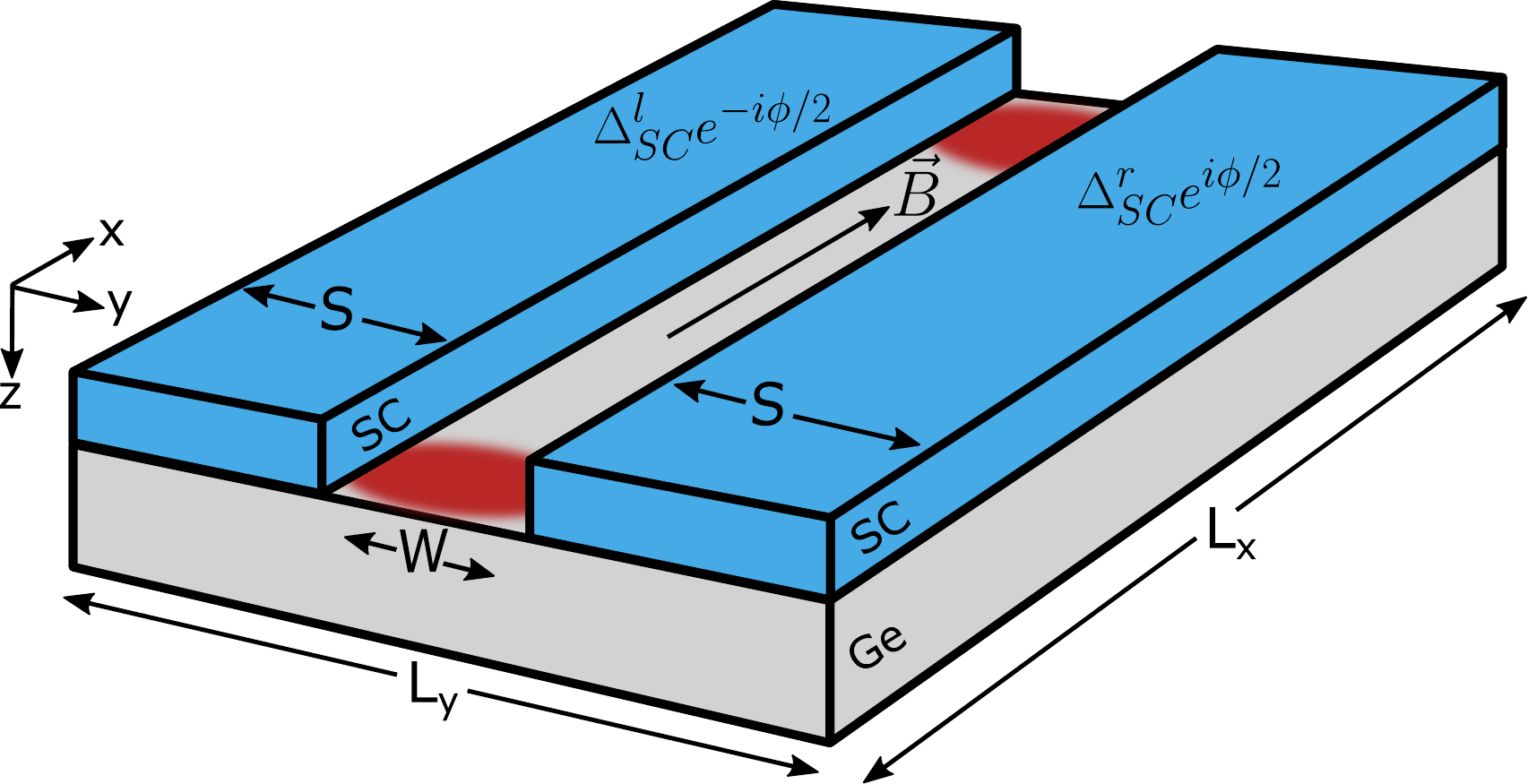}
	\caption{A germanium 2DHG (gray box) of size $L_x \times L_y$ is proximitized by two superconductors (blue boxes) of width $S$. The superconductors induce superconducting gaps in germanium of strength $\Delta_{\text{SC}}^l$ and $\Delta_{\text{SC}}^r$ with a phase difference $\phi$ between them. They are separated by a normal section of width $W$. A magnetic field $\vec{B}$ is applied along the junction. In the topological phase, MBSs are localized at the ends of the normal region (indicated schematically by the red regions), see Sec.~\ref{sec:planar_jj}.}
	\label{fig:setup}
\end{figure}

Majorana bound states (MBSs) are exotic quasiparticles that emerge in topological superconductors~\cite{kitaev2001unpaired,leijnse2012introduction, qi2011topological, beenakker2013search, sato2016majorana, pawlak2019majorana, laubscher2021majorana}. 
Due to their non-Abelian exchange statistics~\cite{invanov2001non}, MBSs are suitable candidates to store and manipulate quantum information in an intrinsically fault-tolerant way~\cite{kitaev2003fault, nayak2008non, elliot2015colloquium}.
Numerous physical systems have been predicted to host MBSs, with prominent examples including one-dimensional semiconductor nanowires with linear spin-orbit interaction (LSOI)~\cite{lutchyn2010majorana, oreg2010helical, stanescu2011majorana, mourik2012signatures, das2012zero, deng2012anomalous}, 
carbon nanotubes and nanoribbons~\cite{klinovaja2012electric, egger2012emerging, klinovaja2013giant, desjardins2019synthetic}, 
chains of magnetic adatoms~\cite{choy2011majorana, nadj2013proposal, braunecker2013interplay, pientka2013topological, klinovaja2013topological, vazifeh2013self, nadj2014observation, ruby2015end, pawlak2016probing, jack2021detecting}, proximitized topological insulators~\cite{fu2008superconducting, fu2009josephson, cook2011majorana, cook2012stability, jaeck2019observation, legg2021majorana}, thin semiconducting films~\cite{sau2010generic, alicea2010majorana, sau2010non}, and planar Josephson junctions~(JJs)~\cite{pientka2017topological, hell2017two, setiawan2019topological, scharf2019tuning, melo2019supercurrent, pientka2017topological, laeven2020enhanced, volpez2020time, paudel2021enhanced, pakizer2021crystalline, pekerten2022anisotropic, melo2022greedy, ren2019topological, fornieri2019evidence, banerjee2022signatures}~(see Fig.~\ref{fig:setup}), which will be the focus of this work. 
Over the last few years, MBSs in planar JJs based on two-dimensional electron gases (2DEGs) with strong LSOI have been the focus of extensive theoretical~\cite{pientka2017topological,hell2017two, setiawan2019topological, scharf2019tuning, melo2019supercurrent, pientka2017topological, laeven2020enhanced, volpez2020time, paudel2021enhanced, pakizer2021crystalline, pekerten2022anisotropic, melo2022greedy} and experimental~\cite{ren2019topological, fornieri2019evidence, banerjee2022signatures} studies. 
Compared to one-dimensional systems such as nanowires, planar JJs have several advantages.  
First, the phase difference between the two superconductors adds an additional control knob to tune the system into the topological phase~\cite{pientka2017topological}. 
Furthermore, two-dimensional systems such as planar JJs offer additional flexibility for the design of braiding setups~\cite{ren2019topological}.
For example, multiterminal Josephson junctions~\cite{fu2008superconducting, sau2010robustness, hell2017coupling, stern2019fractional, lesser2022majorana, schrade2018majorana, zhou2020phase} hint at the promising possibility to implement networks of Majorana qubits.
Additionally, the normal section of the planar JJ can be formed by selectively removing the superconductor~\cite{fornieri2019evidence,zhou2020phase}, allowing for a flexible geometry of the normal section, e.g., zigzag-shaped or otherwise spatially modulated, which has been demonstrated to increase the bulk gap in the topological phase~\cite{laeven2020enhanced, paudel2021enhanced, melo2022greedy}.

Although previous works on planar JJs focused on 2DEGs, it is well known that also two-dimensional hole gases (2DHGs) can offer a large SOI, a critical requirement to host MBSs.
For this reason, we study a planar JJ based on a 2DHG in this work.
In particular, we focus on holes in planar Ge, a promising material for various quantum information applications~\cite{scappucci2020germanium}.
A reason for the large interest in Ge is its strong and tunable SOI~\cite{hao2010strong, maurand2016cmos, watzinger2018germanium, scappucci2020germanium, hendrickx2020single, hendrickx2020fast, froning2021ultrafast, bosco2021squeezed,wang2022ultrafast}. 
Further, the hyperfine interactions in Ge can be minimized by isotopical purification~\cite{itoh1993high, scappucci2020germanium} or by the design of the nanostructure~\cite{bosco2021fully}, and Ge structures are highly compatible with existing CMOS technology~\cite{veldhorst2017silicon, scappucci2020germanium}. 
In addition, proximity induced superconducting gaps in Ge heterostructures have been realized experimentally~\cite{hendrickx2018gate, hendrickx2019ballistic, vigneau2019germanium, aggarwal2021enhancement, tosato2022hard}. Going a step further, a planar JJ in Ge was recently realized in an experiment, demonstrating an induced hard superconducting gap and phase control of the junction~\cite{tosato2022hard}.
Since MBSs require superconductivity, the successful combination of Ge with superconductors makes Ge an ideal platform to host MBSs. 
The possibility to host MBSs in Ge/Si core/shell nanowires has been theoretically investigated in Ref.~\cite{maier2014majorana}.

In this work, we numerically simulate a Ge planar JJ using the effective Hamiltonian of a Ge 2DHG. In contrast to electron gases, holes in planar Ge present a large cubic SOI (CSOI), which is known to cause qualitatively different effects compared to LSOI in a variety of cases. 
For example, CSOI can lead to an anisotropy in the exchange interaction of hole spin qubits even at zero magnetic field~\cite{hetenyi2022anomalous} and to Rabi frequencies that are two orders of magnitude smaller than in systems with LSOI~\cite{bosco2021squeezed}. Furthermore, the spin relaxation time for holes subject to CSOI has a different power-law dependence on the magnetic field compared to electrons~\cite{kroutvar2004optically}. CSOI can also lead to a larger spin Hall conductivity~\cite{schliemann2005spin, moriya2014cubic}, a different spin diffusion behavior~\cite{bleibaum2006spin}, different behavior in the spin-photon interaction~\cite{bosco2022fully}, and a large non-reciprocal transport that is based on spin-to-charge current conversion~\cite{dantas2022}.

Because of the qualitatively different behavior of CSOI compared to LSOI in many instances, it is not clear from the outset that a planar JJ based on a 2DHG with CSOI can host MBSs in the same way as a planar JJ based on a 2DEG with LSOI. Therefore, the main purpose of this paper is to demonstrate that MBSs can emerge in a planar JJ based on a 2DHG with CSOI. Indeed, by calculating the topological phase diagram numerically, we find an extended region of parameter space where the 2DHG JJ hosts MBSs at the junction ends.
The most striking qualitative difference to planar JJ based on a 2DEG with LSOI is that the CSOI leads to a phase diagram that is asymmetric under the inversion of the superconducting phase difference $\phi \rightarrow -\phi$. In contrast, the phase diagram for LSOI is symmetric.

In addition to MBSs, we show that also trivial Andreev bound states (ABSs) can emerge in the JJ. The presence of these states complicates the experimental detection of MBSs since ABSs can imitate the signatures of MBSs, a phenomenon that has been studied thoroughly for nanowires~\cite{kells2012near, moore2018two, moore2018quantized,  huang2018metamorphosis, penaranda2018quantifying, reeg2018zero, vuik2019reproducing, sarma2021disorder, hess2021local}.
A further difficulty in the experimental realization of MBSs is that the magnetic field needed to tune the system into the topological phase is limited by the critical field of the superconductor. Therefore, having a material with a large $g$ factor is advantageous. Unfortunately, the in-plane $g$ factor of a 2DHG in Ge is rather small and typically ranges between $0.2$ and $1.4$~\cite{watzinger2016heavy, lu2017effective, hendrickx2018gate, hofmann2019assessing, gao2020site, hendrickx2020fast, scappucci2020germanium}.
We use experimentally realistic parameters to assess the accessibility of the topological phase within these limitations and propose changes to the experiment to increase the $g$ factor. With our analysis we conclude that, while challenging, these issues could be overcome and planar Ge JJs can realistically host MBSs.

We note that a planar JJ with CSOI has also been studied in Ref.~\cite{alidoust2021cubic}. However, the MBSs studied there are localized in one direction and infinitely extended in the other direction. In contrast, the MBSs we analyze in the current paper are localized in two directions and located at the ends of the junction, see Fig.~\ref{fig:setup}. Furthermore, in  Ref.~\cite{mayer2022intrinsic}, planar JJs in semimetals such as two-dimensional $\beta$-Sn-$\alpha$-Sn-$\beta$-Sn are studied. Similarly to Ge, these materials can be described in terms of
the Luttinger-Kohn Hamiltonian (see Sec.~\ref{sec:effective_ham}). However, the resulting effective Hamiltonian describing the semimetal 2DHG takes on a different form without CSOI terms and without strain, making these systems fundamentally different from the 2DHGs in semiconducting Ge heterostructures we study in this work.

This paper is structured as follows. The effective Hamiltonian of a Ge 2DHG is introduced in Sec.~\ref{sec:effective_ham}. The setup and Hamiltonian of the planar JJ are described in Sec.~\ref{subsec:setup}. We analyze the topological phase diagram of this system in Sec.~\ref{subsec:mbs}. In Sec.~\ref{subsec:quasi_mbs} we study ABSs in the planar JJ. We do numerical simulations with experimentally realistic parameters in Sec.~\ref{subsec:experiment}. A conclusion and outlook are given in Sec.~\ref{sec:conclusion}.
Further details on the system and the numerical calculations are given in the Appendixes~\ref{appsec:anisotropic_spectrum}-\ref{appsec:potential_barrier}.

\section{\label{sec:effective_ham}Effective Hamiltonian for planar germanium}
To study Ge planar JJs, we first derive the effective Hamiltonian describing a 2DHG, following Ref.~\cite{terrazos2021theory}. The holes at the top of the valence band have spin~$3/2$. The valence bands in bulk Ge are well described by the isotropic Luttinger-Kohn Hamiltonian~\cite{luttinger1956quantum, winkler2003spin}:
\begin{eqnarray}
	H_{\text{LK}} = \frac{\hbar^2}{m}\left[
	\left(\gamma_1 + \frac{5 \gamma_s}{2}\right)\frac{\vec{k}^{2}}{2}
	- \gamma_s \left( \vec{k} \cdot \vec{J} \right)^2 
	\right]  - \mu, 
	\label{eq:lk_hamiltonian_isotropic}
\end{eqnarray}
where $m$ is the bare electron mass, $\gamma_s=(\gamma_2+\gamma_3)/2$,
the $\gamma_i$ are the Luttinger parameters~\cite{winkler2003spin},
$\vec{k} = (k_x, k_y, k_z)$ is the vector of momentum,
$\vec{J} = (J_x, J_y, J_z)$ are the spin-$3/2$ operators, 
and $\mu$ is the chemical potential. The holes are divided into heavy holes with spin eigenvalues $J_z = \pm 3/2$ and light holes with $J_z = \pm 1/2$~\cite{scappucci2020germanium}.
While the isotropic Luttinger-Kohn Hamiltonian in Eq.~\eqref{eq:lk_hamiltonian_isotropic} provides in general a good description of the Ge heterostructure, anisotropic corrections arising from the cubic lattice can become relevant in certain cases~\cite{xiong2021emergence, bosco2021squeezed}. The effect of these anisotropic corrections is addressed in Sec.~\ref{subsec:experiment} and App.~\ref{appsec:anisotropic_spectrum}. We find that the qualitative behavior of the phase diagram is not affected by these anisotropic corrections.

\begin{figure}
	\centering
	\includegraphics[width=0.5\columnwidth]{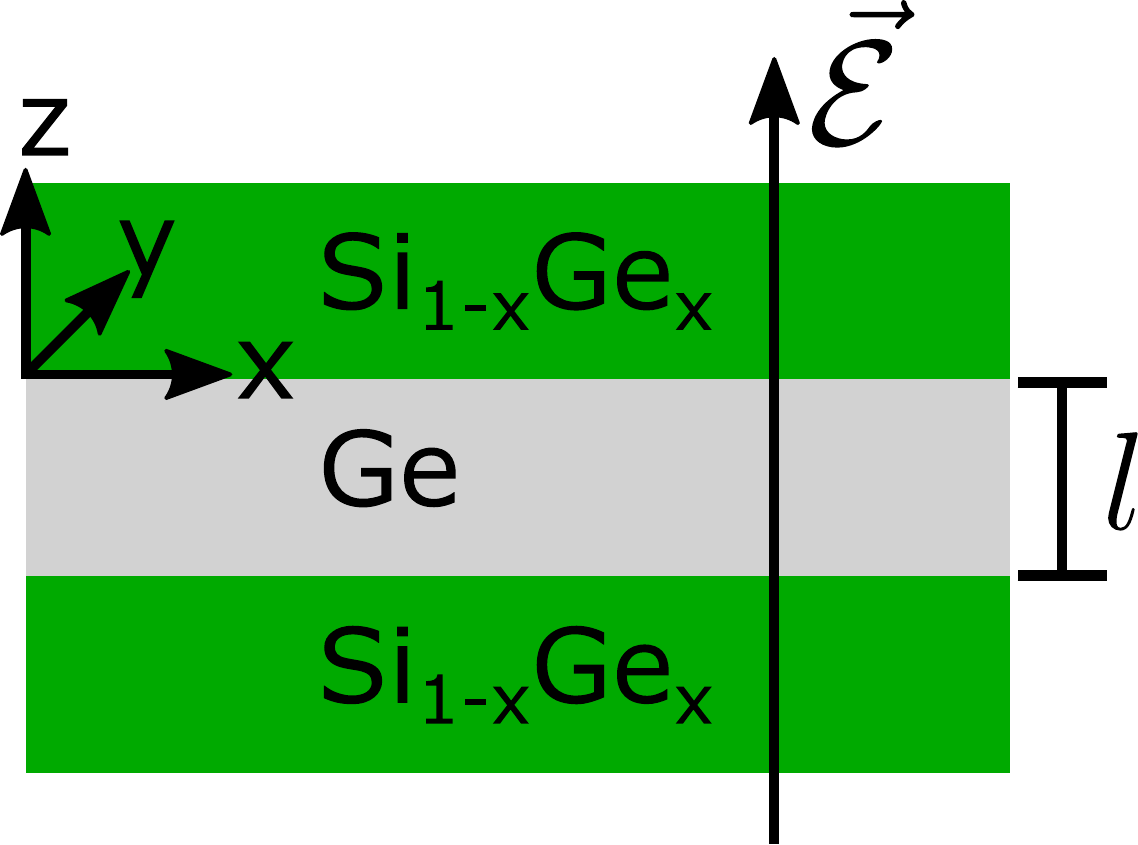}
	\caption{
		Sketch of the germanium heterostructure. A thin germanium layer (gray) of thickness $l$ is sandwiched between two layers of Si$_{1-x}$Ge$_x$ (green) and an electric field of strength $\mathcal{E}$ is applied along the $z$ axis.}
	\label{fig:ge_heterostructure}
\end{figure}

Next, we study the situation where the holes are confined to a two-dimensional plane. The axis normal to this plane is referred to as the confinement axis. Since the material is isotropic, the choice of confinement axis is arbitrary and we choose the $z$ axis, see Fig.~\ref{fig:ge_heterostructure}.
For SOI to be present, the structural inversion symmetry of the system must be broken. This can be achieved, e.g., by applying an electric field of strength $\mathcal{E}$ along the confinement axis~\footnote{Another way to break inversion symmetry is by an inversion asymmetric confinement potential~\cite{bosco2021hole}.}.
Moreover, state-of-the-art Ge heterostructures are sandwiched between two layers of Si$_{1-x}$Ge$_x$ with $x \in [0.7, 0.9]$~\cite{scappucci2020germanium, lodari2022lightly}, see Fig.~\ref{fig:ge_heterostructure}.
Because these materials have different lattice constants, the Ge is strained. For holes this results in the additional energy $E_s>0$ that is typically of a few tens of~meV~\cite{bosco2021squeezed, terrazos2021theory, wang2021optimal}.
By adding these contributions, one obtains~\cite{bosco2021squeezed, bir1974symmetry}:
\begin{eqnarray} 
	H_{\text{2DHG}}\left(k_x, k_y, z\right) =&& H_{\text{LK}}\left(k_x, k_y, -i\hbar \partial_z\right) \nonumber \\
	&&- e \mathcal{E} z - E_s J_z^2  ,
	\label{eq:lk_confined}
\end{eqnarray}
where $H_{\text{LK}}\left(k_x, k_y, -i\hbar \partial_z\right)$ is the Hamiltonian defined in Eq.~\eqref{eq:lk_hamiltonian_isotropic}, but the momentum $k_z$ is replaced by its position-space representation, and $e>0$ is the elementary charge.

We confine the 2DHG by enforcing Dirichlet boundary conditions on the wave function $\psi(x, y, z=0)=0$ and $\psi(x, y, z=-l)=0$, where $l$ is the thickness of the Ge, see Fig.~\ref{fig:ge_heterostructure}. These boundary conditions physically model the large energy gap that arises abruptly at the SiGe-Ge interfaces.
In contrast to Ref.~\cite{terrazos2021theory}, in this work, we consider the case where the electric field is strong such that the hole wave function is confined to the top boundary of the Ge structure within the electric length~\cite{bosco2021squeezed}
\begin{equation}\label{eq:electric_length_scale}
	l_\mathcal{E} = \left(\frac{\hbar^2 \gamma_1}{2m e\mathcal{E}}\right)^{1/3} .
\end{equation}
The electric length determines the characteristic energy
\begin{equation} \label{eq:electric_energy}
	E_\mathcal{E} = \frac{\hbar^2 \gamma_1}{2 m l_\mathcal{E}^2} 
	=
	e \mathcal{E} l_\mathcal{E}  .
\end{equation}
In the limit of a strong electric field, the quantum well is much thicker than the electric length, i.e., $l \approx 15\text{ - } 30$~nm and $l_\mathcal{E}~\approx~8$~nm for $\mathcal{E}=1$~V$/\mu$m. Therefore, we neglect the lower boundary at the well and assume a semi-infinite system in $z$ direction~\cite{bosco2021squeezed}. Thus, we will assume $z~\in~(-\infty,~0]$. 

\begin{figure}
	\centering
	\includegraphics[width=\columnwidth]{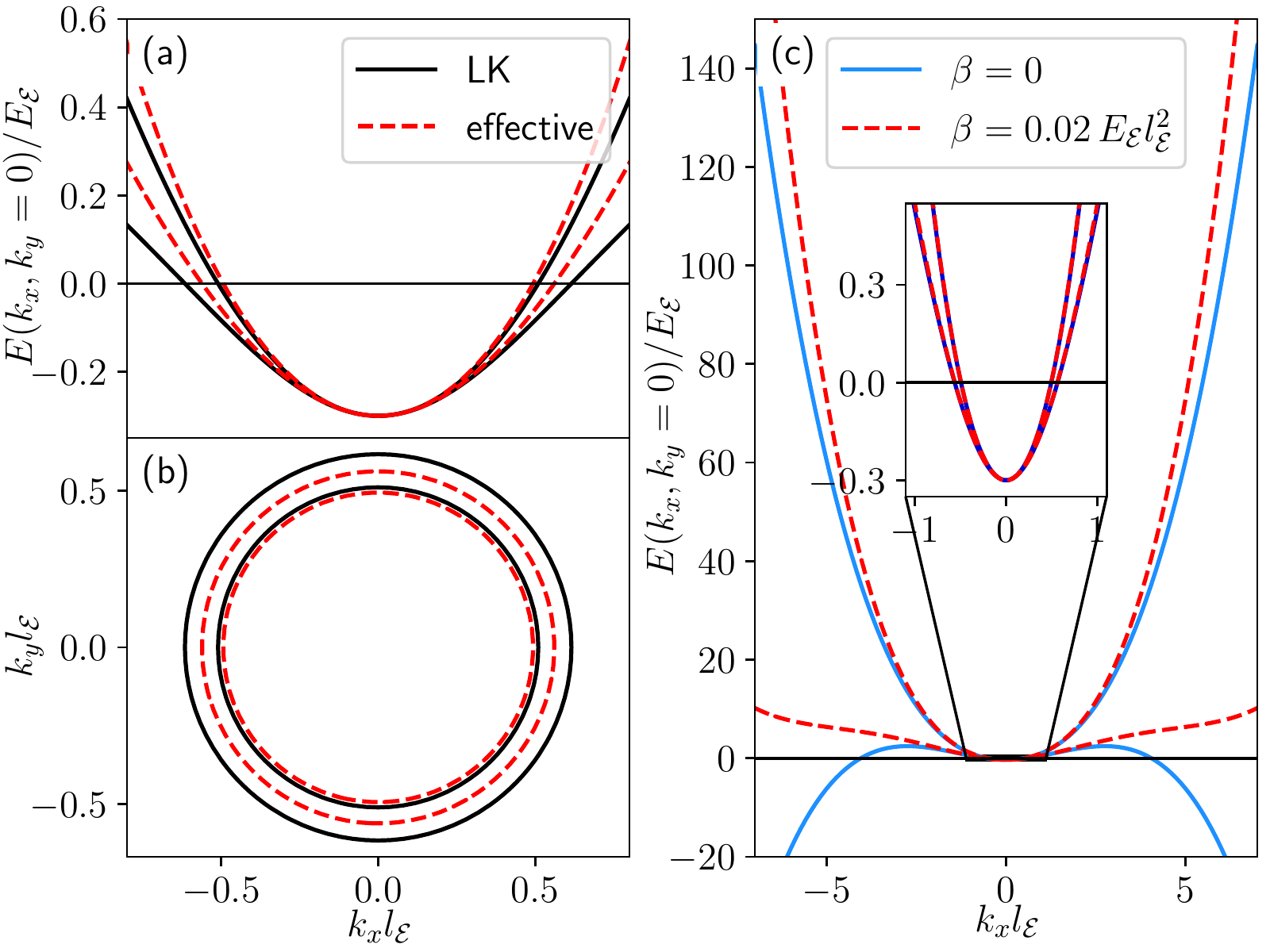}
	\caption{(a): Spectrum $E(k_x, k_y)$ of the 2DHG.
		(b):~Fermi surfaces, i.e., the momenta where $E(k_x, k_y)=0$. The solid black lines in panels~(a) and~(b) correspond to the two lowest energy bands of the Luttinger-Kohn (LK) Hamiltonian confined to two dimensions. The red dashed lines correspond to the energy eigenvalues of the effective Hamiltonian $H_\text{eff}+H_4$ defined in Eqs.~\eqref{eq:effective_hamiltonian_isotropic} and~\eqref{eq:quartic_k}.
		The two models agree best at small momenta. 
		(c):~Spectrum $E(k_x, k_y=0)$ of the effective model with $\beta=0$ (solid blue curve) and $\beta=0.02 \, E_\mathcal{E} l_\mathcal{E}^4$ (dashed red curve). Without the quartic term, the lower energy band bends downwards at large momenta, causing an unphysical Fermi surface. By introducing the quartic term defined in Eq.~\eqref{eq:quartic_k}, this is avoided. The quartic term does not change the low-energy physics noticeably, as is shown in the inset of panel~(c).
		The parameters for the Luttinger-Kohn Hamiltonian are $E_s=0$, $\gamma_1=13.38$, and $\gamma_s=4.965$~\cite{winkler2003spin}. 
		Due to the normalization, the spectrum is independent of the electric field $\mathcal{E}$.	  
		The parameters for the effective model are $\mu=0.3 E_\mathcal{E}$, $\hbar^2/2m^\ast=1.096 E_\mathcal{E} l_\mathcal{E}^2$, $\alpha=0.133 E_\mathcal{E} l_\mathcal{E}^3$.
		These parameters were determined using perturbation theory~\cite{winkler2003spin}.
	}
	\label{fig:spectrum_and_fermi_surface}
\end{figure}

We then discretize the Hamiltonian $H_{\text{2DHG}}$ in Eq.~\eqref{eq:lk_confined} by using the eigenfunctions of a particle with mass $m/\gamma_1$ in a triangular well with slope $-e\mathcal{E}$ as basis states:
\begin{equation} \label{eq:basis_states}
	f_n(z) = \frac{\text{Ai}\left(\text{Ai}_0^{\left(n\right)}-z/l_\mathcal{E}\right)}{\sqrt{l_\mathcal{E}}\text{ Ai}^\prime\left(\text{Ai}_0^{\left(n\right)}\right)}  ,
\end{equation}
where $\text{Ai}(x)$ is the Airy function, $\text{Ai}_0^{(n)}$ is the $n$-th zero of the Airy function with $n=1,2,\dots$ being a quantum number, and $\text{Ai}^\prime(x)$ is the derivative of $\text{Ai}(x)$.
We project the Hamiltonian $H_{\text{2DHG}}$ of Eq.~\eqref{eq:lk_confined} onto the basis states defined in Eq.~\eqref{eq:basis_states} and diagonalize the resulting Hamiltonian to obtain the energy spectrum of the Luttinger-Kohn Hamiltonian confined to two dimensions. 
At $k_x=k_y=0$, the two lowest energy bands are degenerate, as is expected from time-reversal symmetry. The bands split for $k_x^2+k_y^2 \neq 0$ (see Fig.~\ref{fig:spectrum_and_fermi_surface}) resulting in two circular Fermi surfaces.

A low-energy effective Hamiltonian to describe the two lowest bands of Eq.~\eqref{eq:lk_confined} is derived by treating the momentum perturbatively (see Ref.~\cite{winkler2003spin} for an introduction to quasi-degenerate perturbation theory). To third order, the effective Hamiltonian acting on the subspace of heavy holes with the spin quantization axis aligned to the confinement axis ($z$ axis) contains only a kinetic and a CSOI term and is given by~\cite{miserev2017dimensional, terrazos2021theory, marcellina2017spin, michal2021longitudinal}: 
\begin{eqnarray}
	H_\text{eff} = \frac{\hbar^2 k^2}{2m^\ast} - \mu
	+ i \alpha \left(k_+^3 \sigma_- - k_-^3 \sigma_+\right) ,
	\label{eq:effective_hamiltonian_isotropic}
\end{eqnarray}
where $m^\ast$ is the effective mass, $k^2=k_x^2+k_y^2$, $k_\pm = k_x \pm i k_y$, $\sigma_\pm = \sigma_x \pm i \sigma_y$, and $\sigma_i$ are the Pauli matrices. 
In the derivation of the effective Hamiltonian in Eq.~\eqref{eq:effective_hamiltonian_isotropic}, only the two lowest subbands of the Luttinger-Kohn Hamiltonian are considered. This approximation is justified because the gap to the next subband is of the order of the electric energy $E_\mathcal{E}$ [defined in Eq.~\eqref{eq:electric_energy}], typically a few meV to a few tens of meV and much larger than the chemical potential and the superconducting gap.

We emphasize again that the effective Hamiltonian in Eq.~\eqref{eq:effective_hamiltonian_isotropic} comes from a perturbation theory valid at small momenta. However, in Secs.~\ref{sec:planar_jj} and~\ref{subsec:experiment} we intend to use this effective Hamiltonian in a numerical tight-binding simulation, which cannot be restricted to small momenta only. 
It is therefore crucial to check that the large momentum behavior of the effective model does not introduce unphysical effects, see Fig.~\ref{fig:spectrum_and_fermi_surface}(c). Here, the blue curve indicates the energy eigenvalues of Eq.~\eqref{eq:effective_hamiltonian_isotropic}. 
At large momenta, the cubic term dominates over the quadratic term, causing the lower energy band to bend downwards. The CSOI therefore causes an additional \textit{unphysical} Fermi surface at large momenta. This is in striking contrast to LSOI, where the quadratic term dominates at large momenta, preventing the appearance of unphysical Fermi surfaces. We stress again that this high-momentum behavior of CSOI is not physical since the effective model is only valid at low momenta. 
To remove the unphysical Fermi surface, we introduce a fictitious quartic potential of the form
\begin{equation}
	\label{eq:quartic_k}
	H_4 = \beta k^4 = \beta (k_x^2+k_y^2)^2  .
\end{equation}
We need to choose $\beta$ large enough to remove the unphysical Fermi surface but small enough to keep the small momentum physics unchanged. We therefore require
\begin{equation} \label{eq:limit_beta}
	\beta k_F^4 \ll \alpha k_F^3  ,
\end{equation}
where $k_F$ is the largest Fermi momentum of the two physical Fermi surfaces. 
In App.~\ref{appsec:high_momentum_fermi_surface} it is explained in more detail how the value of $\beta$ is chosen in this work.
Moreover, to assess the validity of the fictitious quartic term, we treat the CSOI terms as a perturbation to the quadratic spectrum, see App.~\ref{appsec:perturbation}. In this case, the quartic term is not required. We find that the perturbatively calculated energies agree well with the energies calculated using $H_\text{eff}+H_4$ defined in Eqs.~\eqref{eq:effective_hamiltonian_isotropic} and~\eqref{eq:quartic_k}. Therefore, by adding this quartic term, we can still accurately capture the low-energy physics of the system.

The energy eigenvalues and Fermi surfaces of $H_\text{eff}+H_4$ defined in Eqs.~\eqref{eq:effective_hamiltonian_isotropic} and~\eqref{eq:quartic_k} are shown by the red dashed curves in Fig.~\ref{fig:spectrum_and_fermi_surface}. 
The spectrum of the effective Hamiltonian grows more rapidly with momentum compared to the spectrum of the Luttinger-Kohn Hamiltonian, leading to a larger deviation between the two models the larger the momentum becomes. However, this variation is not caused by the quartic term $H_4$ because it does not significantly change the energies in the momentum range shown in the plot, see the inset in Fig.~\ref{fig:spectrum_and_fermi_surface}(c). Rather, this deviation shows the range of validity of the third-order perturbation theory used to obtain $H_\text{eff}$ and we believe that it has no qualitative effect on the topological properties of the system. We comment further on this deviation in App.~\ref{appsec:anisotropic_spectrum}. 

We also note that the problem with the unphysical Fermi surface cannot be solved by simply expanding the perturbation theory up to fourth order in momentum because we find that the coefficient of the fourth-order term is typically negative. Therefore, one would now require an unphysical $k^6$ term to remove the unphysical Fermi surface. 
We comment further on the fourth-order perturbation theory in App.~\ref{appsec:anisotropic_spectrum}

\section{\label{sec:planar_jj}Germanium Planar Josephson junction}

\subsection{\label{subsec:setup}Setup}
We use the effective Hamiltonian of Eq.~\eqref{eq:effective_hamiltonian_isotropic} to describe a planar JJ, see Fig.~\ref{fig:setup}. The coordinate system is defined such that the 2DHG lies in the $xy$ plane and the $x$ axis points along the junction. The Ge 2DHG has size $L_x \times L_y$ and is described by the real-space version of the effective Hamiltonian defined in Eq.~\eqref{eq:effective_hamiltonian_isotropic} and the quartic term defined in Eq.~\eqref{eq:quartic_k}:
\begin{eqnarray}
	&&\mathcal{H}_\text{eff}\left(x,y\right)  = 
	-\left[\frac{\hbar^2}{2m^\ast} \left(\partial_x^2 + \partial_y^2\right) + \mu \right]\tau_z \nonumber\\ 
	&&\!\!\!\! \qquad + 2 i \alpha  \left[ \partial_y \left(\partial_y^2-3 \partial_x^2\right)\sigma_x 
	+ \partial_x \left(\partial_x^2-3\partial_y^2\right)\sigma_y \tau_z \right] ,
	\label{eq:continuous_effective_hamiltonian} \\
	&&\mathcal{H}_4 = \beta \left( \partial_x^4 + \partial_y^4 + 2\partial_x^2 \partial_y^2 \right)\tau_z , 
	\label{eq:continuous_quartic}
\end{eqnarray}
written in the Nambu basis
\begin{equation}
	\! \Psi(x,y) \! = \! \left( \! \psi_{\uparrow}\left(x,y\right), \psi_{\downarrow}\left(x,y\right), \psi^\dagger_{\uparrow}\left(x,y\right), \psi^\dagger_{\downarrow}\left(x,y\right) \! \right)^T \!\! ,
\end{equation}
where $\psi_{s}^\dagger\left(x,y\right)$ creates a particle with effective spin $s$ at position $(x,y)$.  We note that this definition of the Nambu basis follows a different convention compared to Ref.~\cite{pientka2017topological}.
The $\sigma_i$ ($\tau_i$) are the Pauli matrices acting in spin (particle-hole) space.

The 2DHG is proximitized by two $s$-wave superconductors.
The proximity-induced superconducting pairings have strengths $\Delta_{\text{SC}}^l$ on the left and $\Delta_{\text{SC}}^r$ on the right side, respectively, and there is a relative superconducting phase difference of $\phi$ across the junction. Between the two superconducting regions, there is a narrow gap of width $W$ with no induced superconductivity. 
Therefore, the superconducting term of the Hamiltonian is given by:
\begin{equation}
	\mathcal{H}_{\text{SC}}(y) \!=\! 
	\Delta_{\text{SC}}\left(y\right) \! \frac{i\sigma_y}{2} \! \left( 
	e^{-i \phi\left(y\right)/2} \tau_+
	- e^{i \phi\left(y\right)/2} \tau_-
	\right) ,
	\label{eq:sc_term_continuous}
\end{equation}
with $\tau_\pm = \tau_x \pm i \tau_y$ and
\begin{eqnarray}
	\Delta_{\text{SC}}\left(y\right) &=& \begin{cases}
		\Delta_{\text{SC}}^l & \text{if } 0 \leq y < S \\
		0 & \text{if } S \leq y < S+W \\
		\Delta_{\text{SC}}^r & \text{if } S+W \leq y < L_y 
	\end{cases} ,
	\label{eq:definition_delta_sc_continuous}	\\
	\phi\left(y\right) &=& \begin{cases}
		\phi & \text{if } 0 \leq y < S \\
		-\phi & \text{if } S+W \leq y < L_y  
	\end{cases} ,
	\label{eq:definition_phi_continuous} 
\end{eqnarray}
where $S=(L_y-W)/2$ is the width of the superconducting regions.
A more comprehensive analysis of superconductivity in a hole gas would require considering the pairing between holes with total angular momentum projection $J_z = \pm 3/2$ and $J_z = \pm 1/2$ separately~\cite{maier2014majorana, mao2012hole}. However, the superconducting term in Eq.~\eqref{eq:sc_term_continuous} is the effective lowest-energy pairing term~\cite{maier2014majorana} and we will limit our model to this term only.

In the normal-conducting region between the two superconductors, a magnetic field $\vec{B}$ is applied parallel to the $x$ axis. The resulting Zeeman energy $\Delta_Z$ is given by:
\begin{equation}
	\Delta_Z = \frac{1}{2} g \mu_B |\vec{B}|,
	\label{eq:magnetic_field_to_zeeman}
\end{equation} 
where $g$ is the effective $g$ factor of the holes and $\mu_B$ is the Bohr magneton. This leads to the following Zeeman term in the Hamiltonian:
\begin{equation}
	\mathcal{H}_Z(y) = \Delta_Z\left(y\right) \sigma_x \tau_z,
	\label{eq:zeeman_term_continuous}
\end{equation}
with
\begin{equation}
	\Delta_{Z}\left(y\right) = \begin{cases}
		0 & \text{if } 0 \leq y < S \\
		\Delta_Z & \text{if } S \leq y < S+W \\
		0 & \text{if } S+W \leq y < L_y 
	\end{cases}.
	\label{eq:definition_delta_z_continuous}
\end{equation}

The total Hamiltonian of the system is then given by:
\begin{eqnarray}
	H &=& \frac{1}{2} \int dx \, dy \, \Psi^\dagger(x,y) \mathcal{H}\left(x,y\right) \Psi(x,y) ,\\
	\!\!\!\!\mathcal{H}\left(x,y\right) \! &=& \!
	\mathcal{H}_\text{eff}\left(x,y\right) 
	\! + \!  \mathcal{H}_4\left(x,y\right) 
	\! + \!  \mathcal{H}_{\text{SC}}(y) 
	\! + \!  \mathcal{H}_Z(y) .
	\label{eq:continuous_hamiltonian_all_terms}
\end{eqnarray}

The Hamiltonian in Eq.~\eqref{eq:continuous_hamiltonian_all_terms} is particle-hole symmetric:
\begin{equation}
	\tau_x \mathcal{H}^\ast\left(x,y\right) \tau_x = - \mathcal{H} \left(x,y\right).
	\label{eq:particle_hole_symmetry}
\end{equation}
The usual time-reversal symmetry is broken by the Zeeman term $\Delta_Z$ and the superconducting phase difference $\phi$.
Therefore, the system is in symmetry class D and its topological phase is characterized by an invariant $Q_{\mathbb{Z}_2} = \pm 1$~\cite{chiu2016classification, pientka2017topological, hell2017two, setiawan2019topological}. 
However, if $\Delta_{\text{SC}}^l=\Delta_{\text{SC}}^r$, the system has an additional effective time-reversal symmetry even if $\Delta_Z \neq 0$ and $\phi \neq 0$~\cite{pientka2017topological, hell2017two, setiawan2019topological}:
\begin{eqnarray}
	\mathcal{H}^\ast(L_y-y, -\partial_y) 
	=  \mathcal{H}\left(y, \partial_y\right),
	\label{eq:modified_time_reversal}
\end{eqnarray}
which is reminiscent of a mirroring symmetry.
This brings the system into the BDI symmetry class and its topological phase is characterized by an invariant $Q_\mathbb{Z} \in \mathbb{Z}$~\cite{pientka2017topological, hell2017two, chiu2016classification, setiawan2019topological}. 
The topological invariants of the D and BDI symmetry classes are related as follows~\cite{tewari2012topological}:
\begin{equation}
	Q_{\mathbb{Z}_2} = \left(-1\right)^{Q_\mathbb{Z}} .
\end{equation}
The consequences of the effective time-reversal symmetry are further discussed in Sec.~\ref{subsec:mbs}.

In the following, we study the total Hamiltonian given in Eq.~\eqref{eq:continuous_hamiltonian_all_terms} for two different geometries: the finite and the semi-infinite one. In the finite geometry, it is assumed that the system has finite extent in both $x$ and $y$ directions. 
In the semi-infinite geometry it is assumed that the junction is finite in the $y$ direction but extends infinitely in the $x$ direction, such that $k_x$ is a good quantum number. The discretized Hamiltonians for both geometries are given explicitly in App.~\ref{appsec:tight_binding}.

\subsection{\label{subsec:mbs}Majorana bound states}
The narrow normal section in the planar JJ can host MBSs at its ends, see Fig.~\ref{fig:setup}. This has been shown for planar JJs based on 2DEG with LSOI~\cite{hell2017two, setiawan2019topological, scharf2019tuning, pientka2017topological, volpez2020time, pakizer2021crystalline, pekerten2022anisotropic, ren2019topological, fornieri2019evidence, banerjee2022signatures}. 
In this section, we demonstrate that MBSs also appear in a Ge planar JJ with CSOI.

\begin{figure*}
	\centering
	\includegraphics[width=\linewidth]{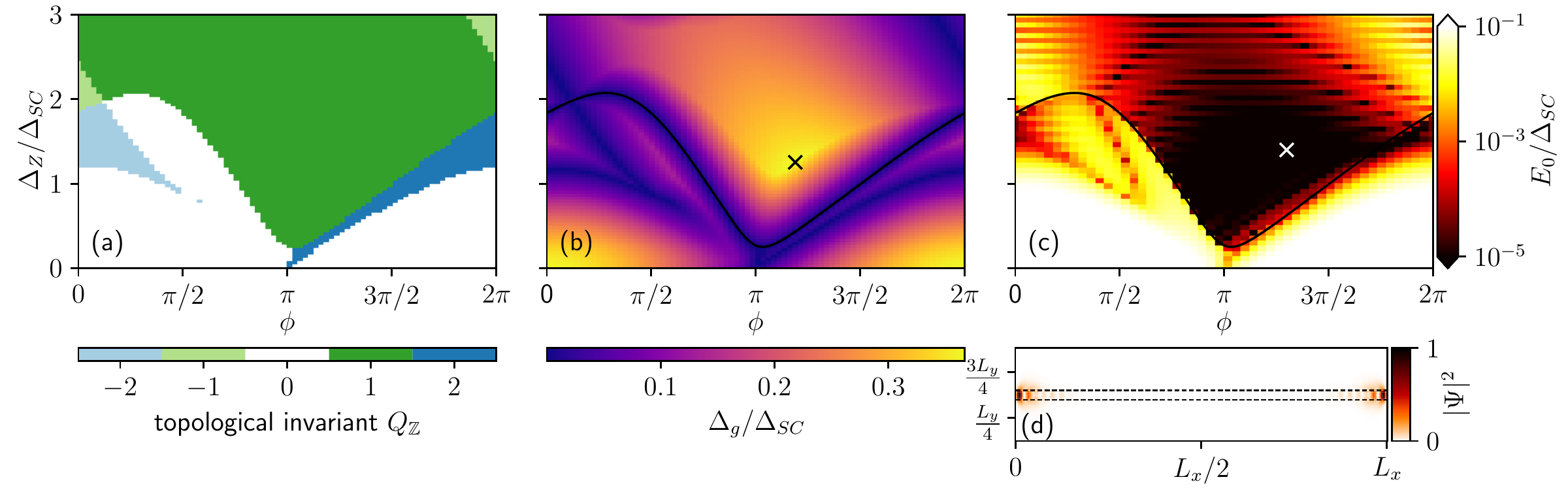}
	\caption{
		(a),~(b), and~(c): Topological invariant $Q_\mathbb{Z}$ of the BDI symmetry class, the bulk gap $\Delta_g$ [defined in Eq.~\eqref{eq:bulk_gap}], and the lowest energy level $E_0$ (calculated in the finite geometry) as functions of the Zeeman energy $\Delta_Z$ and the superconducting phase difference $\phi$.
		The black cross in panel~(b) indicates the maximum bulk gap $\Delta_m$ in the topological phase.
		The black curves in panels~(b) and~(c) indicate where the bulk gap closes at $k_x=0$.
		(d): Probability distribution $|\Psi|^2$ of a MBS (in arbitrary units).
		Boundaries between regions with different values of $Q_\mathbb{Z}$ in panel~(a) coincide with a closing of the bulk gap, i.e., $\Delta_g=0$, in panel~(b). If the parity of $Q_\mathbb{Z}$ changes, the bulk gap closes at $k_x = 0$, whereas otherwise it closes at some finite $k_x \neq 0$.
		Regions where $Q_\mathbb{Z} \neq 0$ coincide with regions where the lowest energy level $E_0$ is close to zero in panel~(c). These bound states are MBSs.
		Used parameters: 
		$L_x=199.5 l_\mathcal{E}$ [in~(c) and~(d)], 
		$L_y = 49.5 l_\mathcal{E}$, $W=4.5 l_\mathcal{E}$, 
		$\hbar^2/2m^\ast = 1.141 E_\mathcal{E} l_\mathcal{E}^2$, $\alpha = 0.39 E_\mathcal{E} l_\mathcal{E}^3$, $\beta=0.14 E_\mathcal{E} l_\mathcal{E}^4$, 
		$\mu = 0.3 E_\mathcal{E}$, $\Delta_{\text{SC}} = \mu/3$. 
		In~(d) $\phi =1.3 \pi$ and $\Delta_Z/\Delta_{\text{SC}}=1.4$ [indicated by the white cross in panel~(c)].
	}
	\label{fig:majorana_normal}
\end{figure*}

Before presenting the result of the numerical simulations, we comment on the parameters used in this section and Sec.~\ref{subsec:quasi_mbs}. For simplicity, we assume that the Ge is unstrained. Furthermore, we assume an electric field of $\mathcal{E}=1$~V$/\mu$m, resulting in $l_\mathcal{E} \approx 8$~nm and $E_\mathcal{E} \approx 8$~meV.
Instead of perturbatively calculated values for $\hbar^2/2m^\ast$ and $\alpha$, we refer to Ref.~\cite[Tab.~III]{marcellina2017spin}, where $\hbar^2/2m^\ast=584 \text{ meV nm}^2 = 1.141 E_\mathcal{E} l_\mathcal{E}^2$ and $\alpha = 800\text{ meV nm}^3 = 0.195 E_\mathcal{E} l_\mathcal{E}^3$. We use the specified value for $\hbar^2/2m^\ast$ but, to keep the localization length of the MBSs small enough for numerical simulations, we set $\alpha = 0.39\, E_\mathcal{E} l_\mathcal{E}^3$. We note that using the perturbative calculation explained in Sec.~\ref{sec:effective_ham} to derive $\hbar^2/2m^\ast$ and $\alpha$ in unstrained Ge gives similar values.
The chemical potential is pinned close to the bottom of the valence band~\cite{scappucci2020germanium} but can be tuned by gate potentials in experiments~\cite{fornieri2019evidence}. The smaller the chemical potential, the better is the fit of the effective model to the Luttinger-Kohn Hamiltonian at the Fermi surface. However, if the chemical potential is too small, there are not enough particles in the system to induce superconductivity. In addition, the CSOI is larger at large momenta. As a tradeoff, we set $\mu=0.3E_\mathcal{E} $. 
For the superconducting gap, we assume $\Delta_{\text{SC}} \equiv \Delta_{\text{SC}}^l=\Delta_{\text{SC}}^r=\mu/3 = 0.8$~meV, 
which, for the convenience of the numerical simulations, is chosen rather large compared to the superconducting gap in aluminum, i.e. $0.34$~meV~\cite[p. 268]{kittel2005kittel}. We note that the proximity effect of the superconductor might renormalize material parameters of the Ge heterostructure~\cite{reeg2018proximity, reeg2018metallization}. 
However, we expect that this does not change the qualitative behavior of the system and therefore do not consider it in this analysis.
We will limit our analysis to relatively small Zeeman fields, i.e., $\Delta_Z/\Delta_{\text{SC}}  \leq 3$.
The exact numerical values for the parameters used in the tight-binding calculations are given in App.~\ref{appsec:tight_binding}. 

As already mentioned in Sec.~\ref{subsec:setup}, if the two superconductors induce the same gap $\Delta_{\text{SC}}=\Delta_{\text{SC}}^l=\Delta_{\text{SC}}^r$, the system is in the BDI symmetry class~\cite{pientka2017topological, hell2017two, setiawan2019topological}, giving the invariant $Q_\mathbb{Z} \in \mathbb{Z}$~\cite{chiu2016classification}. Consequently, it is possible for several MBSs to be present at the same end of the system without them hybridizing. 
Following the procedure explained in Ref.~\cite{pientka2017topological}, we calculate $Q_\mathbb{Z}$ numerically as a function of the Zeeman energy $\Delta_Z$ and the superconducting phase difference $\phi$ in order to obtain  the phase diagram shown in Fig.~\ref{fig:majorana_normal}(a). This phase diagram can be complemented by calculating the bulk gap~$\Delta_g$, which comes from quasi-one-dimensional states located within the normal section. The bulk gap is therefore defined in the semi-infinite geometry as:
\begin{equation}
	\Delta_g = \min_{k_x} \left| E\left(k_x\right) \right|,
	\label{eq:bulk_gap}
\end{equation}
where $E(k_x)$ are the energy eigenvalues of the Hamiltonian in the semi-infinite geometry.
As expected, we find that the boundaries where $Q_\mathbb{Z}$ changes its value correspond to lines where $\Delta_g$ goes to zero, see Fig.~\ref{fig:majorana_normal}(b).
We note that at boundaries where the parity of $Q_\mathbb{Z}$ changes, the bulk gap closes at $k_x=0$. If $Q_\mathbb{Z}$ changes but its parity remains unchanged, the bulk gap closes at some finite value $k_x \neq 0$~\cite{pientka2017topological}. 
We denote the maximum bulk gap in the topological phase as $\Delta_m$. Its value in Fig.~\ref{fig:majorana_normal}(b) is $\Delta_m/\Delta_{\text{SC}} = 0.37$ at $\phi = 1.16\pi$ and $\Delta_Z/\Delta_{\text{SC}} = 1.22$.

Additionally, further insight can be gained by calculating the lowest-energy level $E_0$ of the system in the finite geometry, see Fig.~\ref{fig:majorana_normal}(c).
In the trivial phase, i.e., $Q_\mathbb{Z}=0$, the lowest-energy state is either a bulk state or an ABS~\cite{prada2020andreev, sauls2018andreev}, see also Sec.~\ref{subsec:quasi_mbs}. 
If $Q_\mathbb{Z} \neq 0$, the lowest energy states are the MBSs. In an infinitely long system, MBSs occur at zero energy due to particle-hole symmetry~\cite{laubscher2021majorana}. In a system with finite length, MBSs at opposite ends of the junction overlap and therefore their energy is not exactly zero but vanishes exponentially with the system length $L_x$, sometimes exhibiting oscillations~\cite{prada2012transport, rainis2013towards}.
As a first approximation, $E_0$ can therefore help to distinguish between the trivial and the topological phases: In the topological phase~$E_0$ is close to zero and in the trivial phase~$E_0$ has a finite value. 
However, the localization length of the MBSs depends on the size of the bulk gap, which varies throughout the phase diagram. If the localization length is comparable to the system length~$L_x$, the MBSs significantly overlap with each other, causing $E_0$ to split away from zero. This is the case e.g. in the upper left and right corner of Fig.~\ref{fig:majorana_normal}(c). In this case, using only the energy $E_0$ is not enough to distinguish the topological phase from the trivial phase. In addition, right after the topological phase transition, the bulk gap is still small and the MBSs have very long localization lengths. Therefore, Fig.~\ref{fig:majorana_normal}(c) shows no sharp boundaries between the trivial and topological phases. A further drawback of an $E_0$ diagram is that
a planar JJ can host trivial ABSs that are at, or close to, zero energy, see Sec.~\ref{subsec:quasi_mbs}. A plot showing $E_0$ cannot  distinguish these ABSs from MBSs.
Moreover, such a plot cannot distinguish MBSs in regions with $Q_\mathbb{Z}=\pm 1$ and $\pm 2$ from each other~\footnote{This problem is solved by considering not only the lowest energy level $E_0$ but the four lowest energy levels and counting how many are close to zero. The number of pairs of MBSs corresponds to $|Q_\mathbb{Z}|$.}. 
In addition to the above described properties of Fig.~\ref{fig:majorana_normal}(c), we observe a strong oscillation of the MBS energy as a function of the Zeeman energy $\Delta_Z$. There is also an oscillation of the MBS energy as a function of the superconducting phase difference $\phi$, but this dependence is less pronounced.
For LSOI, the MBS energy was also found to oscillate both with the Zeeman energy $\Delta_Z$ and with the phase difference $\phi$~\cite{banerjee2022signatures, fornieri2019evidence}.

In realistic systems, the effective time-reversal symmetry of Eq.~\eqref{eq:modified_time_reversal} is broken, e.g., by impurities, fabrication imperfections, or having $\Delta_{\text{SC}}^l \neq \Delta_{\text{SC}}^r$. In this case, the MBSs will hybridize locally until either no or one MBS remains on each side of the junction.
Thus, in the remainder of this paper, we will treat our system as belonging to the symmetry class D, where $Q_{\mathbb{Z}_2} = 1$ characterizes the trivial phase with no MBSs and $Q_{\mathbb{Z}_2} = - 1$ characterizes the topological phase with one pair of MBSs. Transitions between these two phases are indicated by a bulk gap closing at $k_x=0$.

\begin{figure}[t]
	\centering
	\includegraphics[width=\columnwidth]{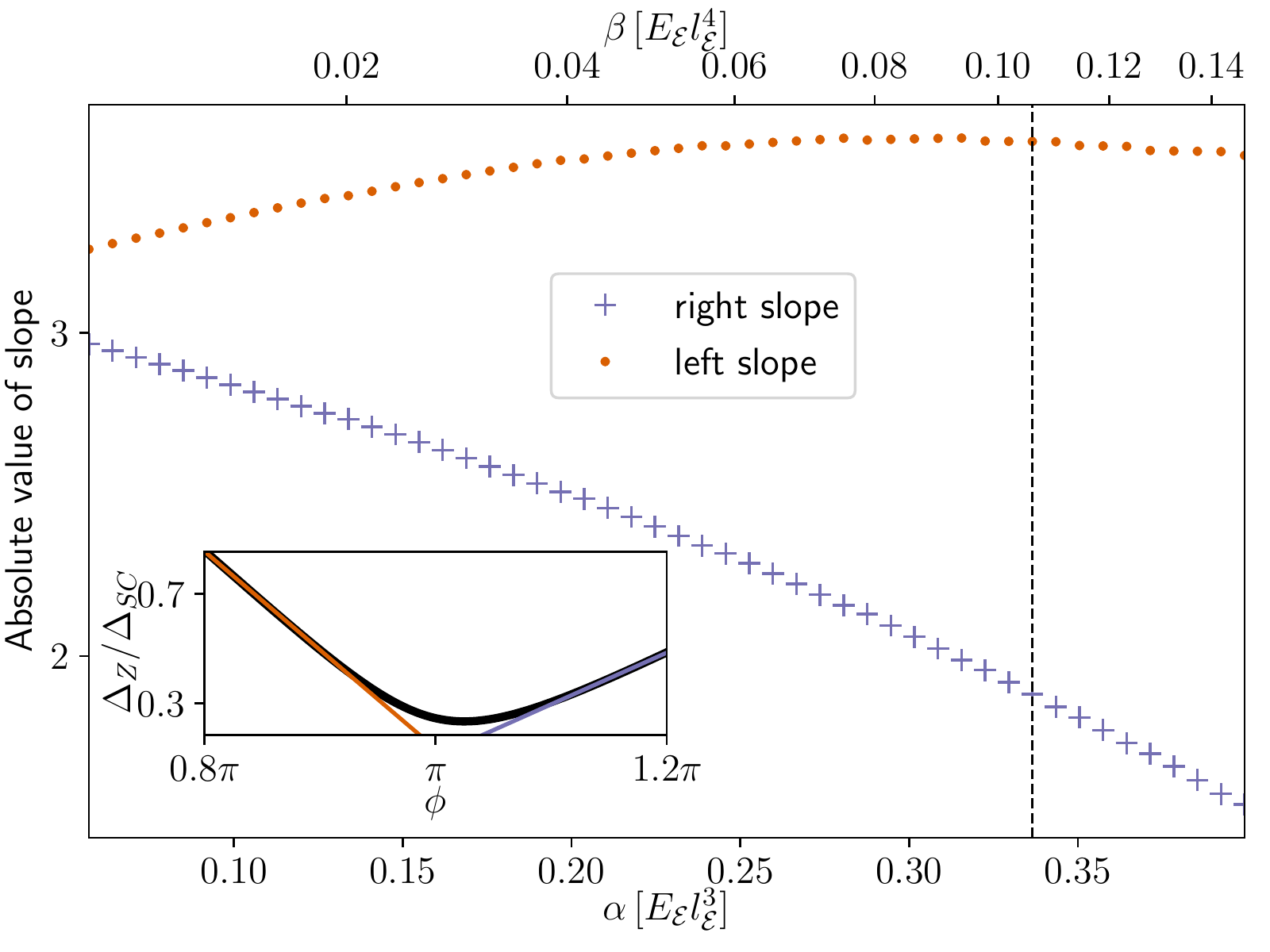}
	\caption{Fitted slope of the $\Delta_Z$ vs. $\phi$ phase transition curve for different strengths $\alpha$ of CSOI. An example of the fit is shown in the inset, where the phase transition curve for $\alpha = 0.34 E_\mathcal{E} l_\mathcal{E}^3$
		(indicated by the vertical dashed black line) and the fitted lines are shown.  The larger the CSOI strength, the more the left and right slopes differ, therefore the more asymmetric the phase diagram becomes.
		The parameters are $L_y = 49.5 l_\mathcal{E}$, $W=4.5 l_\mathcal{E}$, $\hbar^2/2m^\ast = 1.141 E_\mathcal{E} l_\mathcal{E}^2$, $\mu = 0.3 E_\mathcal{E}$, $\Delta_{\text{SC}} = \mu/3$.
		The strength $\beta$ of the quartic term is varied together with the CSOI strength $\alpha$ as follows. For several values of $\alpha$, the corresponding value of $\beta$ is determined following App.~\ref{appsec:high_momentum_fermi_surface}. Using a quadratic fit, we then set $\beta = c_0 + c_1 \alpha + c_2 \alpha^2$ with $c_0=0.0017 E_\mathcal{E} l_\mathcal{E}^4$, $c_1=0.0236 l_\mathcal{E}$ and $c_2 = 0.8507 E_\mathcal{E}^{-1} l_\mathcal{E}^{-2}$.
	}
	\label{fig:fitted_slope_cubic}
\end{figure}

At $\phi \approx \pi$, the Zeeman energy required to enter the topological phase is at its minimum, see Fig.~\ref{fig:majorana_normal}. However, this minimum is not exactly centered at $\phi = \pi$ and the curve indicating the topological phase transition is not symmetric around $\phi=\pi$. To characterize this asymmetry, we notice that the phase transition curve is parabolic near its minimum, which we denote as $\phi^{(m)}$. Further away from $\phi^{(m)}$, the curve is approximately linear, i.e., $\Delta_Z/\Delta_{\text{SC}} = p_0 + p_1 \phi$, for real coefficients $p_0$ and $p_1$. Fitting a line to the phase transition curve for superconducting phase differences $\phi < \phi^{(m)}$ ($\phi > \phi^{(m)}$) gives a slope $p_{1,l}$ ($p_{1,r}$)
\footnote{The technical details of how to fit the slopes of the phase transition curves are as follows. The Zeeman energies $\Delta_{Z,i}$ at which $E(k_x=0)=0$ are determined for 100 evenly spaced values of the superconducting phase difference $\phi_i$, with $i=1, \dots, 100$, $\phi_1=0.8\pi$ and $\phi_{100}=1.2\pi$. The slopes are determined by fitting a linear function to the pairs $(\phi_i, \Delta_{Z,i})$. The minimum Zeeman field is reached at index $M$, i.e., $\phi^{(m)} \approx \phi_M$, which is not exactly at $\pi$. Around this minimum the phase diagram is parabolic. To not distort the linear fit, points around the minimum must be excluded. The left slope is determined using the points $(\phi_i, \Delta_{Z,i})$ with $i=1, \dots, M-26$. The right slope is calculated using the points $(\phi_i, \Delta_{Z,i})$ with $i=M+26, \dots, 100$. }.
The difference between these two slopes is a measure of the asymmetry of the phase diagram. We find that the larger the CSOI strength, the more asymmetric the phase diagram becomes, see Fig.~\ref{fig:fitted_slope_cubic}.
These observations are in contrast to JJs based on 2DEGs with LSOI, where it was found that the phase diagram is symmetric around $\phi = \pi$~\cite{pientka2017topological}.
To explain this difference, it suffices to consider the Hamiltonian in the semi-infinite geometry (defined in App.~\ref{appsec:tight_binding}) at $k_x=0$ because this case determines the curve of the phase transition. We note that $\tilde{H}^\ast(k_x=0, \phi, \alpha) = \tilde{H}(k_x=0, -\phi, -\alpha)$, and, in addition, the Hamiltonian is Hermitian giving us $E(k_x=0, \phi, \alpha) = E(k_x=0, -\phi, -\alpha)$. This relation holds true for both LSOI and CSOI.
The main difference is that LSOI can be gauged away without altering the superconducting term~\cite{braunecker2010spin, pientka2017topological}, therefore giving $E(k_x=0, \phi) = E(k_x=0, -\phi)$, which implies a symmetric phase diagram around $\phi=\pi$. In contrast, CSOI cannot be gauged away and therefore the phase diagram is generally not symmetric.

\begin{figure}
	\centering
	\includegraphics[width=\columnwidth]{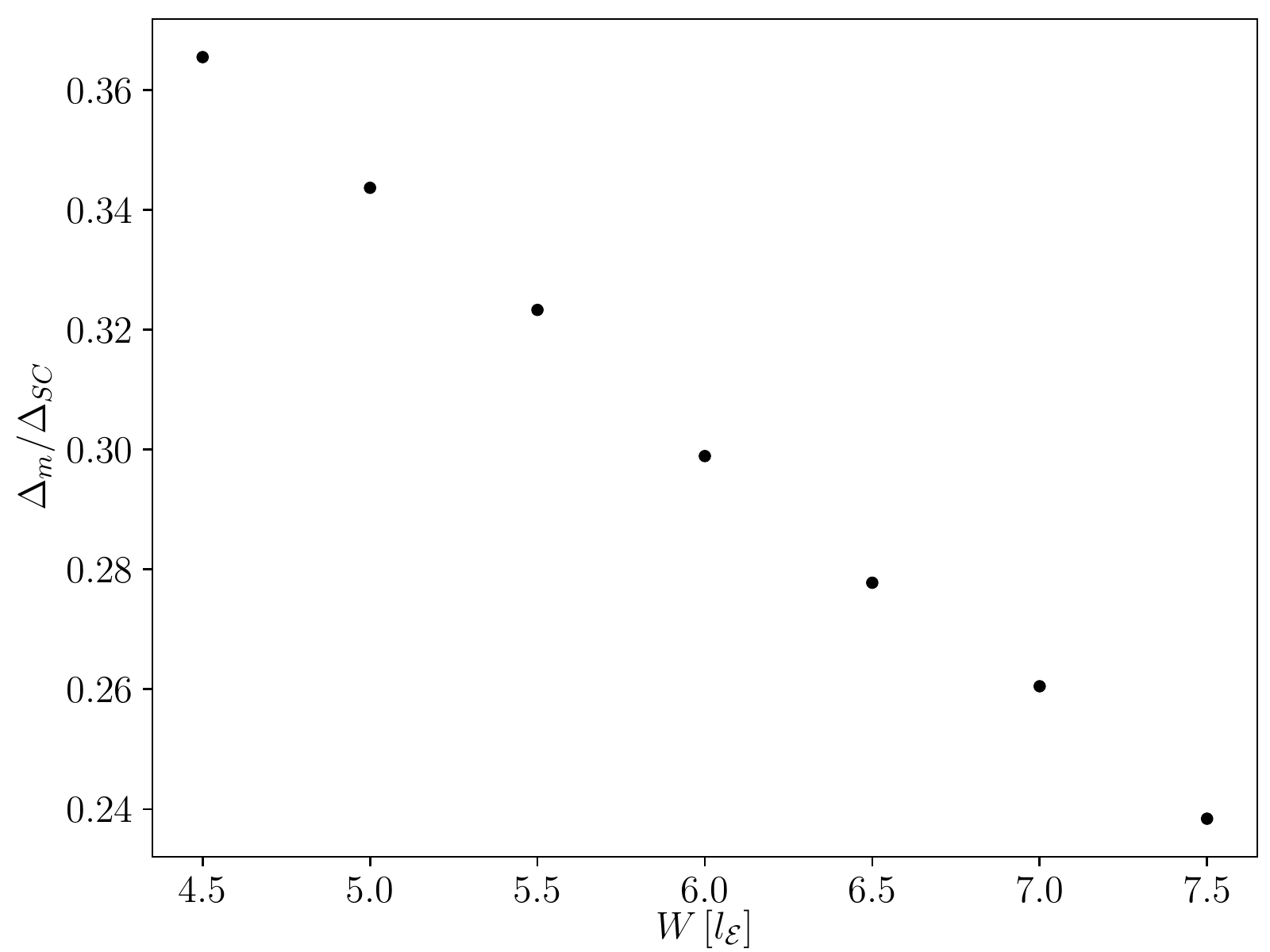}
	\caption{Largest bulk gap in the topological phase $\Delta_m$ as a function of the normal section width $W$. As the normal section becomes wider, $\Delta_m$ decreases. The parameters are
		$N_y = 49.5 l_\mathcal{E}$,
		$\hbar^2/2m^\ast = 1.14 E_\mathcal{E} l_\mathcal{E}^2$,
		$\alpha = 0.39 E_\mathcal{E} l_\mathcal{E}^3$,
		$\beta = 0.14 E_\mathcal{E} l_\mathcal{E}^4$,
		$\mu = 0.3 E_\mathcal{E}$,
		$\Delta_{\text{SC}} = \mu/3$.
	}
	\label{fig:topo_gap_vs_lw}
\end{figure}

Finally, we can also study how the normal section width, which had been kept fixed up to now, influences the topological gap and therefore the stability of the MBSs. Varying the superconducting phase difference $\phi$ and the Zeeman energy $\Delta_Z$, we determine the largest bulk gap in the topological phase $\Delta_m$ throughout the phase diagram [see Fig.~\ref{fig:majorana_normal}(b)].
We find that $\Delta_m$ decreases as the normal section becomes wider, see Fig.~\ref{fig:topo_gap_vs_lw}. A similar trend was also found for 2DEG planar JJs with LSOI in Ref.~\cite{pientka2017topological}.

\subsection{\label{subsec:quasi_mbs}Andreev bound states}

\begin{figure*}
	\centering
	\includegraphics[width=\linewidth]{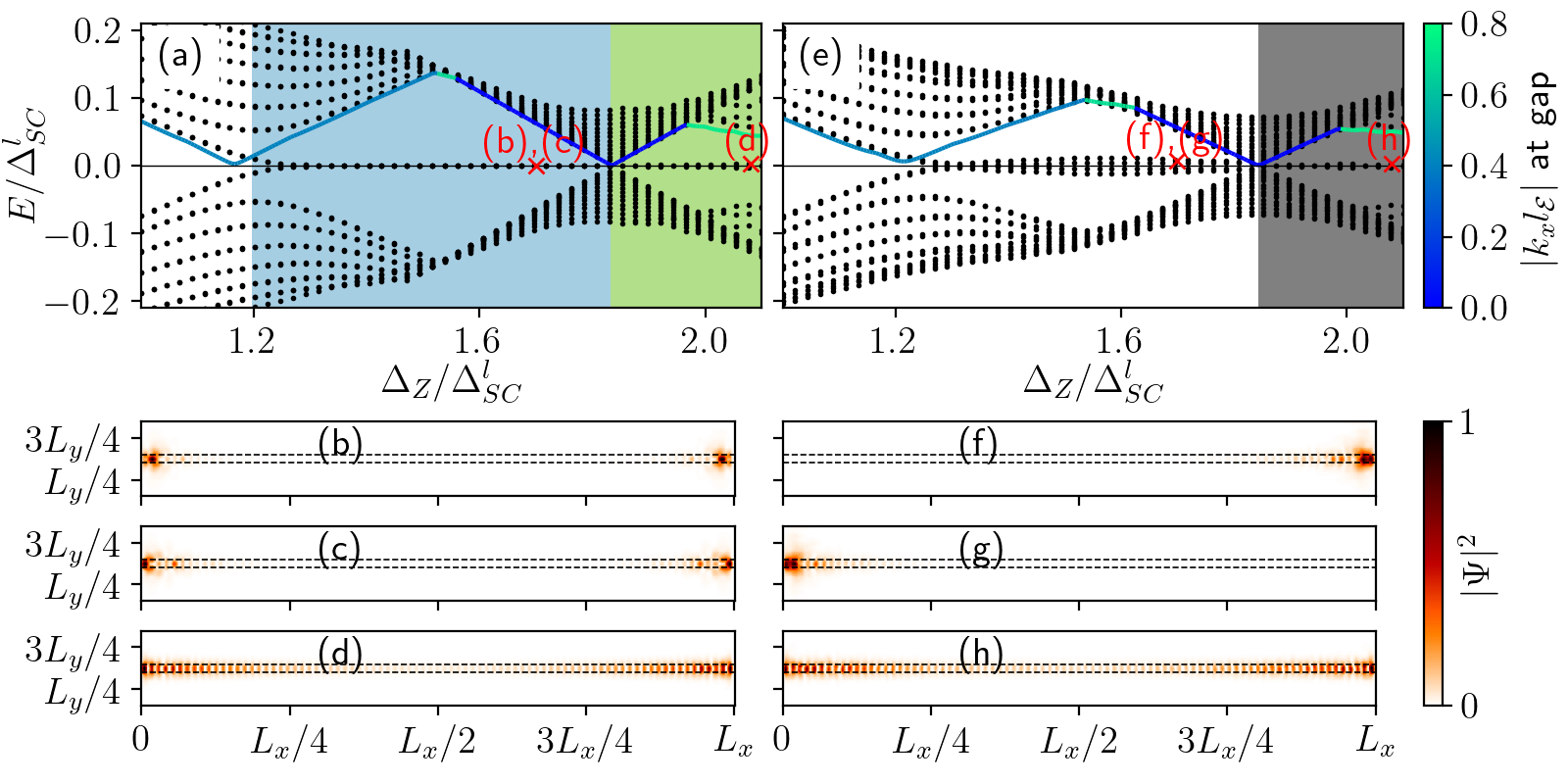}
	\caption{
		(a) and~(e):~Energy spectrum of a planar JJ as a function of the Zeeman energy $\Delta_Z$. The black dots are the energy levels of the system calculated numerically in the finite geometry. The colored lines indicate the bulk gap $\Delta_g$, defined in Eq.~\eqref{eq:bulk_gap}. The line color indicates the value of $k_x$ for which $E(k_x) = \Delta_g$.
		In~(a), the system has the effective time-reversal symmetry defined in Eq.~\eqref{eq:modified_time_reversal}. The background color in~(a) signifies the corresponding BDI topological invariant, $Q_\mathbb{Z}$.  In the white region $Q_\mathbb{Z}=0$ (trivial phase, no zero-energy states), in the green region $Q_\mathbb{Z}=-1$ (two zero-energy states, i.e., one pair of MBSs), and in the blue region $Q_\mathbb{Z}=-2$ (four zero-energy states, i.e., two pairs of MBSs). 
		In~(e) the effective time-reversal symmetry is broken by having $\Delta_{\text{SC}}^l \neq \Delta_{\text{SC}}^r$.
		In the finite geometry we also assume disorder in the chemical potential of the system (see App.~\ref{app:disorder}).
		The background color signifies the corresponding topological invariant $Q_{\mathbb{Z}_2}$ of symmetry class~D. In the white region $Q_{\mathbb{Z}_2}=1$ (trivial phase, no MBSs) and in the gray region $Q_{\mathbb{Z}_2}=-1$ (topological phase, one pair of MBSs). 
		(b)-(d)~and~(f)-(h):~Probability distributions $|\Psi|^2$ of the MBSs [panels~(b), (c), (d), and~(h)] and ABSs [panels~(f) and~(g)] (arbitrary units).
		Although the two pairs of MBSs that appear in the blue region of panel~(a) hybridize when the symmetry is broken, the resulting ABSs are still localized in both $x$ and $y$ direction and their energies are close to zero. 
		The parameters used are 
		$L_x = 349.5 l_\mathcal{E}$, 
		$L_y = 49.5 l_\mathcal{E}$, 
		$W=4.5 l_\mathcal{E}$, 
		$\hbar^2/2m^\ast = 1.141 E_\mathcal{E} l_\mathcal{E}^2$,
		$\mu = 0.3 E_\mathcal{E}$, 
		$\alpha = 0.39 E_\mathcal{E} l_\mathcal{E}^3$, 
		$\beta = 0.14 E_\mathcal{E} l_\mathcal{E}^4$, 
		$\phi = 0$.
		In~(a)-(d) $\Delta_{\text{SC}}^l=\Delta_{\text{SC}}^r=\mu/3$, in (e)-(h) $\Delta_{\text{SC}}^l=\mu/3$ and $\Delta_{\text{SC}}^r=0.6 \Delta_{\text{SC}}^l$. In (b), (c), (f), and (g) $\Delta_Z/\Delta_{\text{SC}}^l=1.7$ and in~(d) and~(h) $\Delta_Z/\Delta_{\text{SC}}^l=2.08$ [these values are indicated by the red crosses in~(a) and~(e)]. 
	}
	\label{fig:quasi_mbs_bdi_remnant}
\end{figure*}

In the previous section we analyzed MBSs, which are zero-energy states that are present in the topological phase. 
However, it is possible to have ABSs that are at, or close to, zero energy in the trivial phase over a wide range of Zeeman energies $\Delta_Z$. 
Such states can mimic the experimental signatures of topological MBSs, as was extensively investigated for nanowires~\cite{kells2012near, moore2018two, moore2018quantized,  huang2018metamorphosis, penaranda2018quantifying, reeg2018zero, vuik2019reproducing, sarma2021disorder, hess2021local}. In this section, we give two examples of how such ABSs can arise in a planar JJ.

One way such ABSs appear in the system is as remnants of the BDI symmetry class. As already explained in Sec.~\ref{subsec:setup}, if $\Delta_{\text{SC}}^l = \Delta_{\text{SC}}^r$, the system has an effective time-reversal symmetry and is in the BDI symmetry class. In this case there are regions in parameter space with a pair of MBSs at each end of the junction, see Fig.~\ref{fig:quasi_mbs_bdi_remnant}(a)-\ref{fig:quasi_mbs_bdi_remnant}(d). When the effective time-reversal symmetry is broken, the two pairs of MBSs hybridize locally and their energies split away from zero. Even though these states are now trivial ABSs, they are still well localized to the ends of the junction and their energies can still be sufficiently close to zero to be misinterpreted as MBSs in an experiment, see Fig.~\ref{fig:quasi_mbs_bdi_remnant}(e)-\ref{fig:quasi_mbs_bdi_remnant}(h).

\begin{figure*}
	\centering
	\includegraphics[width=\linewidth]{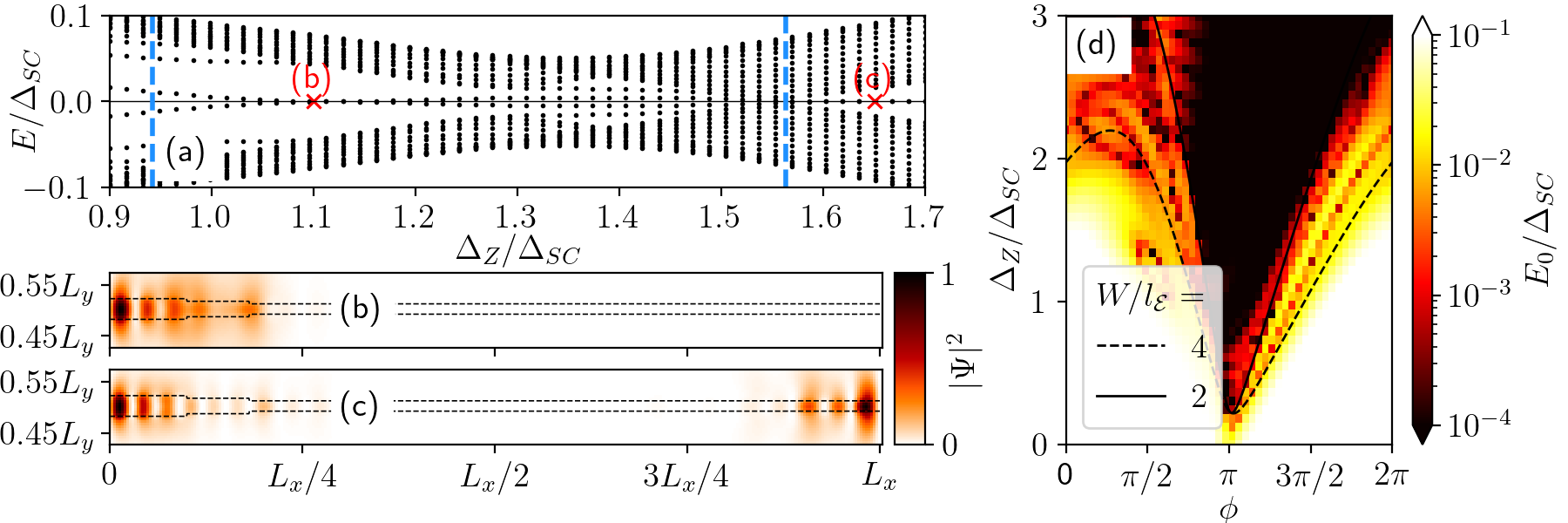}
	\caption{
		The normal section width $W(x)$ of a planar JJ is varied according to Eq.~\eqref{eq:tanh_variation}.
		(a):~Energy spectrum (calculated in the finite geometry) as a function of the Zeeman energy $\Delta_Z$. The blue vertical dashed lines indicate $\Delta_{Z,l}$ and $\Delta_{Z,r}$.
		(b)~and~(c):~Probability distribution $|\Psi|^2$ (arbitrary units) of the lowest-energy states. 
		The horizontal dashed lines indicate the boundaries of the superconductors. 
		(d):~Lowest energy level $E_0$ (calculated in the finite geometry) as a function of the Zeeman energy $\Delta_Z$ and the superconducting phase difference $\phi$. 
		The black dashed (solid) curve indicates $\Delta_{Z,l}$ ($\Delta_{Z,r}$).
		In the region of parameter space where $\Delta_Z \in [\Delta_{Z,l}, \Delta_{Z,r}]$, near-zero-energy ABSs appear [panel~(b)].
		The ABSs are localized to the left side of the junction only. Once the bulk of the system is in the topological phase, MBSs appear [panel~(c)] and they are localized on both ends of the junction.
		The parameters used are: $L_x=149.5 l_\mathcal{E}$, $L_y=99.5 l_\mathcal{E}$ in panels (a)-(c) and $L_y=49.5 l_\mathcal{E}$ in panel (d), 
		$W_{-\infty} = 4.5 l_\mathcal{E}$, $W_{\infty}= 1.5 l_\mathcal{E}$, $\sigma=17.5 l_\mathcal{E}$, $p=21 l_\mathcal{E}$, 
		$\hbar^2/2m^\ast= 1.141 E_\mathcal{E} l_\mathcal{E}^2$, 
		$\mu=0.3 E_\mathcal{E}$, $\Delta_{\text{SC}}=\mu/3$, $\phi=0.8\pi$,
		$\alpha= 0.39 E_\mathcal{E} l_\mathcal{E}^3$, $\beta= 0.14 E_\mathcal{E} l_\mathcal{E}^4$, 
		$\Delta_Z/\Delta_{\text{SC}}=1.1$ [panel~(b), indicated by the red cross in panel (a)], $\Delta_Z/\Delta_{\text{SC}}=1.65$ [panel~(c), indicated by the red cross in panel~(a)]. 
		Note that $W_{\pm \infty}$ and the widths~$W$ used in panel~(d) are related via Eqs.~\eqref{eq:tanh_variation} and~\eqref{eq:rounding_junction_width}, see App.~\ref{appsubsec:finite}. 
	}
	\label{fig:quasi_mbs_junction_width}
\end{figure*}

Near-zero-energy ABSs can also occur if there is disorder at the ends of the junction, resulting in spatially varying system parameters~\cite{moore2018two, moore2018quantized, penaranda2018quantifying, vuik2019reproducing,hess2021local}. As an example, we model a system with a spatially nonuniform normal section width. For a planar JJ with LSOI, it was shown that the normal section width influences the value of $\Delta_Z$ at which the phase transition occurs~\cite{hell2017two, volpez2020time}, and we observe this behavior also in the presence of CSOI. 
We focus on the case when the normal section is wider at the left side of the junction, see Fig.~\ref{fig:quasi_mbs_junction_width}. 
We vary the width of the normal section according to: 
\begin{equation}
	W(x) \!=\! 
	W_{-\infty} \!+\! \frac{W_{\infty}\!-\!W_{-\infty}}{2} \! \left[ \tanh \! \left( \! \frac{x \! -\! p}{\sigma} \!\right) \!+ \!1 \right] ,
	\label{eq:tanh_variation} 
\end{equation}
where $W_{\pm \infty}$ are positive parameters, $\sigma$ characterizes how abruptly the width changes, and $p$ is the point at which $W(p) = \left(W_{-\infty}+W_{\infty}\right)/2$. We choose $p$ to be close to the left side of the junction and $\sigma$ small compared to the MBS localization length. 
In this case, the normal section width is equal to $W(x=L_x)$ throughout the entire system except at the left end of the junction, where the spatial variation of the width acts as a defect. The bulk of the system with normal section width $W(x=L_x)$ enters the topological phase at Zeeman energy $\Delta_{Z,r}$. In contrast, a system with a uniform normal section width of $W(x=0)$ would enter the topological phase at Zeeman energy $\Delta_{Z,l}$.
For the parameters chosen in Fig.~\ref{fig:quasi_mbs_junction_width} one gets $\Delta_{Z,l} < \Delta_{Z,r}$, such that, for $\Delta_Z \in \left[\Delta_{Z,l}, \Delta_{Z,r} \right]$, the bulk of the system is trivial. However, trivial ABSs that are localized to the defect on the left side of the junction emerge. These states have energies close to zero, see Figs.~\ref{fig:quasi_mbs_junction_width}(a) and~\ref{fig:quasi_mbs_junction_width}(b).
For Zeeman energies $\Delta_Z > \Delta_{Z,r}$, the bulk of the system enters the topological phase and MBSs appear on both sides of the junction, see Fig.~\ref{fig:quasi_mbs_junction_width}(c).

To see the relation between the ABSs and $\Delta_{Z,l/r}$ better, consider the phase diagram in Fig.~\ref{fig:quasi_mbs_junction_width}(d). The black solid line indicates $\Delta_{Z,r}$, where the bulk enters the topological phase. The black dashed line indicates $\Delta_{Z,l}$. The region between these two lines is the region in parameter space where the near-zero-energy ABSs appear.

We stress that a nonuniform normal section width is only one possibility of modeling this sort of ABSs. Spatial variations of other parameters can have a similar effect, if the value of the corresponding parameter influences the Zeeman energy $\Delta_Z$ at which the phase transition occurs.

\section{\label{subsec:experiment}Experimental feasibility}
In Sec.~\ref{subsec:mbs} we demonstrated that Ge planar JJs can be brought into the topological phase either by varying the Zeeman energy $\Delta_Z$ or the superconducting phase difference $\phi$.
The model and parameters used were based on a simplified minimal model, which allowed us to characterize the main qualitative features of the phase diagram.
In this section, we analyze if the Ge planar JJ can also host MBSs with a more sophisticated model and experimentally realistic parameters.

So far, we neglected strain in the Ge heterostructure. However, Ge heterostructures are generally strained in experiments, which reduces the strength of the CSOI~\cite{terrazos2021theory}. This changes the shape of the phase transition curve, the bulk gap, and the localization length (in $x$ direction) of the MBSs.
Furthermore, to accurately model the valence band of Ge, we include the Luttinger-Kohn anisotropy, which depends on $\gamma_3-\gamma_2$. In this case, the choice of the confinement axis is no longer arbitrary. If we keep the crystallographic $\langle 001 \rangle$ axis as the confinement axis, the effective Hamiltonian defined in Eq.~\eqref{eq:effective_hamiltonian_isotropic} obtains an additional CSOI term, see App.~\ref{appsec:anisotropic_spectrum}.
Choosing a different confinement axis will give additional CSOI and even LSOI terms~\cite{xiong2021emergence}, if strain is included.  
Moreover, the superconducting gap $\Delta_{\text{SC}}$ is typically smaller than the value $\Delta_{\text{SC}} = 0.8$~meV used in Sec.~\ref{sec:planar_jj}.
Instead, in this section we use $\Delta_{\text{SC}} = 0.486$~meV, a value achieved experimentally in Ref.~\cite{aggarwal2021enhancement} by proximitizing Ge with superconducting aluminum and niobium. 
Further, for simplicity, in the previous section we assumed that the Zeeman energy is zero in the superconducting regions. A non-zero Zeeman energy in the superconducting regions is more realistic and thus we replace Eq.~\eqref{eq:definition_delta_z_continuous} by:
\begin{equation}
	\Delta_{Z}\left(y\right) = \Delta_Z \quad \forall \, y \in \left[0, L_y\right] .
\end{equation}
This choice also decreases the induced superconducting gap in the superconducting regions. 
Generally, a smaller superconducting gap results in a longer localization length of the MBSs and ABSs in $y$ direction. 
Finally, the maximum value of the external magnetic field is limited by the critical magnetic field $B_c$ of the superconductor. Following Ref.~\cite{aggarwal2021enhancement}, we consider $B_c = 1.8$~T.
This poses the most severe limitation on the experimental feasibility of MBSs in Ge planar JJs because the in-plane $g$ factor of holes in Ge is rather small, with experimentally measured values ranging from $0.2$ to $1.4$~\cite{watzinger2016heavy, lu2017effective, hendrickx2018gate, hofmann2019assessing, gao2020site, hendrickx2020fast, scappucci2020germanium}.

In the following, we first analyze the same experimental setup as in Sec.~\ref{sec:planar_jj}, see Fig.~\ref{fig:setup}. Then we propose two different setups that overcome the issues caused by the small $g$ factor. First, a different confinement axis is chosen, giving LSOI and CSOI terms and enhancing the $g$ factor. 
A second method to overcome the small in-plane $g$ factor is to proximitize the Ge with a ferromagnetic insulator, such that the Zeeman term is induced via the ferromagnetic proximity effect~\cite{vaitiekenas2021zero}.
For all three cases we assume 
an electric field $\mathcal{E}=10$~V$/\mu$m, giving $l_\mathcal{E}=3.7$~nm and $E_\mathcal{E}=37$~meV. 
We consider a lightly strained Ge/SiGe heterostructure, where $E_s \approx 9.6$~meV, i.e. $E_s/E_\mathcal{E} = 0.26$ (compare to Ref.~\cite{lodari2022lightly} where $E_s \approx 17$~meV).
All other parameters are given in App.~\ref{appsec:parameters_experiment}.

\subsection*{Case 1: Same setup as in Sec.~\ref{subsec:setup}}
In the first scenario, the setup is the same as described in Sec.~\ref{subsec:setup} but with experimentally realistic parameters. Additionally, we no longer neglect the anisotropy of Ge, which is why the effective Hamiltonian acquires an additional term, see Eq.~\eqref{eq:effective_hamiltonian_anisotropic}. 

\begin{figure}
	\centering
	\includegraphics[width=\columnwidth]{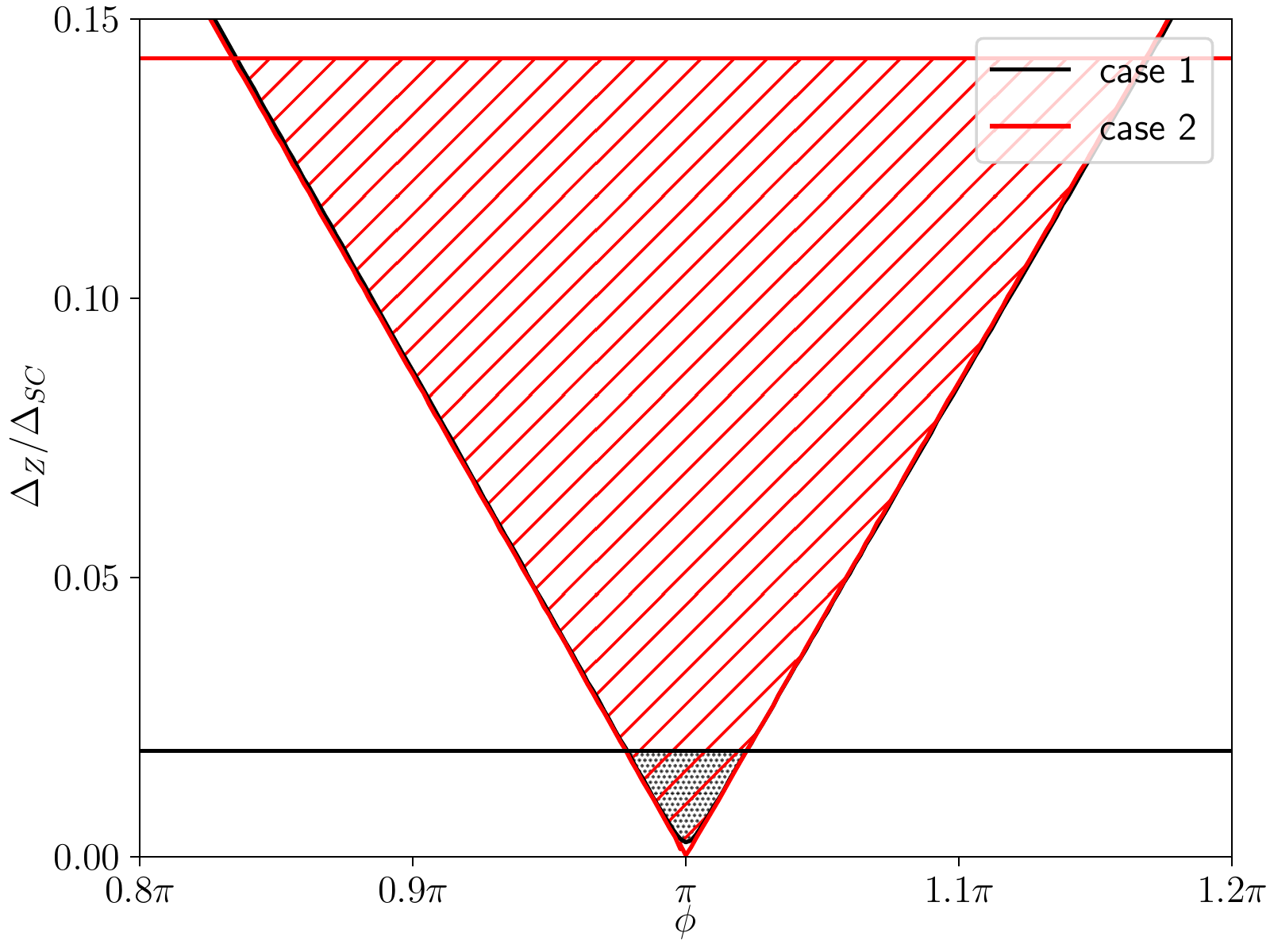}
	\caption{Phase diagram showing the closing of the bulk gap at $k_x=0$, i.e., the line $E(k_x=0)=0$ (calculated in the semi-infinite geometry) as a function of the Zeeman field $\Delta_Z$ and the superconducting phase difference $\phi$ for cases~1 and~2 discussed in Sec.~\ref{subsec:experiment}. In case~1 the confinement axis of the 2DHG is the crystallographic $\langle 001 \rangle$ axis, in case~2 it is the $\langle 110 \rangle$ axis. The horizontal lines indicate the critical Zeeman energies $\Delta_{Z,c}$ for each case. The shaded regions indicate the experimentally accessible topological regions. The parameters are given in App.~\ref{appsec:parameters_experiment}.
 }
	\label{fig:experimental_feasibility_phase_diagram_zeeman_everywhere}
\end{figure}

Calculating the $g$ factor perturbatively~\cite{bosco2021squeezed, bosco2021hole, bosco2021fully}, we find that the Ge 2DHG has a small in-plane $g$ factor of approximately $g \approx 0.18$ consistent with experimental measurements~\cite{watzinger2016heavy, lu2017effective, drichko2018effective, hendrickx2018gate, hofmann2019assessing, gao2020site, hendrickx2020fast, scappucci2020germanium},
giving a critical Zeeman field $\Delta_{Z,c}/\Delta_{\text{SC}} \approx 0.019$. In this case, it is possible to find a region in parameter space where the topological phase is accessible within the limitations discussed in the previous section, but this region in parameter space is rather small and $\phi$ is required to stay close to $\pi$, see Fig.~\ref{fig:experimental_feasibility_phase_diagram_zeeman_everywhere}. 
Since one cannot move far into the topological phase, the MBSs have a large localization length and the bulk gap is at most $\Delta_m/\Delta_{\text{SC}} = 0.004$ in the experimentally accessible topological phase. 
We also note that the exact shape of the phase transition curve, and thus especially its minimum, depends on the normal section width. In Fig.~\ref{fig:experimental_feasibility_phase_diagram_zeeman_everywhere}, we display results for $W=24 l_\mathcal{E}$, however, already $W= 22.5 l_\mathcal{E}$ will result in no experimentally accessible topological phase.
As was shown in Sec.~\ref{subsec:mbs}, the bulk gap in the topological phase can be increased by narrowing the normal section width. However, because the width of the normal section $W$ also affects the shape of the phase transition curve, decreasing $W$ can lead to overall smaller accessible bulk gaps. Because of this, we are not able to find a $W$ that increases the bulk gap in the experimentally accessible topological phase.

\subsection*{Case 2: Different confinement axis}
In case 1, the confinement axis was aligned to the crystallographic $\langle 001 \rangle$ axis. This leads to an effective Hamiltonian that has only CSOI terms and a small effective $g$ factor. 
A different choice of confinement axis can help in different ways.
Using perturbation theory~\cite{bosco2021squeezed, bosco2021hole, bosco2021fully}, we find that confinement along the crystallographic $\langle 110 \rangle$ axis increases the in-plane $g$ factor up to $g \approx 1.33$.
Additionally, in a strained Ge/SiGe quantum well, the effective Hamiltonian for this confinement direction contains LSOI terms~\cite{xiong2021emergence} and additional quadratic and CSOI terms, see Eq.~\eqref{eq:effective_ham_linear_and_cubic}. 
Although the effective Hamiltonian is different compared to case~1, the phase diagram does not change significantly, see Fig.~\ref{fig:experimental_feasibility_phase_diagram_zeeman_everywhere}. However, due to the larger $g$ factor,  the available topological phase space is larger than in case~1. Nevertheless, one is still limited to superconducting phase differences $\phi$ that are close to $\pi$ and the largest accessible bulk gap in the topological phase is $\Delta_m/\Delta_{\text{SC}} = 0.01$.
For the same reason as in case~1, we do not find an increased experimentally accessible bulk gap in the topological phase when decreasing the normal section width $W$.

\subsection*{Case 3: Ferromagnetic proximity effect}
\begin{figure}
	\centering
	\includegraphics[width=\columnwidth]{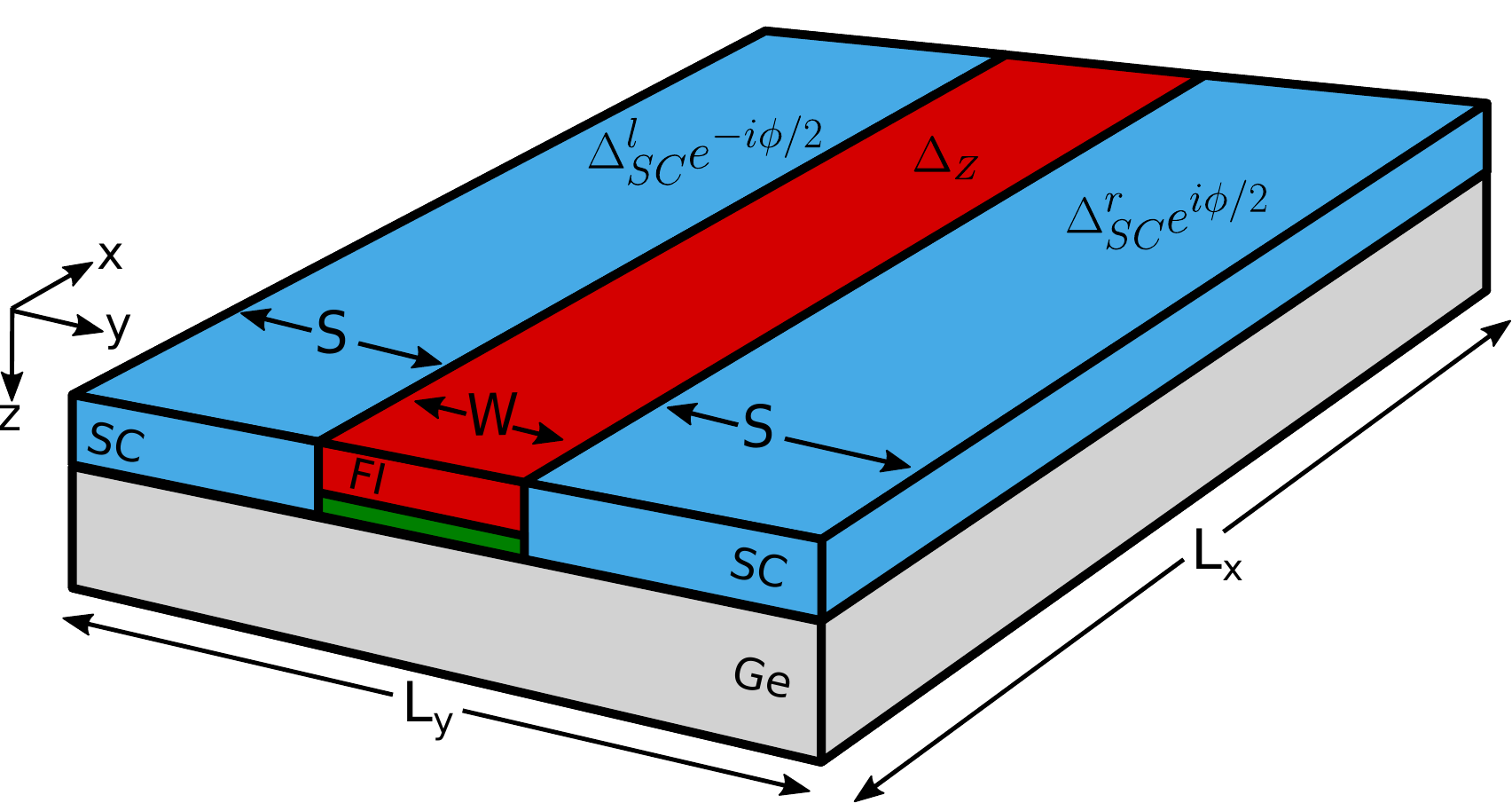}
	\caption{Setup for case~3 of Sec.~\ref{subsec:experiment}. The Zeeman energy~$\Delta_Z$ is induced by proximitizing the Ge (gray) with a ferromagnetic insulator (FI, red), e.g. EuS~\cite{vaitiekenas2021zero}. The strength of the induced Zeeman coupling can be adapted by inserting a thin layer of insulating material (green) between the ferromagnetic insulator and the Ge.}
	\label{fig:adapted_setup_ferromagnet}
\end{figure}

To overcome the limitations posed by the small $g$ factor of Ge, we propose to instead induce a Zeeman energy by proximitizing the Ge with a ferromagnetic insulator, see Fig.~\ref{fig:adapted_setup_ferromagnet}. This approach is motivated by the experimental achievements reported in Ref.~\cite{vaitiekenas2021zero}, where an InAs nanowire is proximitized by both a superconductor (Al) and a ferromagnetic insulator (EuS), inducing both superconducting and Zeeman gaps in the nanowire. In such a setup for Ge, the strength of the induced Zeeman energy is independent of the $g$ factor of Ge and can be adapted by introducing a thin layer of insulating material between the ferromagnetic insulator and the Ge. Once the structure is fabricated, the Zeeman energy is fixed. However, by varying the superconducting phase difference, the system can still be tuned in and out of the topological phase.

We compare the accessible bulk gaps in the topological phase to values for a 2DEG planar JJ with LSOI, with the corresponding Hamiltonian and parameters given in App.~\ref{appsec:parameters_experiment}. The largest accessible bulk gap for $\Delta_Z < \Delta_{\text{SC}}$ is $\Delta_m/\Delta_{\text{SC}}=0.2$ for a normal section width $W=24 l_\mathcal{E}$. Decreasing the width of the normal section to $W= 11.5 l_\mathcal{E}$ increases the bulk gap to $\Delta_m/\Delta_{\text{SC}} = 0.3$. 
Thus, the bulk gaps resulting from CSOI are smaller than the gaps resulting from LSOI. However, finding appropriate parameters for case~3 might bring the bulk gap within the same order of magnitude as the value for LSOI, therefore bringing MBSs in Ge within experimental reach.

To conclude this section, we note that, so far, we did not consider the possible presence of a potential barrier at the interface between the superconducting region and the normal region. Introducing such a barrier changes the shape of the phase diagram~\cite{pientka2017topological}. This is particularly detrimental to case~1 because its experimentally accessible topological phase space is small even without a potential barrier. The effects of a potential barrier are discussed in App.~\ref{appsec:potential_barrier}.

\section{\label{sec:conclusion}Conclusion}
We have numerically simulated a planar JJ comprising a Ge hole gas. 
The Ge 2DHG is described by a discretized effective Hamiltonian containing CSOI terms, where we introduce a fictitious $k^4$ term to remove unphysical Fermi surfaces.
Using this Hamiltonian, we show that a Ge planar JJ with CSOI can host MBSs, as was previously shown for planar JJs based on 2DEGs with LSOI. However, the topological phase diagram in dependence on the Zeeman energy $\Delta_Z$ and the superconducting phase difference $\phi$ is generally asymmetric under $\phi \rightarrow -\phi$ for CSOI, whereas it is symmetric for LSOI.
We also demonstrate that near-zero energy ABSs can be present in the system, either as a remnant of a symmetry or due to spatially nonuniform parameters. These ABSs may imitate the experimental signatures of MBSs. Further, we use experimentally realistic parameters to assess the possibility of entering the topological phase within experimental limitations. The main drawback of using Ge is its small in-plane $g$ factor. We proposed two methods to increase the $g$ factor of the system. 
Alternatively, it was argued in Refs.~\cite{lesser2021three, lesser2022one} that a superconductor-normal-superconductor-normal-superconductor (SNSNS) junction can be brought into the topological phase simply by tuning the relative phases of the superconductors, therefore requiring no Zeeman field. Thus, a Ge-based SNSNS junction would not be disadvantaged by the small in-plane $g$ factor.

We showed that the width of the normal section influences the maximum bulk gap in the topological phase in a Ge planar JJ. This was also shown for 2DEG planar JJs with LSOI~\cite{pientka2017topological}. However, the normal section width is only one parameter that can be optimized to get a maximally large bulk gap. In Refs.~\cite{scharf2019tuning, pakizer2021crystalline, pekerten2022anisotropic} it was demonstrated that the orientation of the in-plane magnetic field relative to the junction and the orientation of the junction relative to the crystal structure of the material influences the bulk gap. Further, it was shown in Refs.~\cite{laeven2020enhanced, paudel2021enhanced, melo2022greedy} that a zigzag shaped or otherwise spatially modulated junction geometry can also increase the bulk gap. 
While these observations were made for 2DEG planar JJs with LSOI, similar modifications might also help to make MBSs in Ge planar JJs experimentally more accessible. It would be interesting to see to which extent these modifications change the bulk gap in Ge planar JJs.

Our analysis shows that a Ge planar JJ is a promising platform to host MBSs. Due to its many favorable properties, Ge could also prove to be suitable platform for other types of MBSs, e.g., MBSs in thin semiconducting films~\cite{sau2010generic, alicea2010majorana, sau2010non}.

\begin{acknowledgments}
We thank Richard Hess and Henry F. Legg for helpful discussions.
This work was supported by the Swiss National Science Foundation, NCCR QSIT, and NCCR SPIN (Grant No. 51NF40-180604). This project received funding from the European Union’s Horizon 2020 research and innovation program (ERC Starting Grant, Grant Agreement No.~757725).
\end{acknowledgments}

\appendix

\section{\label{appsec:anisotropic_spectrum}Anisotropic spectrum}
In this appendix, the anisotropic corrections to the Luttinger-Kohn Hamiltonian are introduced and we study how they affect the effective Hamiltonian. 
The anisotropic Luttinger-Kohn Hamiltonian is given by~\cite{luttinger1956quantum, winkler2003spin, bosco2021squeezed}:
\begin{eqnarray}
	H_{\text{LK}}  & = &    \frac{\hbar^2}{m}  \Bigg[ 
	\left(\gamma_1 +  \frac{5 \gamma_2}{2} \! \right)  \frac{\vec{k}^2}{2}
	 - \gamma_2 \left( 
	k_x^2 J_x^2 + k_y^2 J_y^2 + k_z^2 J_z^2
	\right) 
	\nonumber \\ &&
	-  2\gamma_3 \! \left(
	k_x k_y \! \left\{J_x, J_y \right\}
	 \!+\! k_y k_z \! \left\{J_y, J_z \right\} 
	 \!+\! k_z k_x \! \left\{J_z, J_x \right\}
	\right)
	\Bigg] \nonumber\\
	&&- \mu 
	-E_s J_z^2 ,
	\label{eq:lk_hamiltonian_anisotropic}
\end{eqnarray}
where $\vec{k}=(k_x, k_y, k_z)$ is the momentum vector, the $\gamma_i$ are the Luttinger parameters~\cite{winkler2003spin}, and the anti-commutator is defined as $\{A,B\}=(AB+BA)/2$. We have added strain explicitly in this Hamiltonian to emphasize its importance for the following discussion.
The coordinate axes are aligned to the main crystallographic axes of Ge~\cite{winkler2003spin, marcellina2017spin}. 
For an isotropic material $\gamma_2=\gamma_3=\gamma_s$, which will give Eq.~\eqref{eq:lk_hamiltonian_isotropic}. For Ge, the isotropic approximation is appropriate because the difference $\gamma_3-\gamma_2$ is about 10\% of $\gamma_1$~\cite{marcellina2017spin}. 
In the following, we consider what the effect on the effective Hamiltonian is if one does not want to neglect the anisotropy.
Because the system is anisotropic, the choice of the confinement axis is not arbitrary. Choosing the crystallographic~$\langle 001\rangle$~axis as the confinement direction and following the steps discussed in Sec.~\ref{sec:effective_ham} gives the following anisotropic effective Hamiltonian~\cite{terrazos2021theory, marcellina2017spin}:
\begin{eqnarray}
	H_{\text{eff},a} = H_\text{eff}
	+ i \alpha_a \left(k_+^2 k_- \sigma_+ - k_-^2 k_+ \sigma_- \right),
	\label{eq:effective_hamiltonian_anisotropic}  
\end{eqnarray}
where $H_\text{eff}$ is the isotropic effective Hamiltonian defined in Eq.~\eqref{eq:effective_hamiltonian_isotropic} and $|\alpha_a/\alpha| = |\gamma_3-\gamma_2|/(\gamma_3+\gamma_2)$~\cite{marcellina2017spin}. Therefore, $\alpha_a=0$ in the isotropic case, which reproduces Eq.~\eqref{eq:effective_hamiltonian_isotropic}. 

\begin{figure}
	\centering
	\includegraphics[width=0.7\columnwidth]{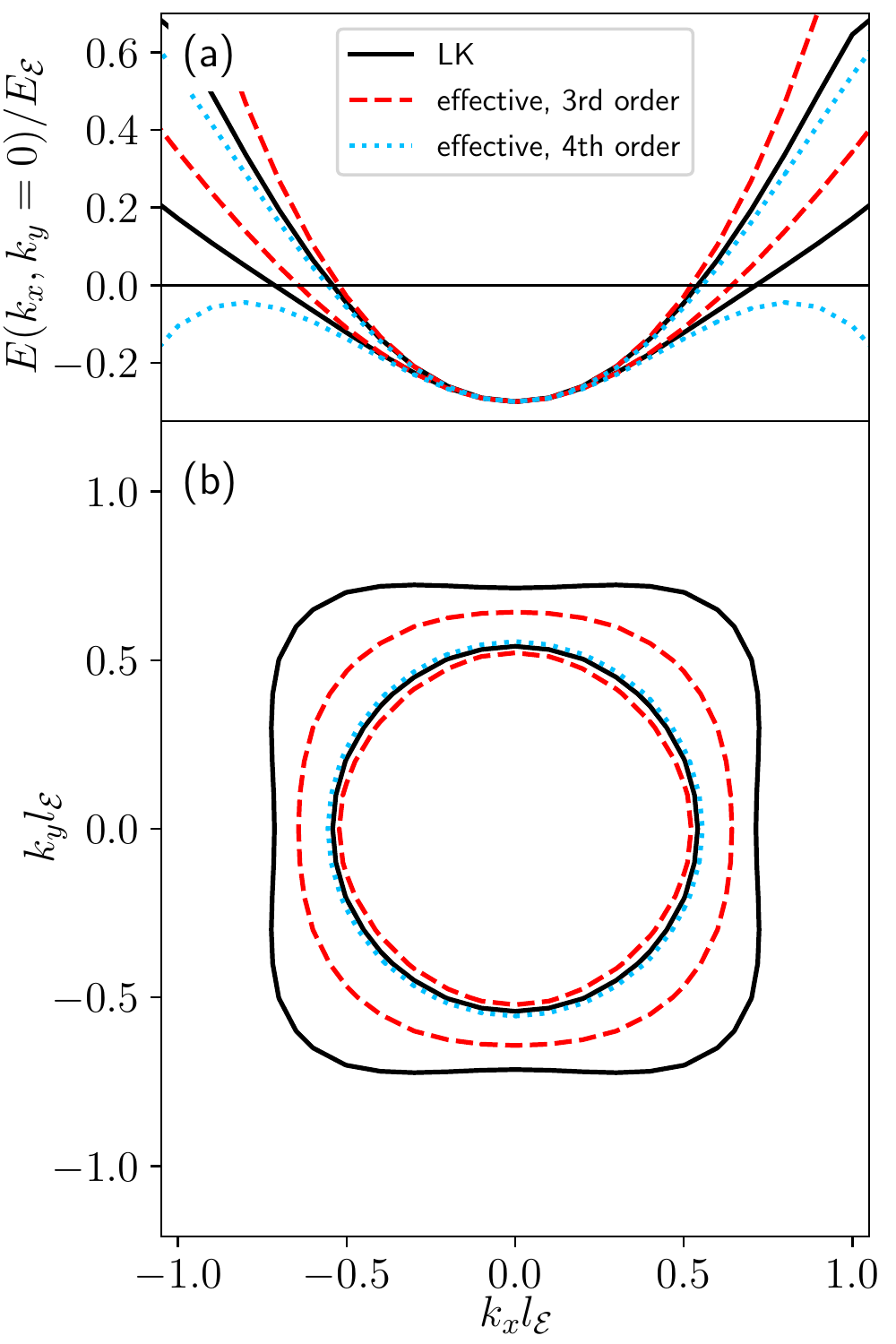}
	\caption{
		(a): Spectrum $E(k_x, k_y=0)$ of the 2DHG in anisotropic Ge [see Eqs.~\eqref{eq:lk_hamiltonian_anisotropic} and ~\eqref{eq:effective_hamiltonian_anisotropic}].
		(b): Fermi surface. 
		The black solid lines correspond to the two lowest energy bands of the Luttinger-Kohn (LK) Hamiltonian of Eq.~\eqref{eq:lk_hamiltonian_anisotropic} confined to two dimensions. The red dashed lines are the energy eigenvalues of the effective Hamiltonian $H_\text{eff,a}+H_4$ defined in Eqs.~\eqref{eq:effective_hamiltonian_anisotropic} and~\eqref{eq:quartic_k}. The blue dotted line corresponds to the effective Hamiltonian expanded up to fourth order in perturbation theory, which corresponds to $H_{\text{eff}, a}^{(4)}$ defined in Eq.~\eqref{eq:fourth_order_effective_ham}.
		The parameters for the Luttinger-Kohn Hamiltonian are $E_s=0$, $\gamma_1=13.38$, $\gamma_2=4.24$, and $\gamma_3=5.69$~\cite{winkler2003spin}. Due to the normalization, the spectrum is independent of the electric field $\mathcal{E}$.
		The parameters for the effective model are determined using perturbation theory~\cite{winkler2003spin} and they are: $\hbar^2/2m^\ast = 0.906 E_\mathcal{E} l_\mathcal{E}^2$, $\alpha=0.189 E_\mathcal{E} l_\mathcal{E}^3$, $\alpha_a/\alpha=0.146$, $\mu=0.3 E_\mathcal{E}$, $\beta = 0.06 E_\mathcal{E} l_\mathcal{E}^4$, $\eta_1 = -0.480 E_\mathcal{E} l_\mathcal{E}^4$, and $\eta_2 = 0.046 E_\mathcal{E} l_\mathcal{E}^4$.}
	\label{fig:spectrum_anisotropic}
\end{figure}

The spectrum of the anisotropic effective Hamiltonian is compared to the Luttinger-Kohn Hamiltonian in Fig.~\ref{fig:spectrum_anisotropic}. As was the case for the isotropic case (see Fig.~\ref{fig:spectrum_and_fermi_surface}), the effective Hamiltonian deviates from the Luttinger-Kohn Hamiltonian at momenta away from $k=0$. 
We stress again that this deviation is not caused by the fictitious quartic term defined in Eq.~\eqref{eq:quartic_k}, but is rather a measure of the range of validity of the third-order perturbation theory used to obtain the effective model. To improve the fit of the effective model to the Luttinger-Kohn Hamiltonian, one may expand it up to fourth order. The effective Hamiltonian is then given by:
\begin{equation}
	H_{\text{eff},a}^{(4)} = H_{\text{eff},a} + \eta_1 k^4 + \eta_2 \left(k_+^4+k_-^4\right),
	\label{eq:fourth_order_effective_ham}
\end{equation}
where $H_{\text{eff},a}$ is defined in Eq.~\eqref{eq:effective_hamiltonian_anisotropic}. 
The spectrum of this Hamiltonian is also shown in Fig.~\ref{fig:spectrum_anisotropic}. 
We stress that the $\eta_1$ term in Eq.~\eqref{eq:fourth_order_effective_ham} and the fictitious quartic term proportional to $\beta$ defined in Eq.~\eqref{eq:quartic_k} have the same form, but the $\eta_1$ term has a physical origin, while the $\beta$ term is fictitious. 
However, we typically find that the $\eta_2$ term is negligible compared to the $\eta_1$ term and that $\eta_1<0$. Therefore, a fictitious $k^6$ term would have to be introduced to remove unphysical Fermi surfaces. 
Since the fourth order terms in $H_{\text{eff},a}^{(4)}$ do not introduce any dependency on spin, they do not qualitatively change the SOI behavior of the system. We conclude that going to third order perturbation theory is sufficient to capture the qualitative behavior of the system. We note that for the parameters used in Sec.~\ref{subsec:experiment}, where we make quantitative predictions, the fit between the Luttinger-Kohn Hamiltonian and the third-order effective theory is even better, see App.~\ref{appsec:parameters_experiment}.

Confining the Ge along a different axis changes the effective Hamiltonian. If the confinement direction is the crystallographic $\langle 110 \rangle$ axis and strain is included, the effective Hamiltonian has linear and cubic SOI terms~\cite{xiong2021emergence} and the quadratic term receives anisotropic corrections. Using the same method as described in Sec.~\ref{sec:effective_ham}, we derive the effective Hamiltonian:
\begin{widetext}
\begin{eqnarray}
	H_\text{eff} &=&
	\frac{i a_1}{2} \left(k_- \sigma_+ - k_+ \sigma_-\right)
	+ \frac{i a_2}{2} \left(k_+ \sigma_+ - k_- \sigma_-\right)
	+ b_1 \left(k_-^2+k_+^2\right) + b_2 k_- k_+
	+ \frac{i c_1}{2} \left(k_-^3 \sigma_+ - k_+^3\sigma_-\right)
	\nonumber \\
	&&+ \frac{i c_2}{2} k^2 \left(k_- \sigma_+ - k_+ \sigma_-\right)
	+ \frac{i c_3}{2} k^2 \left(k_+ \sigma_+ - k_- \sigma_-\right) 
	+ \frac{i c_4}{2} \left(k_+^3 \sigma_+ - k_-^3 \sigma_-\right) ,
	\label{eq:effective_ham_linear_and_cubic}
\end{eqnarray}
\end{widetext}
where $a_i$, $b_i$, and $c_i$ are parameters that are calculated from the original Luttinger-Kohn Hamiltonian perturbatively. This Hamiltonian is used for case~2 in Sec.~\ref{subsec:experiment}. We stress that the LSOI terms in Eq.~\eqref{eq:effective_ham_linear_and_cubic} are specific to 2DHGs and not to be confused with the LSOI terms arising in 2DEGs.

\section{\label{appsec:high_momentum_fermi_surface}High momentum Fermi surfaces}
The effective Hamiltonian defined in Eq.~\eqref{eq:effective_hamiltonian_isotropic} contains a quadratic term and a CSOI term. At large momenta, the CSOI term dominates over the quadratic term, leading to unphysical Fermi surfaces. Because a numerical tight-binding calculation cannot be restricted to small momenta, these Fermi surfaces pose a problem. To eliminate them, a fictitious quartic term of strength $\beta$ is introduced [see Eq.~\eqref{eq:quartic_k}]. This appendix explains how we determine $\beta$.

\begin{figure}
	\centering
	\includegraphics[width=0.5\textwidth]{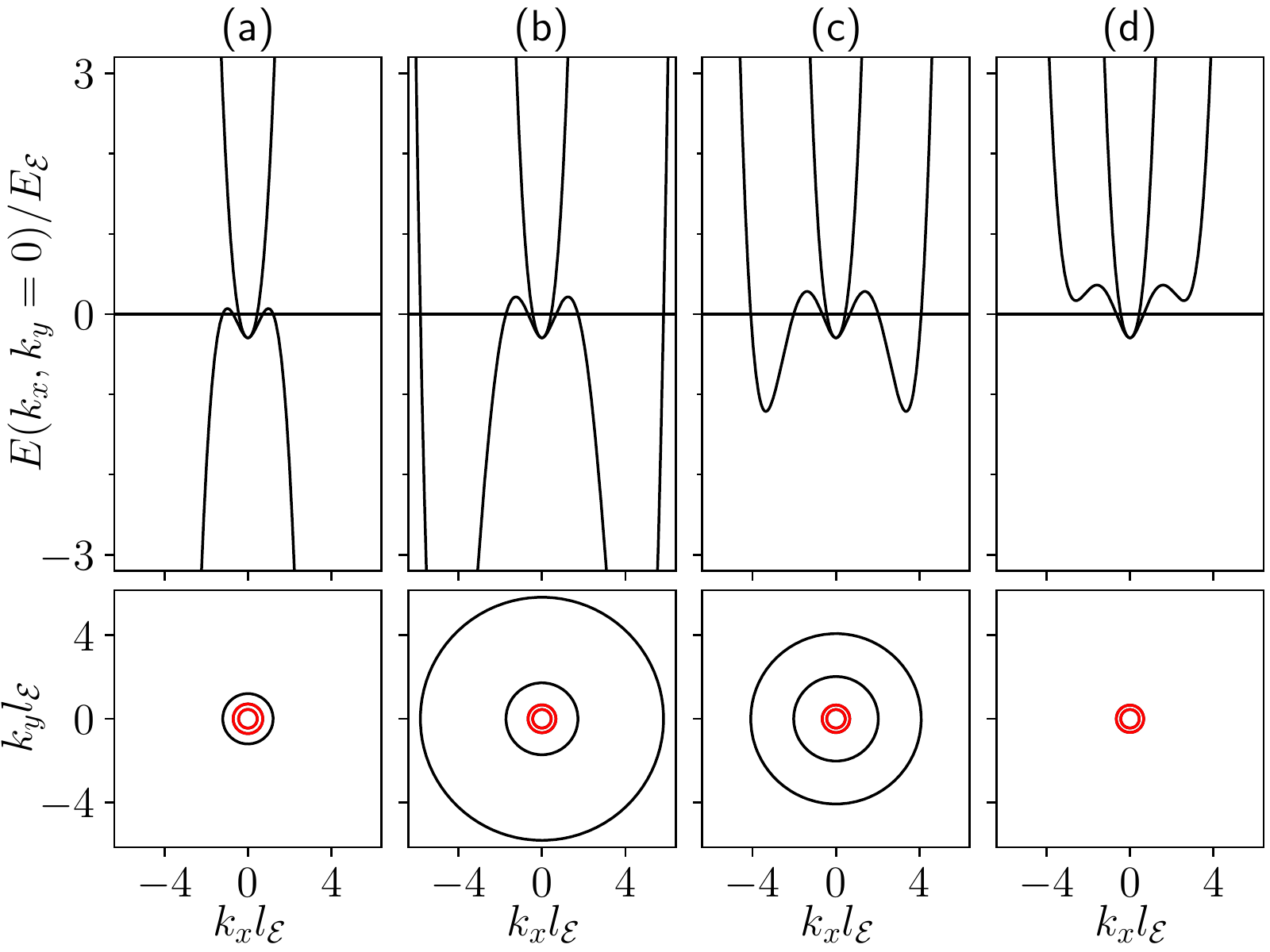}
	\caption{
		Top row: Spectrum $E(k_x, k_y=0)$, calculated using $H_\text{eff}+H_4$ from Eqs.~\eqref{eq:effective_hamiltonian_isotropic} and~\eqref{eq:quartic_k}.
		Bottom row: Fermi surface, i.e., the momenta where $E(k_x, k_y)=0$. The red curves indicate the physical Fermi surfaces, the black curves indicate the unphysical ones.
		In each column, the strength $\beta$ of the quartic term is different: 
		$\beta=0$ in (a), 
		$\beta=0.10 E_\mathcal{E} l_\mathcal{E}^4$ in (b), 
		$\beta=0.12 E_\mathcal{E} l_\mathcal{E}^4$ in (c),
		and $\beta=0.14 E_\mathcal{E} l_\mathcal{E}^4$ in (d).
		The  other parameters are: $\hbar^2/2m^\ast=1.141 E_\mathcal{E} l_\mathcal{E}^2$, $\alpha=0.39 E_\mathcal{E} l_\mathcal{E}^3$, $\mu=0.3 E_\mathcal{E}$. 
		The Fermi momenta on the two physical (red) Fermi surfaces are $k_{F,1} l_\mathcal{E} \approx 0.44 $ and $k_{F,2} l_\mathcal{E} \approx 0.70$. Therefore, $\beta=0.14 E_\mathcal{E} l_\mathcal{E}^4$ 
		gives 
		$\alpha k_{F,2}^3 = 0.13 E_\mathcal{E} \approx 4.3 \beta k_{F,2}^4$, which satisfies the upper boundary set in  Eq.~\eqref{eq:limit_beta}. 
	}
	\label{fig:high_momentum_fermi_surface}
\end{figure}

To not significantly alter the low-momentum behavior, $\beta$ should be kept as small as possible, yet large enough to eliminate unphysical Fermi surfaces. A rough estimate of the upper bound for $\beta$ is given in Eq.~\eqref{eq:limit_beta}. To get a lower bound for $\beta$, we need to plot the Fermi surfaces of the effective Hamiltonian. The energy eigenvalues of the Hamiltonian $H_\text{eff}+H_4$ defined in Eqs.~\eqref{eq:effective_hamiltonian_isotropic} and~\eqref{eq:quartic_k} with different values of $\beta$ are shown in the top row of Fig.~\ref{fig:high_momentum_fermi_surface}. The corresponding Fermi surfaces are shown in the bottom row of Fig.~\ref{fig:high_momentum_fermi_surface}. The red curves correspond to the physical Fermi surfaces while the black curves are the unphysical ones. As $\beta$ is increased, the unphysical Fermi surfaces change their shape, while the physical Fermi surfaces do not change noticeably. Once $\beta$ is large enough, the unphysical Fermi surfaces disappear. This is the value of $\beta$ that we want to use for the numerical simulations.

\section{\label{appsec:perturbation}Perturbation calculation to verify the quartic term}
To test the validity of the quartic term defined in Eq.~\eqref{eq:quartic_k}, we treat the CSOI term as a perturbation to the quadratic spectrum, following the quasi-degenerate perturbation theory explained in App.~B of Ref.~\cite{winkler2003spin}. 
 In this case, the quartic term is not required. For simplicity, we consider the Hamiltonian in the semi-infinite geometry at $k_x=0$, as this is the Hamiltonian that defines the phase diagram. The unperturbed Hamiltonian is:
\begin{equation}
	\mathcal{H}_0(y) = \mathcal{H}_\text{kin}(y) + \mathcal{H}_{\text{SC}}(y) + \mathcal{H}_{Z}(y) ,
\end{equation}
where $\mathcal{H}_{\text{SC}}(y)$ is defined in Eq.~\eqref{eq:sc_term_continuous}, $\mathcal{H}_{Z}(y)$ is defined in Eq.~\eqref{eq:zeeman_term_continuous}, and $\mathcal{H}_\text{kin}(y)$ is:
\begin{equation}
	\mathcal{H}_\text{kin}(y) = -\left(\frac{\hbar^2 \partial_y^2}{2m^\ast} + \mu\right) \tau_z .
\end{equation}
We then consider the cubic SOI term as a perturbation: 
\begin{equation}
	\mathcal{H}^\prime(y) = 2i\alpha \partial_y^3 \sigma_x 
\end{equation}
and we expand in powers of $\alpha$ using quasi-degenerate perturbation theory~\cite{winkler2003spin}. 

\begin{figure}
	\centering
	\includegraphics[width=0.5\textwidth]{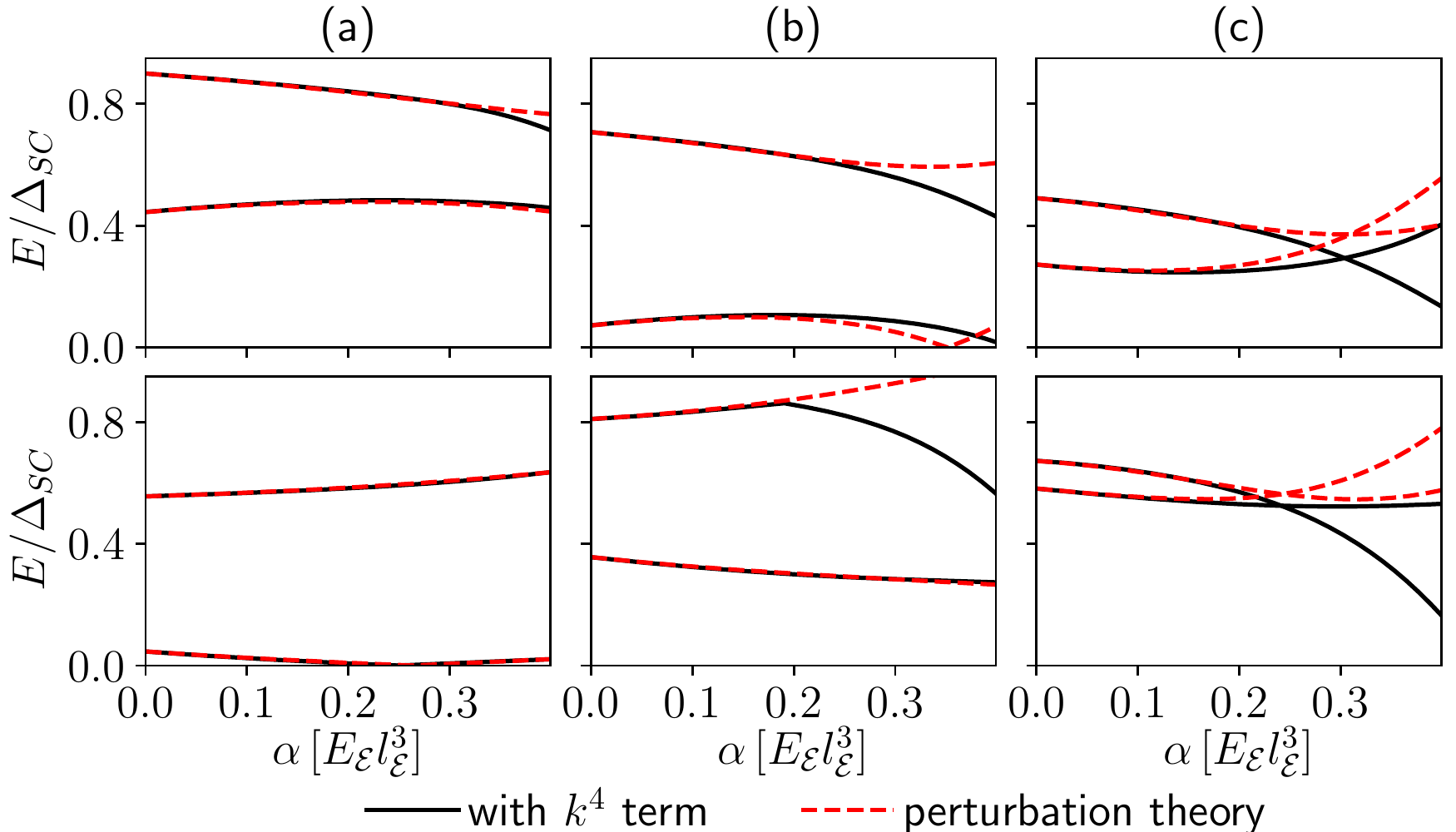}
	\caption{Energy levels of the Hamiltonian given in Eq.~\eqref{eq:continuous_hamiltonian_all_terms} that includes the fictitious quartic term (black solid curve) compared to the energy levels calculated using perturbation theory to third order in the CSOI strength $\alpha$ (red dashed curve). Both calculations are done in the semi-infinite geometry at $k_x=0$.
		The parameters are: $L_y = 49.5 l_\mathcal{E}$, $W=4.5 l_\mathcal{E}$,
		$\hbar^2/2m^\ast = 1.141 E_\mathcal{E} l_\mathcal{E}^2$, $\mu = 0.3 E_\mathcal{E}$, and $\Delta_{\text{SC}} = \mu/3$.
		In the top row $\phi=\pi/4$, in the bottom row $\phi=3\pi/4$. In column (a) $\Delta_Z/\Delta_{\text{SC}}=1$, in column (b) $\Delta_Z/\Delta_{\text{SC}}=2$, and in column (c) $\Delta_Z/\Delta_{\text{SC}}=3$. The values for $\beta$ change together with $\alpha$, as explained in the caption of Fig.~\ref{fig:fitted_slope_cubic}.
	}
	\label{fig:perturbation_theory}
\end{figure}

We compare the perturbatively calculated energy levels to the eigenvalues of the Hamiltonian of Eq.~\eqref{eq:continuous_hamiltonian_all_terms}, which includes the fictitious quartic term. At small CSOI strengths, the two results agree well with each other, see Fig.~\ref{fig:perturbation_theory}. This demonstrates that the fictitious quartic term is suitable to remove the unphysical Fermi surfaces while not changing the low-energy behavior.
In the numerical simulations of Sec.~\ref{sec:planar_jj}, we use $\alpha = 0.39 E_\mathcal{E} l_\mathcal{E}^3$. As expected, perturbation theory for such a large CSOI strength can deviate significantly from the energy eigenvalues of the Hamiltonian defined in Eq.~\eqref{eq:continuous_hamiltonian_all_terms}. 
However, we use such a large CSOI strength in order to obtain MBSs with localization length short enough to use reasonable system sizes. We verified that there are MBSs in the system even for $\alpha = 0.11 E_\mathcal{E} l_\mathcal{E}^3$, for which perturbation theory agrees well with the eigenvalues of the Hamiltonian defined in Eq.~\eqref{eq:continuous_hamiltonian_all_terms}. 
In contrast, the strengths of the CSOI used in Sec.~\ref{subsec:experiment} are smaller, therefore we expect the energy eigenvalues to agree better with the perturbatively calculated energies.

\section{\label{appsec:tight_binding}Discretized Hamiltonian}
For our numerical calculations, we discretize the Hamiltonian of Eq.~\eqref{eq:continuous_hamiltonian_all_terms} on a square lattice with effective lattice constant $a$, resulting in an effective tight-binding Hamiltonian.
Throughout Secs.~\ref{sec:planar_jj} and~\ref{subsec:experiment} we set $a=l_\mathcal{E}/2$. However, note that $l_\mathcal{E}$ in Sec.~\ref{sec:planar_jj} is different from $l_\mathcal{E}$ in Sec.~\ref{subsec:experiment}. 

\subsection{\label{appsubsec:finite}Finite Geometry}
The discrete Nambu basis in the finite geometry is:
\begin{equation}
	\label{eq:nambu_finite}
	c_{n,m} = (c_{\uparrow, n, m}, c_{\downarrow, n, m}, c_{\uparrow, n, m}^\dagger, c_{\downarrow, n, m}^\dagger)^T ,
\end{equation}
where $c_{s,n,m}^\dagger$ creates a particle with effective spin $s$ at position $(x,y)=(na, ma)$ and $n, m \in \mathbb{Z}$. Using this basis, the tight-binding Hamiltonian of the planar JJ in the finite geometry is:
\begin{equation}
	\bar{H} = \frac{1}{2} \left(
	\bar{H}_\text{eff} + \bar{H}_4 
	+ \bar{H}_{\text{SC}} + \bar{H}_{Z} \right)
	\label{eq:tb_finite_geometry}
\end{equation}
with
\begin{widetext}
\begin{eqnarray}
	\bar{H}_\text{eff} + \bar{H}_4 &&=
	\sum_{n=1}^{N_x-1} \sum_{m=0}^{N_y-1} 
	c_{n, m}^\dagger \left( -\frac{t}{a^2} - \frac{4i\alpha}{a^3}\sigma_y 
	-\frac{8\beta}{a^4} \right) \tau_z  c_{n-1,m}
	+ \sum_{n=0}^{N_x-1} \sum_{m=1}^{N_y-1} 
	c_{n, m}^\dagger \left( -\frac{t}{a^2} \tau_z  - \frac{4i\alpha}{a^3}\sigma_x 
	-\frac{8\beta }{a^4} \tau_z\right)  c_{n,m-1}
	\nonumber \\
	&&+ \sum_{n=2}^{N_x-1} \sum_{m=0}^{N_y-1} 
	c_{n, m}^\dagger \left(- \frac{i\alpha}{a^3}  \sigma_y 
	+ \frac{\beta}{a^4}\right) \tau_z c_{n-2,m}
	+ \sum_{n=0}^{N_x-1} \sum_{m=2}^{N_y-1} 
	c_{n, m}^\dagger \left( - \frac{i\alpha}{a^3} \sigma_x 
	+ \frac{\beta}{a^4} \tau_z \right) c_{n,m-2}
	\nonumber \\
	&&+   \sum_{n=1}^{N_x-1} \sum_{m=1}^{N_y-1} 
	c_{n, m}^\dagger \left[ \frac{3i\alpha}{a^3} \left(\sigma_x 
	+ \sigma_y \tau_z \right)  + \frac{2\beta}{a^4} \tau_z \right] c_{n-1,m-1} 
	+ \sum_{n=0}^{N_x-2} \sum_{m=1}^{N_y-1} 
	c_{n, m}^\dagger \left[  \frac{3i\alpha}{a^3} \left(\sigma_x 
	-\sigma_y \tau_z \right) + \frac{2\beta}{a^4} \tau_z \right] c_{n+1,m-1}
	\nonumber\\
	&&+ \sum_{n=0}^{N_x-1} \sum_{m=0}^{N_y-1} 
	c_{n, m}^\dagger \left(
	\frac{2t}{a^2} - \frac{\mu}{2} 
	+ \frac{10 \beta}{a^4}
	\right) \tau_z c_{n,m}
	+ \text{H.c.} ,
	\label{eq:tb_effective_ham}
\end{eqnarray}
\end{widetext}
where $N_x = 1+L_x/a$ and $N_y=1+L_y/a$ are the number of lattice points in $x$ and $y$ direction respectively, and $t=\hbar^2/2m^\ast$. Here,
$\sigma_{\mu}$ and $\tau_\nu$, $\mu, \nu=x,y,z$, are Pauli matrices acting in spin and particle-hole space, respectively.
In Sec.~\ref{sec:planar_jj} we set $\hbar^2/2m^\ast = 1.141 E_\mathcal{E} l_\mathcal{E}^2$, which leads to $t= 4.56 E_\mathcal{E} a^2$. The other system parameters can be expressed in terms of $a$ and $t$ as $\alpha = 0.39 E_\mathcal{E} l_\mathcal{E}^3 = 0.68 t a$, $\beta = 0.14 E_\mathcal{E} l_\mathcal{E}^4 = 0.49 ta^2$, and $\mu = 0.3 E_\mathcal{E} = 0.065 t a^{-2}$.

The superconducting term $\bar{H}_{\text{SC}}$ and the Zeeman term $\bar{H}_{Z}$ are:
\begin{eqnarray}
	\bar{H}_{\text{SC}} &=& \sum_{n=0}^{N_x-1} \sum_{m=0}^{N_y-1} 
	 \Delta_{SC,m}
	c_{n, m}^\dagger
	 \frac{i \sigma_y}{2} \nonumber \\
	 && \hspace{1cm} \times\left( 
	 e^{-i \phi_m/2} \tau_+
	 - e^{i \phi_m/2} \tau_- 
	 \right) 
	 c_{n,m} ,
	  \label{eq:tb_sc}\\
	\bar{H}_Z &=& \sum_{n=0}^{N_x-1} \sum_{m=0}^{N_y-1} 
	\Delta_{Z,m}
	c_{n, m}^\dagger 
	 \sigma_x \tau_z
	c_{n,m} , 
	\label{eq:tb_z}
\end{eqnarray}
where the parameters $\Delta_{{\text{SC}}, m}$, $\phi_m$, and $\Delta_{Z,m}$ are position-dependent:
\begin{eqnarray}
	\Delta_{SC, m} &=& 
	\begin{cases}
		\Delta_{\text{SC}}^l,& \text{if } m < \frac{N_y-N_W}{2}\\
		\Delta_{\text{SC}}^r, & \text{if } m \geq \frac{N_y+N_W}{2}\\
		0,              & \text{otherwise}
	\end{cases} ,
	\label{eq:delta_sc_definition} \\
	\phi_m &=& 
	\begin{cases}
		\phi,& \text{if } m < \frac{N_y-N_W}{2} \\
		-\phi,              & \text{if } m \geq \frac{N_y+N_W}{2}
	\end{cases} ,
	\label{eq:phi_definition} \\
	\Delta_{Z,m} &=& 
	\begin{cases}
		\Delta_Z,& \text{if } \frac{N_y-N_W}{2} \leq m < \frac{N_y+N_W}{2} \\
		0,              & \text{otherwise}
	\end{cases} ,
	\label{eq:delta_z_definition}
\end{eqnarray}
where $N_W=1+W/a$ is the number of lattice points in $y$ direction within the normal section. 
In Sec.~\ref{subsec:quasi_mbs}, the width of the normal section is varied as follows:
\begin{eqnarray}
	\label{eq:rounding_junction_width}
	\!\!\!\!\! N_W(n) \!=\!\! \left\lfloor \!  \frac{N_y\!+\! \left[1\!+\! \frac{W(n a)}{a}\right]}{2} \! \right\rfloor
	\!\!-\!\! \left\lfloor \! \frac{N_y\!-\! \left[1\!+\! \frac{W(n a)}{a}\right]}{2}  \! \right\rfloor \!\!,  \label{eq:nw_variation} 
\end{eqnarray}
where $N_W(n)$ gives the number of lattice points (in $y$ direction) at position $x=n a$, $\lfloor \,\, \rfloor$~is the rounding down operator, and $W(x)$ is given by Eq.~\eqref{eq:tanh_variation}. 
The values for $W$ in Fig.~\ref{fig:quasi_mbs_junction_width} are derived as follows. Using the parameters given in the caption of Fig.~\ref{fig:quasi_mbs_junction_width}, $N_y = 1+L_y/a$, and Eqs.~\eqref{eq:tanh_variation} and~\eqref{eq:rounding_junction_width}, one gets that at the left end of the junction, the number of lattice points in the normal section is given by $N_W(n=0) = 9$, which corresponds to a junction width $W= a(N_W-1) = 4l_\mathcal{E}$. On the right side, the index $n$ has the value $n=N_x-1 = L_x/a = 299$, which gives $N_W(n=299) = 5$, therefore giving $W=2 l_\mathcal{E}$.

\subsection{\label{app:disorder} Disorder}
In Fig.~\ref{fig:quasi_mbs_bdi_remnant}, disorder is added to the system. We model disorder by a position-dependent chemical potential of the form:
\begin{equation}
	\mu_{n,m} = \mu_0 + \delta \mu_{n,m},
\end{equation}
where $n$ and $m$ label the position in the system $(x,y) = (na, ma)$, $\mu_0$  is a constant chemical potential and $\delta \mu_{m,n}$ are random fluctuations drawn from a normal distribution with mean value zero. For Fig.~\ref{fig:quasi_mbs_bdi_remnant}, the parameters used for the chemical potential are $\mu_0 = 0.3 E_\mathcal{E}$ and the standard deviation of the normal distribution of $\delta \mu_{n,m}$ is $0.2 \mu_0$.

\subsection{Semi-infinite geometry}
In the semi-infinite geometry, it is assumed that the system is infinitely extended in $x$ direction. In this case, $k_x$ is a good quantum number and the Nambu basis is:
\begin{equation}
	\label{eq:nambu_semi_inf}
	c_{k_x,m} = \left(c_{\uparrow, k_x,m}, c_{\downarrow, k_x,m}, c_{\uparrow, -k_x,m}^\dagger, c_{\downarrow, -k_x,m}^\dagger\right)^T ,
\end{equation}
where $m \in \mathbb{Z}$ and $c_{s, k_x,m}^\dagger$ creates a particle at position $y=ma$ with effective spin $s$ and momentum $k_x$ in $x$ direction.  The tight-binding Hamiltonian in the semi-infinite geometry is:
\begin{eqnarray}
	\tilde{H} =  \! \frac{1}{2} \int_{-\infty}^\infty \!\! dk_x \!
	\left[ 
	\tilde{H}_\text{eff} \!\left(k_x\right) 
	+ \tilde{H}_4 \!\left(k_x\right) 
	+ \tilde{H}_{Z}  \!\left(k_x\right) 
	+ \tilde{H}_{\text{SC}} \! \left(k_x\right) \!
	\right] \! , \nonumber \\
	\label{eq:semi_inf_geometry_integral}
\end{eqnarray}
with
\begin{widetext}
\begin{eqnarray}
	\tilde{H}_\text{eff}\left(k_x\right) &&+ \tilde{H}_4\left(k_x\right) = 
	\sum_{m=0}^{N_y-1}
	c_{k_x, m}^\dagger
	\left\{
	\frac{t\left[2-\cos\left(ak_x\right)\right]}{a^2}
	-\frac{\mu}{2}
	+\frac{\beta}{a^4} \left[10-8\cos\left(ak_x\right)+\cos\left(2ak_x\right)\right]
	\right. \nonumber \\ &&  \left.
	-\frac{\alpha}{a^3} \left[4 \sin\left(ak_x\right) + \sin\left(2ak_x\right) \right]\sigma_y 
	\right\} \tau_z c_{k_x, m} 
	+\sum_{m=1}^{N_y-1} c_{k_x,m}^\dagger 
	\left\{
	-\frac{t}{a^2} \tau_z 
	+\frac{4\beta}{a^4} \left[-2+\cos\left(ak_x\right)\right]  \tau_z
	+ \frac{6 \alpha }{a^3} \sin\left(ak_x\right) \sigma_y \tau_z 
	\right. \nonumber \\&&  \left.
	+ \frac{2i \alpha}{a^3} \left[3 \cos\left(ak_x\right)-2\right] \sigma_x 
	\right\} c_{k_x, m-1} 
	+\sum_{m=2}^{N_y-1} c_{k_x,m}^\dagger
	\left(
	\frac{-i \alpha}{a^3} \sigma_x
	+\frac{\beta}{a^4} \tau_z
	\right) c_{k_x, m-2}
	+ \text{H.c.}
	,
	\label{eq:h_eff_semi_infinite}  
\end{eqnarray}

\begin{eqnarray}
	\tilde{H}_{\text{SC}}\left(k_x\right) &=&  \sum_{m=0}^{N_y-1} 
	\Delta_{SC,m}
	c_{k_x,m}^\dagger 
	\frac{i \sigma_y}{2} \left(
	e^{-i \phi_m/2} \tau_+
	- e^{i \phi_m/2} \tau_-
	\right) 
	c_{k_x,m}, \\
	\tilde{H}_Z\left(k_x\right) &=& \sum_{m=0}^{N_y-1} 
	\Delta_{Z,m}
	c_{k_x,m}^\dagger 
	\sigma_x \tau_z
	c_{k_x,m},
\end{eqnarray}
where $\Delta_{SC,m}$ is defined in Eq.~\eqref{eq:delta_sc_definition}, $\phi_m$ in Eq.~\eqref{eq:phi_definition}, and $\Delta_{Z,m}$ in Eq.~\eqref{eq:delta_z_definition}.

\subsection{Anisotropic term}
The discretized version of the anisotropic contribution in Eq.~\eqref{eq:effective_hamiltonian_anisotropic} in the finite geometry is given by:
\begin{eqnarray}
	\bar{H}_a =
	\frac{i\alpha_a}{a^3}
	\Bigg[&&
	- 4 \sum_{n=1}^{N_x-1} \sum_{m=0}^{N_y-1}  c_{n, m}^\dagger  \sigma_y \tau_z c_{n-1,m}
	- 4 \sum_{n=0}^{N_x-1} \sum_{m=1}^{N_y-1}  c_{n, m}^\dagger  \sigma_x 
	c_{n,m-1}
	+  \sum_{n=2}^{N_x-1} \sum_{m=0}^{N_y-1}  c_{n, m}^\dagger  \sigma_y \tau_z c_{n-2,m} \nonumber\\
	&&+ \sum_{n=0}^{N_x-1} \sum_{m=2}^{N_y-1}  c_{n, m}^\dagger  \sigma_x 
	c_{n,m-2}
	+\sum_{n=1}^{N_x-1} \sum_{m=1}^{N_y-1}  c_{n, m}^\dagger  \left(\sigma_x 
	+\sigma_y \tau_z \right)    c_{n-1,m-1} \nonumber \\
	&&+ \sum_{n=0}^{N_x-2} \sum_{m=1}^{N_y-1}  c_{n, m}^\dagger  \left(\sigma_x 
	-\sigma_y \tau_z\right)   c_{n+1,m-1} 
	\Bigg] 
	+ \text{H.c.}, 
	\label{eq:tb_kin_cubic_quartic}
\end{eqnarray}
where $c_{n,m}$ is the Nambu basis defined in Eq.~\eqref{eq:nambu_finite}.
In the semi-infinite geometry, this term is:
\begin{eqnarray}
	\tilde{H}_a\left(k_x\right) &=& 
	\frac{\alpha_a}{a^3} \Bigg\{
	\sum_{m=0}^{N_y-1}
	c_{k_x, m}^\dagger 
	\left[ 
	-4\sin\left(ak_x\right) + \sin\left(2ak_x\right)  
	\right] \sigma_y \tau_z c_{k_x, m}  \nonumber \\
	&& \hspace{-0.2cm}+ \sum_{m=1}^{N_y-1} 2 c_{k_x,m}^\dagger 
	 \left\{
	\left[
	\cos\left(ak_x\right)-2\right] i  \sigma_x 
	+ \sin\left(ak_x\right) \sigma_y \tau_z  \right\}  c_{k_x, m-1} 
	+\sum_{m=2}^{N_y-1} c_{k_x,m}^\dagger
	i \sigma_x 
	c_{k_x, m-2} \Bigg\}
	+ \text{H.c.}
	,
	\label{eq:anisotropic_semi_infinite}
\end{eqnarray}
\end{widetext}
where the $c_{k_x,m}$ is the Nambu basis defined in Eq.~\eqref{eq:nambu_semi_inf}.

\begin{figure}
	\centering
	\includegraphics[width=\columnwidth]{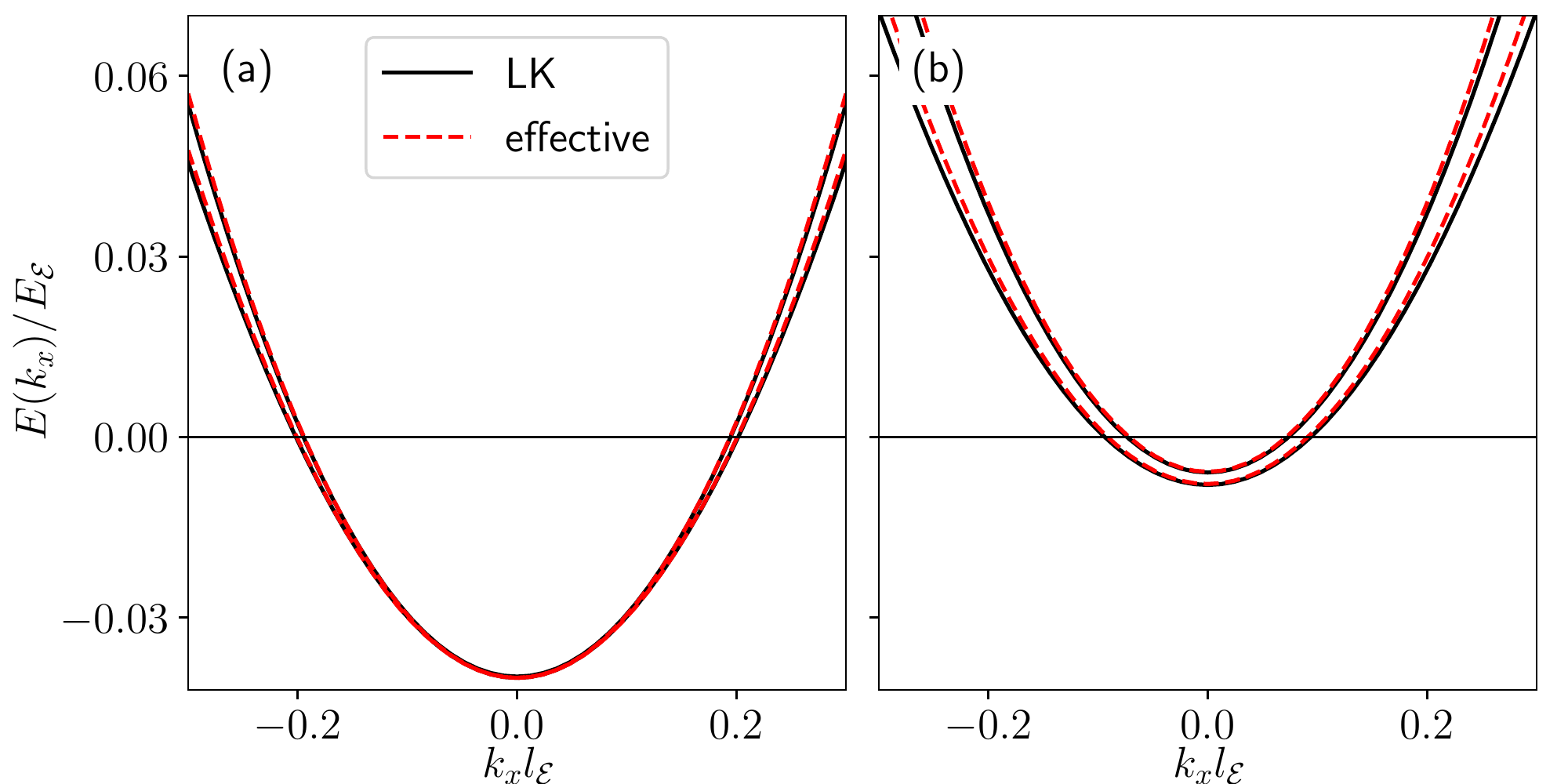}
	\caption{
		{Spectrum $E(k_x)$ of the 2DHG with parameters used for case~1 in Sec.~\ref{subsec:experiment}. The momentum $k_y$ is fixed to $k_y=0$ in~(a) and $k_y l_\mathcal{E}=0.18$ in~(b). The solid black line corresponds to the two lowest energy bands of the Luttinger-Kohn (LK) Hamiltonian confined to two dimensions [see Eq.~\eqref{eq:lk_confined}]. The red dashed lines correspond to the energy eigenvalues of the effective model [see Eqs.~\eqref{eq:effective_hamiltonian_isotropic} and~\eqref{eq:quartic_k}]. Due to the fact that the chemical potential is rather small, the two models agree well at the Fermi level.
			The Luttinger parameters used for the LK Hamiltonian are the same as in Fig.~\ref{fig:spectrum_anisotropic}, the strain is $E_s/E_\mathcal{E}=0.26$, and the chemical potential is $\mu=0.04 E_\mathcal{E}$. The parameters used for the effective model are given in the column ``case~1'' in Tab.~\ref{tab:coefficients_experimental}.
		}
	}
	\label{fig:spectrum_case_1}
\end{figure}

\section{\label{appsec:parameters_experiment}Experimentally realistic parameters} 
Table~\ref{tab:coefficients_experimental} lists the values used for the numerical calculations in Sec.~\ref{subsec:experiment}. 
The momentum space spectrum for case~1 is shown in Fig~\ref{fig:spectrum_case_1}. We choose a relatively small chemical potential, so the results obtained by using the Luttinger-Kohn Hamiltonian and the effective model agree very well at the Fermi level. We checked that this is also true for the parameters used in case~2.

\begin{table*}
	\begin{tabular}{|c|c|c|c|}
		\hline
		& case 1 & case 2 
		& 2DEG LSOI\\
		\hline 
		confinement axis & $\langle 001 \rangle$& $\langle 110 \rangle$ 
		& 
		\\
		\hline
		strain $E_s/E_\mathcal{E}$ & 0.26 & 0.26 
		&  
		\\
		\hline
		axis of magnetic field & $x$ & $x$ 
		& $x$\\
		\hline
		$g$ factor & 0.18 & 
		& \\
		\hline
		critical magnetic field $B_c$ [T]& 1.8 & 1.8 
		& \\
		\hline
		$\Delta_{Z,c}/\Delta_{\text{SC}}$ & 0.0193 & 
		& 1\\
		\hline
		$\Delta_{\text{SC}}/E_\mathcal{E}$ &  0.0131 & 0.0131  
		& 0.0131\\
		\hline 
		$\mu/E_\mathcal{E}$ & 0.04 & 0.04 
		& 0.1 \\
		\hline
		linear coefficients  $[E_\mathcal{E} l_\mathcal{E}]$ & 0 &  $a_1=-0.00115$, $a_2=-0.00785$ 
		& $\alpha_{lin}=0.075$ \\
		\hline
		quadratic coefficients $[E_\mathcal{E} l_\mathcal{E}^2]$ & $\hbar^2/2m^\ast=1.025$   & $b_1=-0.100$, $b_2=1.176$  
		& $\hbar^2/2m^\ast=3$\\
		\hline
		cubic coefficients $[E_\mathcal{E} l_\mathcal{E}^3]$ & $\alpha= 0.102$,     & $c_1=-0.00278$, $c_2=-0.0195$,  
		& \\
		&$\alpha_a/\alpha=0.146$ &  $c_3=0.0219$, $c_4=0.171$ 
		& \\
		\hline
		$\beta$ $[E_\mathcal{E} l_\mathcal{E}^4]$ & 0.01375  &  0.009375 
		& \\
		\hline
		$W$ $[l_\mathcal{E}]$ & 24 & 28 
		& 24 \\
		\hline
		$L_y$ $[l_\mathcal{E}]$ & 299.5 & 299.5 
		& 399.5 \\
		\hline
	\end{tabular}
	\caption{Parameters for the two cases presented in Sec.~\ref{subsec:experiment} and for a 2DEG system with LSOI. The linear, quadratic, and cubic coefficients as well as the $g$ factors are calculated using perturbation theory~\cite{winkler2003spin, michal2021longitudinal, marcellina2017spin, bosco2021squeezed, bosco2021hole, bosco2021fully}. 
		We set $\mu \approx 3 \Delta_{\text{SC}}$. 
	The parameters are given in units of the electric energy $E_\mathcal{E}=37$~meV and electric length $l_\mathcal{E}=3.7$~nm.
	The normal section in all three cases is between $W\approx 90$~nm and $W\approx 100$~nm, which is comparable to the 80~nm used in Ref.~\cite{fornieri2019evidence}. 
	For cases~1 and~2, the proximitized regions are approximately $510$~nm wide and for the 2DEG it is approximately $690$~nm. 
	The critical magnetic field and superconducting gap are taken from Ref.~\cite{aggarwal2021enhancement}, where a combination of aluminum and niobium is used to induce superconductivity.
	For the 2DEG LSOI system, the Hamiltonian is $H_{lin} = \hbar^2 k^2/2m^\ast - \mu + \alpha_{lin} \left( k_x \sigma_y - k_y \sigma_x \right)$.
	The quadratic and linear strength coefficients are experimental values taken from Ref.~\cite{fornieri2019evidence}. 
	We assume the same values for $\Delta_{\text{SC}}$ as in cases~1 and~2 and set the critical Zeeman field $\Delta_{Z,c}$ to be equal to $\Delta_{\text{SC}}$.
	Although $E_\mathcal{E}$ and $l_\mathcal{E}$ are not relevant scales for a 2DEG system, we still give the parameters in these units for better comparison to the 2DHG Ge system.
	} 
	\label{tab:coefficients_experimental}
\end{table*}

\section{\label{appsec:potential_barrier}Potential barrier}
A potential barrier at the interface between the superconducting region and the normal region changes the transparency of the junction. Although this does not qualitatively change the behavior of the system, it affects borders of the phase diagram~\cite{pientka2017topological}. In this section, we study the effect of a potential barrier on the systems described in Sec.~\ref{subsec:experiment}. 
To account for a potential barrier of width $W_{\text{b}}$, we introduce a Hamiltonian $\mathcal{H}_{\text{b}}$ that is added to the existing Hamiltonian defined in Eq.~\eqref{eq:continuous_hamiltonian_all_terms}:
\begin{eqnarray}
	\label{eq:potential_barrier}
	\mathcal{H}_{\text{b}}(x,y) &=& \mu_{\text{b}}(x,y) \tau_z , \\
	\mu_{\text{b}}(x,y) &=& \begin{cases}
		\mu_{\text{b}} & \!\!\! \text{if } S-\frac{W_{\text{b}}}{2} < y < S + \frac{W_{\text{b}}}{2} \text{ or } \\
		& \!\!\!  S \! + \! W \! - \! \frac{W_{\text{b}}}{2} \! < \! y \! < \! S \! + \! W \! + \! \frac{W_{\text{b}}}{2},\\
		0 & \!\!\! \text{otherwise. } \\
	\end{cases}.
\end{eqnarray}

\begin{figure}
	\centering
	\includegraphics[width=\columnwidth]{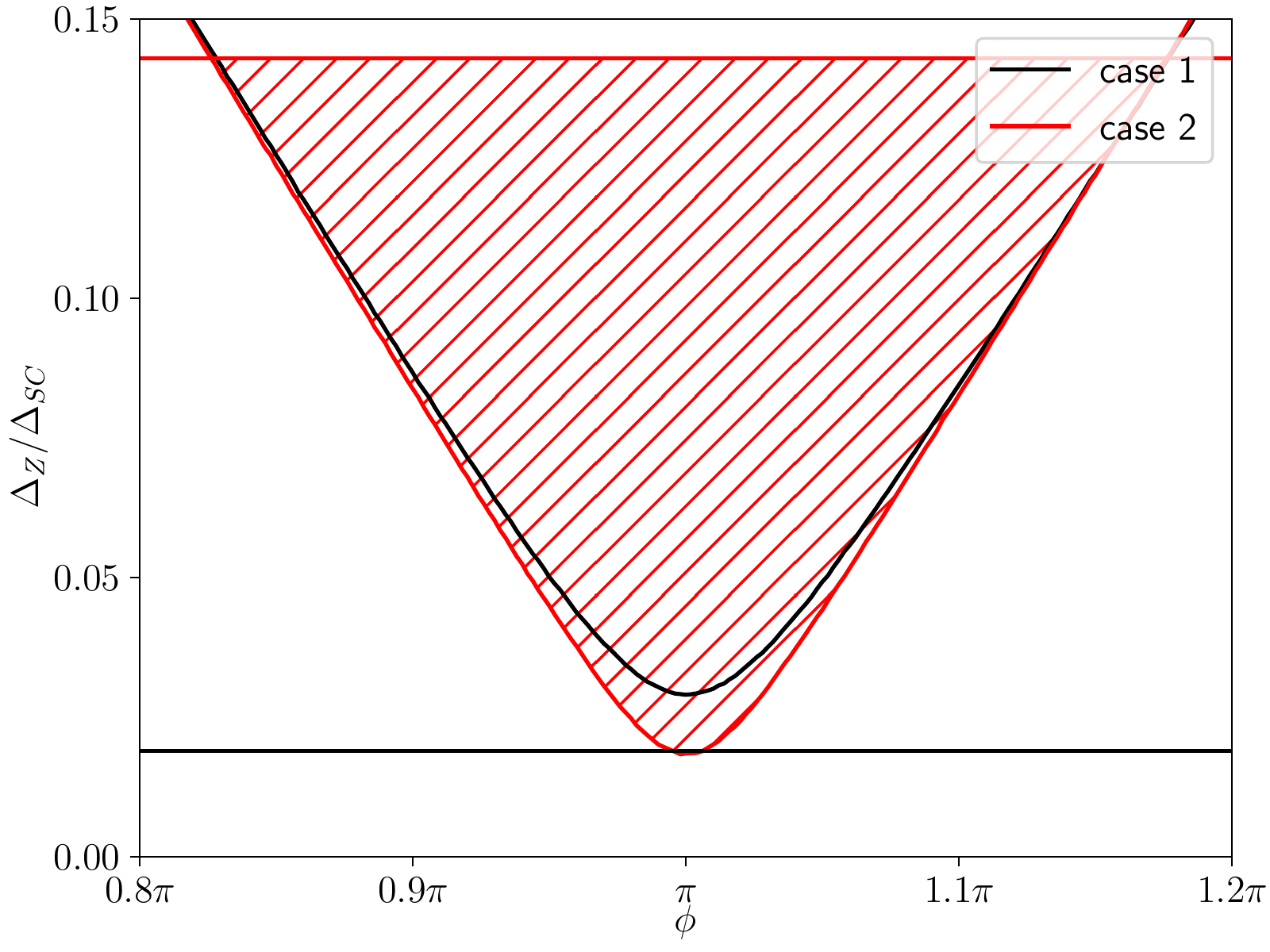}
	\caption{
		Phase diagram analogous to Fig.~\ref{fig:experimental_feasibility_phase_diagram_zeeman_everywhere}, with the only difference that a potential barrier [see Eq.~\eqref{eq:potential_barrier}] is included here. Due to the potential barrier, the topological phase can no longer be reached within the experimental limitations of case~1. Case~2 still has an experimentally accessible topological phase.
		The barrier height is $\mu_{\text{b}} = 0.06 E_\mathcal{E}$ and the barrier width is $W_{\text{b}} = l_\mathcal{E}$. All other parameters are as in Fig.~\ref{fig:experimental_feasibility_phase_diagram_zeeman_everywhere}
}
	\label{fig:experimental_feasibility_potential_barrier}
\end{figure}

The potential barrier changes the shape of the phase transition curve. Most crucially, it increases the minimally required Zeeman energy $\Delta_Z$ to reach the topological phase, see Fig.~\ref{fig:experimental_feasibility_potential_barrier}. This effect does not have a noticeably detrimental effect on the available topological phase space for case~2. However, even without a potential barrier, case~1 has a significantly smaller available topological phase space compared to case~2. A high enough potential barrier can therefore make the topological phase unreachable in case~1, see Fig.~\ref{fig:experimental_feasibility_potential_barrier}. 

\clearpage
\bibliography{bibliography}

\begin{thebibliography}{129}%
\makeatletter
\providecommand \@ifxundefined [1]{%
 \@ifx{#1\undefined}
}%
\providecommand \@ifnum [1]{%
 \ifnum #1\expandafter \@firstoftwo
 \else \expandafter \@secondoftwo
 \fi
}%
\providecommand \@ifx [1]{%
 \ifx #1\expandafter \@firstoftwo
 \else \expandafter \@secondoftwo
 \fi
}%
\providecommand \natexlab [1]{#1}%
\providecommand \enquote  [1]{``#1''}%
\providecommand \bibnamefont  [1]{#1}%
\providecommand \bibfnamefont [1]{#1}%
\providecommand \citenamefont [1]{#1}%
\providecommand \href@noop [0]{\@secondoftwo}%
\providecommand \href [0]{\begingroup \@sanitize@url \@href}%
\providecommand \@href[1]{\@@startlink{#1}\@@href}%
\providecommand \@@href[1]{\endgroup#1\@@endlink}%
\providecommand \@sanitize@url [0]{\catcode `\\12\catcode `\$12\catcode
  `\&12\catcode `\#12\catcode `\^12\catcode `\_12\catcode `\%12\relax}%
\providecommand \@@startlink[1]{}%
\providecommand \@@endlink[0]{}%
\providecommand \url  [0]{\begingroup\@sanitize@url \@url }%
\providecommand \@url [1]{\endgroup\@href {#1}{\urlprefix }}%
\providecommand \urlprefix  [0]{URL }%
\providecommand \Eprint [0]{\href }%
\providecommand \doibase [0]{https://doi.org/}%
\providecommand \selectlanguage [0]{\@gobble}%
\providecommand \bibinfo  [0]{\@secondoftwo}%
\providecommand \bibfield  [0]{\@secondoftwo}%
\providecommand \translation [1]{[#1]}%
\providecommand \BibitemOpen [0]{}%
\providecommand \bibitemStop [0]{}%
\providecommand \bibitemNoStop [0]{.\EOS\space}%
\providecommand \EOS [0]{\spacefactor3000\relax}%
\providecommand \BibitemShut  [1]{\csname bibitem#1\endcsname}%
\let\auto@bib@innerbib\@empty
\bibitem [{\citenamefont {Kitaev}(2001)}]{kitaev2001unpaired}%
  \BibitemOpen
  \bibfield  {author} {\bibinfo {author} {\bibfnamefont {A.~Y.}\ \bibnamefont
  {Kitaev}},\ }\bibfield  {title} {\bibinfo {title} {{Unpaired Majorana
  fermions in quantum wires}},\ }\href
  {https://doi.org/10.1070/1063-7869/44/10s/s29} {\bibfield  {journal}
  {\bibinfo  {journal} {Physics-Uspekhi}\ }\textbf {\bibinfo {volume} {44}},\
  \bibinfo {pages} {131} (\bibinfo {year} {2001})}\BibitemShut {NoStop}%
\bibitem [{\citenamefont {Leijnse}\ and\ \citenamefont
  {Flensberg}(2012)}]{leijnse2012introduction}%
  \BibitemOpen
  \bibfield  {author} {\bibinfo {author} {\bibfnamefont {M.}~\bibnamefont
  {Leijnse}}\ and\ \bibinfo {author} {\bibfnamefont {K.}~\bibnamefont
  {Flensberg}},\ }\bibfield  {title} {\bibinfo {title} {{Introduction to
  topological superconductivity and Majorana fermions}},\ }\href
  {https://doi.org/10.1088/0268-1242/27/12/124003} {\bibfield  {journal}
  {\bibinfo  {journal} {Semiconductor Science and Technology}\ }\textbf
  {\bibinfo {volume} {27}},\ \bibinfo {pages} {124003} (\bibinfo {year}
  {2012})}\BibitemShut {NoStop}%
\bibitem [{\citenamefont {Qi}\ and\ \citenamefont
  {Zhang}(2011)}]{qi2011topological}%
  \BibitemOpen
  \bibfield  {author} {\bibinfo {author} {\bibfnamefont {X.-L.}\ \bibnamefont
  {Qi}}\ and\ \bibinfo {author} {\bibfnamefont {S.-C.}\ \bibnamefont {Zhang}},\
  }\bibfield  {title} {\bibinfo {title} {{Topological insulators and
  superconductors}},\ }\href {https://doi.org/10.1103/RevModPhys.83.1057}
  {\bibfield  {journal} {\bibinfo  {journal} {Rev. Mod. Phys.}\ }\textbf
  {\bibinfo {volume} {83}},\ \bibinfo {pages} {1057} (\bibinfo {year}
  {2011})}\BibitemShut {NoStop}%
\bibitem [{\citenamefont {Beenakker}(2013)}]{beenakker2013search}%
  \BibitemOpen
  \bibfield  {author} {\bibinfo {author} {\bibfnamefont {C.}~\bibnamefont
  {Beenakker}},\ }\bibfield  {title} {\bibinfo {title} {{Search for Majorana
  Fermions in Superconductors}},\ }\href
  {https://doi.org/10.1146/annurev-conmatphys-030212-184337} {\bibfield
  {journal} {\bibinfo  {journal} {Annual Review of Condensed Matter Physics}\
  }\textbf {\bibinfo {volume} {4}},\ \bibinfo {pages} {113} (\bibinfo {year}
  {2013})}\BibitemShut {NoStop}%
\bibitem [{\citenamefont {Sato}\ and\ \citenamefont
  {Fujimoto}(2016)}]{sato2016majorana}%
  \BibitemOpen
  \bibfield  {author} {\bibinfo {author} {\bibfnamefont {M.}~\bibnamefont
  {Sato}}\ and\ \bibinfo {author} {\bibfnamefont {S.}~\bibnamefont
  {Fujimoto}},\ }\bibfield  {title} {\bibinfo {title} {{Majorana fermions and
  topology in superconductors}},\ }\href
  {https://journals.jps.jp/doi/full/10.7566/JPSJ.85.072001} {\bibfield
  {journal} {\bibinfo  {journal} {Journal of the Physical Society of Japan}\
  }\textbf {\bibinfo {volume} {85}},\ \bibinfo {pages} {072001} (\bibinfo
  {year} {2016})}\BibitemShut {NoStop}%
\bibitem [{\citenamefont {Pawlak}\ \emph {et~al.}(2019)\citenamefont {Pawlak},
  \citenamefont {Hoffman}, \citenamefont {Klinovaja}, \citenamefont {Loss},\
  and\ \citenamefont {Meyer}}]{pawlak2019majorana}%
  \BibitemOpen
  \bibfield  {author} {\bibinfo {author} {\bibfnamefont {R.}~\bibnamefont
  {Pawlak}}, \bibinfo {author} {\bibfnamefont {S.}~\bibnamefont {Hoffman}},
  \bibinfo {author} {\bibfnamefont {J.}~\bibnamefont {Klinovaja}}, \bibinfo
  {author} {\bibfnamefont {D.}~\bibnamefont {Loss}},\ and\ \bibinfo {author}
  {\bibfnamefont {E.}~\bibnamefont {Meyer}},\ }\bibfield  {title} {\bibinfo
  {title} {{Majorana fermions in magnetic chains}},\ }\href
  {https://doi.org/https://doi.org/10.1016/j.ppnp.2019.04.004} {\bibfield
  {journal} {\bibinfo  {journal} {Progress in Particle and Nuclear Physics}\
  }\textbf {\bibinfo {volume} {107}},\ \bibinfo {pages} {1} (\bibinfo {year}
  {2019})}\BibitemShut {NoStop}%
\bibitem [{\citenamefont {Laubscher}\ and\ \citenamefont
  {Klinovaja}(2021)}]{laubscher2021majorana}%
  \BibitemOpen
  \bibfield  {author} {\bibinfo {author} {\bibfnamefont {K.}~\bibnamefont
  {Laubscher}}\ and\ \bibinfo {author} {\bibfnamefont {J.}~\bibnamefont
  {Klinovaja}},\ }\bibfield  {title} {\bibinfo {title} {{Majorana bound states
  in semiconducting nanostructures}},\ }\href
  {https://doi.org/10.1063/5.0055997} {\bibfield  {journal} {\bibinfo
  {journal} {Journal of Applied Physics}\ }\textbf {\bibinfo {volume} {130}},\
  \bibinfo {pages} {081101} (\bibinfo {year} {2021})}\BibitemShut {NoStop}%
\bibitem [{\citenamefont {Ivanov}(2001)}]{invanov2001non}%
  \BibitemOpen
  \bibfield  {author} {\bibinfo {author} {\bibfnamefont {D.~A.}\ \bibnamefont
  {Ivanov}},\ }\bibfield  {title} {\bibinfo {title} {{Non-Abelian Statistics of
  Half-Quantum Vortices in $\mathit{p}$-Wave Superconductors}},\ }\href
  {https://doi.org/10.1103/PhysRevLett.86.268} {\bibfield  {journal} {\bibinfo
  {journal} {Phys. Rev. Lett.}\ }\textbf {\bibinfo {volume} {86}},\ \bibinfo
  {pages} {268} (\bibinfo {year} {2001})}\BibitemShut {NoStop}%
\bibitem [{\citenamefont {Kitaev}(2003)}]{kitaev2003fault}%
  \BibitemOpen
  \bibfield  {author} {\bibinfo {author} {\bibfnamefont {A.}~\bibnamefont
  {Kitaev}},\ }\bibfield  {title} {\bibinfo {title} {{Fault-tolerant quantum
  computation by anyons}},\ }\href
  {https://doi.org/https://doi.org/10.1016/S0003-4916(02)00018-0} {\bibfield
  {journal} {\bibinfo  {journal} {Annals of Physics}\ }\textbf {\bibinfo
  {volume} {303}},\ \bibinfo {pages} {2} (\bibinfo {year} {2003})}\BibitemShut
  {NoStop}%
\bibitem [{\citenamefont {Nayak}\ \emph {et~al.}(2008)\citenamefont {Nayak},
  \citenamefont {Simon}, \citenamefont {Stern}, \citenamefont {Freedman},\ and\
  \citenamefont {Das~Sarma}}]{nayak2008non}%
  \BibitemOpen
  \bibfield  {author} {\bibinfo {author} {\bibfnamefont {C.}~\bibnamefont
  {Nayak}}, \bibinfo {author} {\bibfnamefont {S.~H.}\ \bibnamefont {Simon}},
  \bibinfo {author} {\bibfnamefont {A.}~\bibnamefont {Stern}}, \bibinfo
  {author} {\bibfnamefont {M.}~\bibnamefont {Freedman}},\ and\ \bibinfo
  {author} {\bibfnamefont {S.}~\bibnamefont {Das~Sarma}},\ }\bibfield  {title}
  {\bibinfo {title} {{Non-Abelian anyons and topological quantum
  computation}},\ }\href {https://doi.org/10.1103/RevModPhys.80.1083}
  {\bibfield  {journal} {\bibinfo  {journal} {Rev. Mod. Phys.}\ }\textbf
  {\bibinfo {volume} {80}},\ \bibinfo {pages} {1083} (\bibinfo {year}
  {2008})}\BibitemShut {NoStop}%
\bibitem [{\citenamefont {Elliott}\ and\ \citenamefont
  {Franz}(2015)}]{elliot2015colloquium}%
  \BibitemOpen
  \bibfield  {author} {\bibinfo {author} {\bibfnamefont {S.~R.}\ \bibnamefont
  {Elliott}}\ and\ \bibinfo {author} {\bibfnamefont {M.}~\bibnamefont
  {Franz}},\ }\bibfield  {title} {\bibinfo {title} {{Colloquium: Majorana
  fermions in nuclear, particle, and solid-state physics}},\ }\href
  {https://doi.org/10.1103/RevModPhys.87.137} {\bibfield  {journal} {\bibinfo
  {journal} {Rev. Mod. Phys.}\ }\textbf {\bibinfo {volume} {87}},\ \bibinfo
  {pages} {137} (\bibinfo {year} {2015})}\BibitemShut {NoStop}%
\bibitem [{\citenamefont {Lutchyn}\ \emph {et~al.}(2010)\citenamefont
  {Lutchyn}, \citenamefont {Sau},\ and\ \citenamefont
  {Das~Sarma}}]{lutchyn2010majorana}%
  \BibitemOpen
  \bibfield  {author} {\bibinfo {author} {\bibfnamefont {R.~M.}\ \bibnamefont
  {Lutchyn}}, \bibinfo {author} {\bibfnamefont {J.~D.}\ \bibnamefont {Sau}},\
  and\ \bibinfo {author} {\bibfnamefont {S.}~\bibnamefont {Das~Sarma}},\
  }\bibfield  {title} {\bibinfo {title} {{Majorana Fermions and a Topological
  Phase Transition in Semiconductor-Superconductor Heterostructures}},\ }\href
  {https://doi.org/10.1103/PhysRevLett.105.077001} {\bibfield  {journal}
  {\bibinfo  {journal} {Phys. Rev. Lett.}\ }\textbf {\bibinfo {volume} {105}},\
  \bibinfo {pages} {077001} (\bibinfo {year} {2010})}\BibitemShut {NoStop}%
\bibitem [{\citenamefont {Oreg}\ \emph {et~al.}(2010)\citenamefont {Oreg},
  \citenamefont {Refael},\ and\ \citenamefont {von Oppen}}]{oreg2010helical}%
  \BibitemOpen
  \bibfield  {author} {\bibinfo {author} {\bibfnamefont {Y.}~\bibnamefont
  {Oreg}}, \bibinfo {author} {\bibfnamefont {G.}~\bibnamefont {Refael}},\ and\
  \bibinfo {author} {\bibfnamefont {F.}~\bibnamefont {von Oppen}},\ }\bibfield
  {title} {\bibinfo {title} {{Helical Liquids and Majorana Bound States in
  Quantum Wires}},\ }\href {https://doi.org/10.1103/PhysRevLett.105.177002}
  {\bibfield  {journal} {\bibinfo  {journal} {Phys. Rev. Lett.}\ }\textbf
  {\bibinfo {volume} {105}},\ \bibinfo {pages} {177002} (\bibinfo {year}
  {2010})}\BibitemShut {NoStop}%
\bibitem [{\citenamefont {Stanescu}\ \emph {et~al.}(2011)\citenamefont
  {Stanescu}, \citenamefont {Lutchyn},\ and\ \citenamefont
  {Das~Sarma}}]{stanescu2011majorana}%
  \BibitemOpen
  \bibfield  {author} {\bibinfo {author} {\bibfnamefont {T.~D.}\ \bibnamefont
  {Stanescu}}, \bibinfo {author} {\bibfnamefont {R.~M.}\ \bibnamefont
  {Lutchyn}},\ and\ \bibinfo {author} {\bibfnamefont {S.}~\bibnamefont
  {Das~Sarma}},\ }\bibfield  {title} {\bibinfo {title} {{Majorana fermions in
  semiconductor nanowires}},\ }\href
  {https://doi.org/10.1103/PhysRevB.84.144522} {\bibfield  {journal} {\bibinfo
  {journal} {Phys. Rev. B}\ }\textbf {\bibinfo {volume} {84}},\ \bibinfo
  {pages} {144522} (\bibinfo {year} {2011})}\BibitemShut {NoStop}%
\bibitem [{\citenamefont {Mourik}\ \emph {et~al.}(2012)\citenamefont {Mourik},
  \citenamefont {Zuo}, \citenamefont {Frolov}, \citenamefont {Plissard},
  \citenamefont {Bakkers},\ and\ \citenamefont
  {Kouwenhoven}}]{mourik2012signatures}%
  \BibitemOpen
  \bibfield  {author} {\bibinfo {author} {\bibfnamefont {V.}~\bibnamefont
  {Mourik}}, \bibinfo {author} {\bibfnamefont {K.}~\bibnamefont {Zuo}},
  \bibinfo {author} {\bibfnamefont {S.~M.}\ \bibnamefont {Frolov}}, \bibinfo
  {author} {\bibfnamefont {S.~R.}\ \bibnamefont {Plissard}}, \bibinfo {author}
  {\bibfnamefont {E.~P. A.~M.}\ \bibnamefont {Bakkers}},\ and\ \bibinfo
  {author} {\bibfnamefont {L.~P.}\ \bibnamefont {Kouwenhoven}},\ }\bibfield
  {title} {\bibinfo {title} {{Signatures of Majorana Fermions in Hybrid
  Superconductor-Semiconductor Nanowire Devices}},\ }\href
  {https://doi.org/10.1126/science.1222360} {\bibfield  {journal} {\bibinfo
  {journal} {Science}\ }\textbf {\bibinfo {volume} {336}},\ \bibinfo {pages}
  {1003} (\bibinfo {year} {2012})}\BibitemShut {NoStop}%
\bibitem [{\citenamefont {Das}\ \emph {et~al.}(2012)\citenamefont {Das},
  \citenamefont {Ronen}, \citenamefont {Most}, \citenamefont {Oreg},
  \citenamefont {Heiblum},\ and\ \citenamefont {Shtrikman}}]{das2012zero}%
  \BibitemOpen
  \bibfield  {author} {\bibinfo {author} {\bibfnamefont {A.}~\bibnamefont
  {Das}}, \bibinfo {author} {\bibfnamefont {Y.}~\bibnamefont {Ronen}}, \bibinfo
  {author} {\bibfnamefont {Y.}~\bibnamefont {Most}}, \bibinfo {author}
  {\bibfnamefont {Y.}~\bibnamefont {Oreg}}, \bibinfo {author} {\bibfnamefont
  {M.}~\bibnamefont {Heiblum}},\ and\ \bibinfo {author} {\bibfnamefont
  {H.}~\bibnamefont {Shtrikman}},\ }\bibfield  {title} {\bibinfo {title}
  {{Zero-bias peaks and splitting in an Al--InAs nanowire topological
  superconductor as a signature of Majorana fermions}},\ }\href
  {https://doi.org/10.1038/nphys2479} {\bibfield  {journal} {\bibinfo
  {journal} {Nature Physics}\ }\textbf {\bibinfo {volume} {8}},\ \bibinfo
  {pages} {887} (\bibinfo {year} {2012})}\BibitemShut {NoStop}%
\bibitem [{\citenamefont {Deng}\ \emph {et~al.}(2012)\citenamefont {Deng},
  \citenamefont {Yu}, \citenamefont {Huang}, \citenamefont {Larsson},
  \citenamefont {Caroff},\ and\ \citenamefont {Xu}}]{deng2012anomalous}%
  \BibitemOpen
  \bibfield  {author} {\bibinfo {author} {\bibfnamefont {M.~T.}\ \bibnamefont
  {Deng}}, \bibinfo {author} {\bibfnamefont {C.~L.}\ \bibnamefont {Yu}},
  \bibinfo {author} {\bibfnamefont {G.~Y.}\ \bibnamefont {Huang}}, \bibinfo
  {author} {\bibfnamefont {M.}~\bibnamefont {Larsson}}, \bibinfo {author}
  {\bibfnamefont {P.}~\bibnamefont {Caroff}},\ and\ \bibinfo {author}
  {\bibfnamefont {H.~Q.}\ \bibnamefont {Xu}},\ }\bibfield  {title} {\bibinfo
  {title} {{Anomalous Zero-Bias Conductance Peak in a Nb–InSb Nanowire–Nb
  Hybrid Device}},\ }\href {https://doi.org/10.1021/nl303758w} {\bibfield
  {journal} {\bibinfo  {journal} {Nano Letters}\ }\textbf {\bibinfo {volume}
  {12}},\ \bibinfo {pages} {6414} (\bibinfo {year} {2012})}\BibitemShut
  {NoStop}%
\bibitem [{\citenamefont {Klinovaja}\ \emph {et~al.}(2012)\citenamefont
  {Klinovaja}, \citenamefont {Gangadharaiah},\ and\ \citenamefont
  {Loss}}]{klinovaja2012electric}%
  \BibitemOpen
  \bibfield  {author} {\bibinfo {author} {\bibfnamefont {J.}~\bibnamefont
  {Klinovaja}}, \bibinfo {author} {\bibfnamefont {S.}~\bibnamefont
  {Gangadharaiah}},\ and\ \bibinfo {author} {\bibfnamefont {D.}~\bibnamefont
  {Loss}},\ }\bibfield  {title} {\bibinfo {title} {{Electric-field-induced
  Majorana Fermions in Armchair Carbon Nanotubes}},\ }\href
  {https://doi.org/10.1103/PhysRevLett.108.196804} {\bibfield  {journal}
  {\bibinfo  {journal} {Phys. Rev. Lett.}\ }\textbf {\bibinfo {volume} {108}},\
  \bibinfo {pages} {196804} (\bibinfo {year} {2012})}\BibitemShut {NoStop}%
\bibitem [{\citenamefont {Egger}\ and\ \citenamefont
  {Flensberg}(2012)}]{egger2012emerging}%
  \BibitemOpen
  \bibfield  {author} {\bibinfo {author} {\bibfnamefont {R.}~\bibnamefont
  {Egger}}\ and\ \bibinfo {author} {\bibfnamefont {K.}~\bibnamefont
  {Flensberg}},\ }\bibfield  {title} {\bibinfo {title} {{Emerging Dirac and
  Majorana fermions for carbon nanotubes with proximity-induced pairing and
  spiral magnetic field}},\ }\href {https://doi.org/10.1103/PhysRevB.85.235462}
  {\bibfield  {journal} {\bibinfo  {journal} {Phys. Rev. B}\ }\textbf {\bibinfo
  {volume} {85}},\ \bibinfo {pages} {235462} (\bibinfo {year}
  {2012})}\BibitemShut {NoStop}%
\bibitem [{\citenamefont {{Klinovaja, Jelena and Loss,
  Daniel}}(2013)}]{klinovaja2013giant}%
  \BibitemOpen
  \bibfield  {author} {\bibinfo {author} {\bibnamefont {{Klinovaja, Jelena and
  Loss, Daniel}}},\ }\bibfield  {title} {\bibinfo {title} {{Giant Spin-Orbit
  Interaction Due to Rotating Magnetic Fields in Graphene Nanoribbons}},\
  }\href {https://doi.org/10.1103/PhysRevX.3.011008} {\bibfield  {journal}
  {\bibinfo  {journal} {Phys. Rev. X}\ }\textbf {\bibinfo {volume} {3}},\
  \bibinfo {pages} {011008} (\bibinfo {year} {2013})}\BibitemShut {NoStop}%
\bibitem [{\citenamefont {Desjardins}\ \emph {et~al.}(2019)\citenamefont
  {Desjardins}, \citenamefont {Contamin}, \citenamefont {Delbecq},
  \citenamefont {Dartiailh}, \citenamefont {Bruhat}, \citenamefont {Cubaynes},
  \citenamefont {Viennot}, \citenamefont {Mallet}, \citenamefont {Rohart},
  \citenamefont {Thiaville} \emph {et~al.}}]{desjardins2019synthetic}%
  \BibitemOpen
  \bibfield  {author} {\bibinfo {author} {\bibfnamefont {M.}~\bibnamefont
  {Desjardins}}, \bibinfo {author} {\bibfnamefont {L.}~\bibnamefont
  {Contamin}}, \bibinfo {author} {\bibfnamefont {M.}~\bibnamefont {Delbecq}},
  \bibinfo {author} {\bibfnamefont {M.}~\bibnamefont {Dartiailh}}, \bibinfo
  {author} {\bibfnamefont {L.}~\bibnamefont {Bruhat}}, \bibinfo {author}
  {\bibfnamefont {T.}~\bibnamefont {Cubaynes}}, \bibinfo {author}
  {\bibfnamefont {J.}~\bibnamefont {Viennot}}, \bibinfo {author} {\bibfnamefont
  {F.}~\bibnamefont {Mallet}}, \bibinfo {author} {\bibfnamefont
  {S.}~\bibnamefont {Rohart}}, \bibinfo {author} {\bibfnamefont
  {A.}~\bibnamefont {Thiaville}}, \emph {et~al.},\ }\bibfield  {title}
  {\bibinfo {title} {{Synthetic spin--orbit interaction for Majorana
  devices}},\ }\href {https://doi.org/10.1038/s41563-019-0457-6} {\bibfield
  {journal} {\bibinfo  {journal} {{Nature Materials}}\ }\textbf {\bibinfo
  {volume} {18}},\ \bibinfo {pages} {1060} (\bibinfo {year}
  {2019})}\BibitemShut {NoStop}%
\bibitem [{\citenamefont {Choy}\ \emph {et~al.}(2011)\citenamefont {Choy},
  \citenamefont {Edge}, \citenamefont {Akhmerov},\ and\ \citenamefont
  {Beenakker}}]{choy2011majorana}%
  \BibitemOpen
  \bibfield  {author} {\bibinfo {author} {\bibfnamefont {T.-P.}\ \bibnamefont
  {Choy}}, \bibinfo {author} {\bibfnamefont {J.~M.}\ \bibnamefont {Edge}},
  \bibinfo {author} {\bibfnamefont {A.~R.}\ \bibnamefont {Akhmerov}},\ and\
  \bibinfo {author} {\bibfnamefont {C.~W.~J.}\ \bibnamefont {Beenakker}},\
  }\bibfield  {title} {\bibinfo {title} {{Majorana fermions emerging from
  magnetic nanoparticles on a superconductor without spin-orbit coupling}},\
  }\href {https://doi.org/10.1103/PhysRevB.84.195442} {\bibfield  {journal}
  {\bibinfo  {journal} {Phys. Rev. B}\ }\textbf {\bibinfo {volume} {84}},\
  \bibinfo {pages} {195442} (\bibinfo {year} {2011})}\BibitemShut {NoStop}%
\bibitem [{\citenamefont {Nadj-Perge}\ \emph {et~al.}(2013)\citenamefont
  {Nadj-Perge}, \citenamefont {Drozdov}, \citenamefont {Bernevig},\ and\
  \citenamefont {Yazdani}}]{nadj2013proposal}%
  \BibitemOpen
  \bibfield  {author} {\bibinfo {author} {\bibfnamefont {S.}~\bibnamefont
  {Nadj-Perge}}, \bibinfo {author} {\bibfnamefont {I.~K.}\ \bibnamefont
  {Drozdov}}, \bibinfo {author} {\bibfnamefont {B.~A.}\ \bibnamefont
  {Bernevig}},\ and\ \bibinfo {author} {\bibfnamefont {A.}~\bibnamefont
  {Yazdani}},\ }\bibfield  {title} {\bibinfo {title} {{Proposal for realizing
  Majorana fermions in chains of magnetic atoms on a superconductor}},\ }\href
  {https://doi.org/10.1103/PhysRevB.88.020407} {\bibfield  {journal} {\bibinfo
  {journal} {Phys. Rev. B}\ }\textbf {\bibinfo {volume} {88}},\ \bibinfo
  {pages} {020407} (\bibinfo {year} {2013})}\BibitemShut {NoStop}%
\bibitem [{\citenamefont {Braunecker}\ and\ \citenamefont
  {Simon}(2013)}]{braunecker2013interplay}%
  \BibitemOpen
  \bibfield  {author} {\bibinfo {author} {\bibfnamefont {B.}~\bibnamefont
  {Braunecker}}\ and\ \bibinfo {author} {\bibfnamefont {P.}~\bibnamefont
  {Simon}},\ }\bibfield  {title} {\bibinfo {title} {{Interplay between
  Classical Magnetic Moments and Superconductivity in Quantum One-Dimensional
  Conductors: Toward a Self-Sustained Topological Majorana Phase}},\ }\href
  {https://doi.org/10.1103/PhysRevLett.111.147202} {\bibfield  {journal}
  {\bibinfo  {journal} {Phys. Rev. Lett.}\ }\textbf {\bibinfo {volume} {111}},\
  \bibinfo {pages} {147202} (\bibinfo {year} {2013})}\BibitemShut {NoStop}%
\bibitem [{\citenamefont {Pientka}\ \emph {et~al.}(2013)\citenamefont
  {Pientka}, \citenamefont {Glazman},\ and\ \citenamefont {von
  Oppen}}]{pientka2013topological}%
  \BibitemOpen
  \bibfield  {author} {\bibinfo {author} {\bibfnamefont {F.}~\bibnamefont
  {Pientka}}, \bibinfo {author} {\bibfnamefont {L.~I.}\ \bibnamefont
  {Glazman}},\ and\ \bibinfo {author} {\bibfnamefont {F.}~\bibnamefont {von
  Oppen}},\ }\bibfield  {title} {\bibinfo {title} {{Topological superconducting
  phase in helical Shiba chains}},\ }\href
  {https://doi.org/10.1103/PhysRevB.88.155420} {\bibfield  {journal} {\bibinfo
  {journal} {Phys. Rev. B}\ }\textbf {\bibinfo {volume} {88}},\ \bibinfo
  {pages} {155420} (\bibinfo {year} {2013})}\BibitemShut {NoStop}%
\bibitem [{\citenamefont {Klinovaja}\ \emph {et~al.}(2013)\citenamefont
  {Klinovaja}, \citenamefont {Stano}, \citenamefont {Yazdani},\ and\
  \citenamefont {Loss}}]{klinovaja2013topological}%
  \BibitemOpen
  \bibfield  {author} {\bibinfo {author} {\bibfnamefont {J.}~\bibnamefont
  {Klinovaja}}, \bibinfo {author} {\bibfnamefont {P.}~\bibnamefont {Stano}},
  \bibinfo {author} {\bibfnamefont {A.}~\bibnamefont {Yazdani}},\ and\ \bibinfo
  {author} {\bibfnamefont {D.}~\bibnamefont {Loss}},\ }\bibfield  {title}
  {\bibinfo {title} {{Topological Superconductivity and Majorana Fermions in
  RKKY Systems}},\ }\href {https://doi.org/10.1103/PhysRevLett.111.186805}
  {\bibfield  {journal} {\bibinfo  {journal} {Phys. Rev. Lett.}\ }\textbf
  {\bibinfo {volume} {111}},\ \bibinfo {pages} {186805} (\bibinfo {year}
  {2013})}\BibitemShut {NoStop}%
\bibitem [{\citenamefont {Vazifeh}\ and\ \citenamefont
  {Franz}(2013)}]{vazifeh2013self}%
  \BibitemOpen
  \bibfield  {author} {\bibinfo {author} {\bibfnamefont {M.~M.}\ \bibnamefont
  {Vazifeh}}\ and\ \bibinfo {author} {\bibfnamefont {M.}~\bibnamefont
  {Franz}},\ }\bibfield  {title} {\bibinfo {title} {{Self-Organized Topological
  State with Majorana Fermions}},\ }\href
  {https://doi.org/10.1103/PhysRevLett.111.206802} {\bibfield  {journal}
  {\bibinfo  {journal} {Phys. Rev. Lett.}\ }\textbf {\bibinfo {volume} {111}},\
  \bibinfo {pages} {206802} (\bibinfo {year} {2013})}\BibitemShut {NoStop}%
\bibitem [{\citenamefont {Nadj-Perge}\ \emph {et~al.}(2014)\citenamefont
  {Nadj-Perge}, \citenamefont {Drozdov}, \citenamefont {Li}, \citenamefont
  {Chen}, \citenamefont {Jeon}, \citenamefont {Seo}, \citenamefont {MacDonald},
  \citenamefont {Bernevig},\ and\ \citenamefont
  {Yazdani}}]{nadj2014observation}%
  \BibitemOpen
  \bibfield  {author} {\bibinfo {author} {\bibfnamefont {S.}~\bibnamefont
  {Nadj-Perge}}, \bibinfo {author} {\bibfnamefont {I.~K.}\ \bibnamefont
  {Drozdov}}, \bibinfo {author} {\bibfnamefont {J.}~\bibnamefont {Li}},
  \bibinfo {author} {\bibfnamefont {H.}~\bibnamefont {Chen}}, \bibinfo {author}
  {\bibfnamefont {S.}~\bibnamefont {Jeon}}, \bibinfo {author} {\bibfnamefont
  {J.}~\bibnamefont {Seo}}, \bibinfo {author} {\bibfnamefont {A.~H.}\
  \bibnamefont {MacDonald}}, \bibinfo {author} {\bibfnamefont {B.~A.}\
  \bibnamefont {Bernevig}},\ and\ \bibinfo {author} {\bibfnamefont
  {A.}~\bibnamefont {Yazdani}},\ }\bibfield  {title} {\bibinfo {title}
  {{Observation of Majorana fermions in ferromagnetic atomic chains on a
  superconductor}},\ }\href
  {https://www.science.org/doi/abs/10.1126/science.1259327} {\bibfield
  {journal} {\bibinfo  {journal} {Science}\ }\textbf {\bibinfo {volume} {346}}
  (\bibinfo {year} {2014})}\BibitemShut {NoStop}%
\bibitem [{\citenamefont {Ruby}\ \emph {et~al.}(2015)\citenamefont {Ruby},
  \citenamefont {Pientka}, \citenamefont {Peng}, \citenamefont {von Oppen},
  \citenamefont {Heinrich},\ and\ \citenamefont {Franke}}]{ruby2015end}%
  \BibitemOpen
  \bibfield  {author} {\bibinfo {author} {\bibfnamefont {M.}~\bibnamefont
  {Ruby}}, \bibinfo {author} {\bibfnamefont {F.}~\bibnamefont {Pientka}},
  \bibinfo {author} {\bibfnamefont {Y.}~\bibnamefont {Peng}}, \bibinfo {author}
  {\bibfnamefont {F.}~\bibnamefont {von Oppen}}, \bibinfo {author}
  {\bibfnamefont {B.~W.}\ \bibnamefont {Heinrich}},\ and\ \bibinfo {author}
  {\bibfnamefont {K.~J.}\ \bibnamefont {Franke}},\ }\bibfield  {title}
  {\bibinfo {title} {{End States and Subgap Structure in Proximity-Coupled
  Chains of Magnetic Adatoms}},\ }\href
  {https://doi.org/10.1103/PhysRevLett.115.197204} {\bibfield  {journal}
  {\bibinfo  {journal} {Phys. Rev. Lett.}\ }\textbf {\bibinfo {volume} {115}},\
  \bibinfo {pages} {197204} (\bibinfo {year} {2015})}\BibitemShut {NoStop}%
\bibitem [{\citenamefont {Pawlak}\ \emph {et~al.}(2016)\citenamefont {Pawlak},
  \citenamefont {Kisiel}, \citenamefont {Klinovaja}, \citenamefont {Meier},
  \citenamefont {Kawai}, \citenamefont {Glatzel}, \citenamefont {Loss},\ and\
  \citenamefont {Meyer}}]{pawlak2016probing}%
  \BibitemOpen
  \bibfield  {author} {\bibinfo {author} {\bibfnamefont {R.}~\bibnamefont
  {Pawlak}}, \bibinfo {author} {\bibfnamefont {M.}~\bibnamefont {Kisiel}},
  \bibinfo {author} {\bibfnamefont {J.}~\bibnamefont {Klinovaja}}, \bibinfo
  {author} {\bibfnamefont {T.}~\bibnamefont {Meier}}, \bibinfo {author}
  {\bibfnamefont {S.}~\bibnamefont {Kawai}}, \bibinfo {author} {\bibfnamefont
  {T.}~\bibnamefont {Glatzel}}, \bibinfo {author} {\bibfnamefont
  {D.}~\bibnamefont {Loss}},\ and\ \bibinfo {author} {\bibfnamefont
  {E.}~\bibnamefont {Meyer}},\ }\bibfield  {title} {\bibinfo {title} {{Probing
  atomic structure and Majorana wavefunctions in mono-atomic Fe chains on
  superconducting Pb surface}},\ }\href {https://doi.org/10.1038/npjqi.2016.35}
  {\bibfield  {journal} {\bibinfo  {journal} {npj Quantum Information}\
  }\textbf {\bibinfo {volume} {2}},\ \bibinfo {pages} {16035} (\bibinfo {year}
  {2016})}\BibitemShut {NoStop}%
\bibitem [{\citenamefont {J{\"a}ck}\ \emph {et~al.}(2021)\citenamefont
  {J{\"a}ck}, \citenamefont {Xie},\ and\ \citenamefont
  {Yazdani}}]{jack2021detecting}%
  \BibitemOpen
  \bibfield  {author} {\bibinfo {author} {\bibfnamefont {B.}~\bibnamefont
  {J{\"a}ck}}, \bibinfo {author} {\bibfnamefont {Y.}~\bibnamefont {Xie}},\ and\
  \bibinfo {author} {\bibfnamefont {A.}~\bibnamefont {Yazdani}},\ }\bibfield
  {title} {\bibinfo {title} {{Detecting and distinguishing Majorana zero modes
  with the scanning tunnelling microscope}},\ }\href
  {https://doi.org/10.1038/s42254-021-00328-z} {\bibfield  {journal} {\bibinfo
  {journal} {Nature Reviews Physics}\ }\textbf {\bibinfo {volume} {3}},\
  \bibinfo {pages} {541} (\bibinfo {year} {2021})}\BibitemShut {NoStop}%
\bibitem [{\citenamefont {Fu}\ and\ \citenamefont
  {Kane}(2008)}]{fu2008superconducting}%
  \BibitemOpen
  \bibfield  {author} {\bibinfo {author} {\bibfnamefont {L.}~\bibnamefont
  {Fu}}\ and\ \bibinfo {author} {\bibfnamefont {C.~L.}\ \bibnamefont {Kane}},\
  }\bibfield  {title} {\bibinfo {title} {{Superconducting Proximity Effect and
  Majorana Fermions at the Surface of a Topological Insulator}},\ }\href
  {https://doi.org/10.1103/PhysRevLett.100.096407} {\bibfield  {journal}
  {\bibinfo  {journal} {Phys. Rev. Lett.}\ }\textbf {\bibinfo {volume} {100}},\
  \bibinfo {pages} {096407} (\bibinfo {year} {2008})}\BibitemShut {NoStop}%
\bibitem [{\citenamefont {Fu}\ and\ \citenamefont
  {Kane}(2009)}]{fu2009josephson}%
  \BibitemOpen
  \bibfield  {author} {\bibinfo {author} {\bibfnamefont {L.}~\bibnamefont
  {Fu}}\ and\ \bibinfo {author} {\bibfnamefont {C.~L.}\ \bibnamefont {Kane}},\
  }\bibfield  {title} {\bibinfo {title} {{Josephson current and noise at a
  superconductor/quantum-spin-Hall-insulator/superconductor junction}},\ }\href
  {https://doi.org/10.1103/PhysRevB.79.161408} {\bibfield  {journal} {\bibinfo
  {journal} {Phys. Rev. B}\ }\textbf {\bibinfo {volume} {79}},\ \bibinfo
  {pages} {161408} (\bibinfo {year} {2009})}\BibitemShut {NoStop}%
\bibitem [{\citenamefont {Cook}\ and\ \citenamefont
  {Franz}(2011)}]{cook2011majorana}%
  \BibitemOpen
  \bibfield  {author} {\bibinfo {author} {\bibfnamefont {A.}~\bibnamefont
  {Cook}}\ and\ \bibinfo {author} {\bibfnamefont {M.}~\bibnamefont {Franz}},\
  }\bibfield  {title} {\bibinfo {title} {{Majorana fermions in a
  topological-insulator nanowire proximity-coupled to an $s$-wave
  superconductor}},\ }\href {https://doi.org/10.1103/PhysRevB.84.201105}
  {\bibfield  {journal} {\bibinfo  {journal} {Phys. Rev. B}\ }\textbf {\bibinfo
  {volume} {84}},\ \bibinfo {pages} {201105} (\bibinfo {year}
  {2011})}\BibitemShut {NoStop}%
\bibitem [{\citenamefont {Cook}\ \emph {et~al.}(2012)\citenamefont {Cook},
  \citenamefont {Vazifeh},\ and\ \citenamefont {Franz}}]{cook2012stability}%
  \BibitemOpen
  \bibfield  {author} {\bibinfo {author} {\bibfnamefont {A.~M.}\ \bibnamefont
  {Cook}}, \bibinfo {author} {\bibfnamefont {M.~M.}\ \bibnamefont {Vazifeh}},\
  and\ \bibinfo {author} {\bibfnamefont {M.}~\bibnamefont {Franz}},\ }\bibfield
   {title} {\bibinfo {title} {{Stability of Majorana fermions in
  proximity-coupled topological insulator nanowires}},\ }\href
  {https://doi.org/10.1103/PhysRevB.86.155431} {\bibfield  {journal} {\bibinfo
  {journal} {Phys. Rev. B}\ }\textbf {\bibinfo {volume} {86}},\ \bibinfo
  {pages} {155431} (\bibinfo {year} {2012})}\BibitemShut {NoStop}%
\bibitem [{\citenamefont {Jäck}\ \emph {et~al.}(2019)\citenamefont {Jäck},
  \citenamefont {Xie}, \citenamefont {Li}, \citenamefont {Jeon}, \citenamefont
  {Bernevig},\ and\ \citenamefont {Yazdani}}]{jaeck2019observation}%
  \BibitemOpen
  \bibfield  {author} {\bibinfo {author} {\bibfnamefont {B.}~\bibnamefont
  {Jäck}}, \bibinfo {author} {\bibfnamefont {Y.}~\bibnamefont {Xie}}, \bibinfo
  {author} {\bibfnamefont {J.}~\bibnamefont {Li}}, \bibinfo {author}
  {\bibfnamefont {S.}~\bibnamefont {Jeon}}, \bibinfo {author} {\bibfnamefont
  {B.~A.}\ \bibnamefont {Bernevig}},\ and\ \bibinfo {author} {\bibfnamefont
  {A.}~\bibnamefont {Yazdani}},\ }\bibfield  {title} {\bibinfo {title}
  {{Observation of a Majorana zero mode in a topologically protected edge
  channel}},\ }\href {https://doi.org/10.1126/science.aax1444} {\bibfield
  {journal} {\bibinfo  {journal} {Science}\ }\textbf {\bibinfo {volume}
  {364}},\ \bibinfo {pages} {1255} (\bibinfo {year} {2019})}\BibitemShut
  {NoStop}%
\bibitem [{\citenamefont {Legg}\ \emph {et~al.}(2021)\citenamefont {Legg},
  \citenamefont {Loss},\ and\ \citenamefont {Klinovaja}}]{legg2021majorana}%
  \BibitemOpen
  \bibfield  {author} {\bibinfo {author} {\bibfnamefont {H.~F.}\ \bibnamefont
  {Legg}}, \bibinfo {author} {\bibfnamefont {D.}~\bibnamefont {Loss}},\ and\
  \bibinfo {author} {\bibfnamefont {J.}~\bibnamefont {Klinovaja}},\ }\bibfield
  {title} {\bibinfo {title} {{Majorana bound states in topological insulators
  without a vortex}},\ }\href {https://doi.org/10.1103/PhysRevB.104.165405}
  {\bibfield  {journal} {\bibinfo  {journal} {Phys. Rev. B}\ }\textbf {\bibinfo
  {volume} {104}},\ \bibinfo {pages} {165405} (\bibinfo {year}
  {2021})}\BibitemShut {NoStop}%
\bibitem [{\citenamefont {Sau}\ \emph {et~al.}(2010{\natexlab{a}})\citenamefont
  {Sau}, \citenamefont {Lutchyn}, \citenamefont {Tewari},\ and\ \citenamefont
  {Das~Sarma}}]{sau2010generic}%
  \BibitemOpen
  \bibfield  {author} {\bibinfo {author} {\bibfnamefont {J.~D.}\ \bibnamefont
  {Sau}}, \bibinfo {author} {\bibfnamefont {R.~M.}\ \bibnamefont {Lutchyn}},
  \bibinfo {author} {\bibfnamefont {S.}~\bibnamefont {Tewari}},\ and\ \bibinfo
  {author} {\bibfnamefont {S.}~\bibnamefont {Das~Sarma}},\ }\bibfield  {title}
  {\bibinfo {title} {{Generic New Platform for Topological Quantum Computation
  Using Semiconductor Heterostructures}},\ }\href
  {https://doi.org/10.1103/PhysRevLett.104.040502} {\bibfield  {journal}
  {\bibinfo  {journal} {Phys. Rev. Lett.}\ }\textbf {\bibinfo {volume} {104}},\
  \bibinfo {pages} {040502} (\bibinfo {year} {2010}{\natexlab{a}})}\BibitemShut
  {NoStop}%
\bibitem [{\citenamefont {Alicea}(2010)}]{alicea2010majorana}%
  \BibitemOpen
  \bibfield  {author} {\bibinfo {author} {\bibfnamefont {J.}~\bibnamefont
  {Alicea}},\ }\bibfield  {title} {\bibinfo {title} {{Majorana fermions in a
  tunable semiconductor device}},\ }\href
  {https://doi.org/10.1103/PhysRevB.81.125318} {\bibfield  {journal} {\bibinfo
  {journal} {Phys. Rev. B}\ }\textbf {\bibinfo {volume} {81}},\ \bibinfo
  {pages} {125318} (\bibinfo {year} {2010})}\BibitemShut {NoStop}%
\bibitem [{\citenamefont {Sau}\ \emph {et~al.}(2010{\natexlab{b}})\citenamefont
  {Sau}, \citenamefont {Tewari}, \citenamefont {Lutchyn}, \citenamefont
  {Stanescu},\ and\ \citenamefont {Das~Sarma}}]{sau2010non}%
  \BibitemOpen
  \bibfield  {author} {\bibinfo {author} {\bibfnamefont {J.~D.}\ \bibnamefont
  {Sau}}, \bibinfo {author} {\bibfnamefont {S.}~\bibnamefont {Tewari}},
  \bibinfo {author} {\bibfnamefont {R.~M.}\ \bibnamefont {Lutchyn}}, \bibinfo
  {author} {\bibfnamefont {T.~D.}\ \bibnamefont {Stanescu}},\ and\ \bibinfo
  {author} {\bibfnamefont {S.}~\bibnamefont {Das~Sarma}},\ }\bibfield  {title}
  {\bibinfo {title} {{Non-Abelian quantum order in spin-orbit-coupled
  semiconductors: Search for topological Majorana particles in solid-state
  systems}},\ }\href {https://doi.org/10.1103/PhysRevB.82.214509} {\bibfield
  {journal} {\bibinfo  {journal} {Phys. Rev. B}\ }\textbf {\bibinfo {volume}
  {82}},\ \bibinfo {pages} {214509} (\bibinfo {year}
  {2010}{\natexlab{b}})}\BibitemShut {NoStop}%
\bibitem [{\citenamefont {Pientka}\ \emph {et~al.}(2017)\citenamefont
  {Pientka}, \citenamefont {Keselman}, \citenamefont {Berg}, \citenamefont
  {Yacoby}, \citenamefont {Stern},\ and\ \citenamefont
  {Halperin}}]{pientka2017topological}%
  \BibitemOpen
  \bibfield  {author} {\bibinfo {author} {\bibfnamefont {F.}~\bibnamefont
  {Pientka}}, \bibinfo {author} {\bibfnamefont {A.}~\bibnamefont {Keselman}},
  \bibinfo {author} {\bibfnamefont {E.}~\bibnamefont {Berg}}, \bibinfo {author}
  {\bibfnamefont {A.}~\bibnamefont {Yacoby}}, \bibinfo {author} {\bibfnamefont
  {A.}~\bibnamefont {Stern}},\ and\ \bibinfo {author} {\bibfnamefont {B.~I.}\
  \bibnamefont {Halperin}},\ }\bibfield  {title} {\bibinfo {title}
  {{Topological Superconductivity in a Planar Josephson Junction}},\ }\href
  {https://doi.org/10.1103/PhysRevX.7.021032} {\bibfield  {journal} {\bibinfo
  {journal} {Phys. Rev. X}\ }\textbf {\bibinfo {volume} {7}},\ \bibinfo {pages}
  {021032} (\bibinfo {year} {2017})}\BibitemShut {NoStop}%
\bibitem [{\citenamefont {Hell}\ \emph
  {et~al.}(2017{\natexlab{a}})\citenamefont {Hell}, \citenamefont {Leijnse},\
  and\ \citenamefont {Flensberg}}]{hell2017two}%
  \BibitemOpen
  \bibfield  {author} {\bibinfo {author} {\bibfnamefont {M.}~\bibnamefont
  {Hell}}, \bibinfo {author} {\bibfnamefont {M.}~\bibnamefont {Leijnse}},\ and\
  \bibinfo {author} {\bibfnamefont {K.}~\bibnamefont {Flensberg}},\ }\bibfield
  {title} {\bibinfo {title} {{Two-Dimensional Platform for Networks of Majorana
  Bound States}},\ }\href {https://doi.org/10.1103/PhysRevLett.118.107701}
  {\bibfield  {journal} {\bibinfo  {journal} {Phys. Rev. Lett.}\ }\textbf
  {\bibinfo {volume} {118}},\ \bibinfo {pages} {107701} (\bibinfo {year}
  {2017}{\natexlab{a}})}\BibitemShut {NoStop}%
\bibitem [{\citenamefont {Setiawan}\ \emph {et~al.}(2019)\citenamefont
  {Setiawan}, \citenamefont {Stern},\ and\ \citenamefont
  {Berg}}]{setiawan2019topological}%
  \BibitemOpen
  \bibfield  {author} {\bibinfo {author} {\bibfnamefont {F.}~\bibnamefont
  {Setiawan}}, \bibinfo {author} {\bibfnamefont {A.}~\bibnamefont {Stern}},\
  and\ \bibinfo {author} {\bibfnamefont {E.}~\bibnamefont {Berg}},\ }\bibfield
  {title} {\bibinfo {title} {{Topological superconductivity in planar Josephson
  junctions: Narrowing down to the nanowire limit}},\ }\href
  {https://doi.org/10.1103/PhysRevB.99.220506} {\bibfield  {journal} {\bibinfo
  {journal} {Phys. Rev. B}\ }\textbf {\bibinfo {volume} {99}},\ \bibinfo
  {pages} {220506} (\bibinfo {year} {2019})}\BibitemShut {NoStop}%
\bibitem [{\citenamefont {Scharf}\ \emph {et~al.}(2019)\citenamefont {Scharf},
  \citenamefont {Pientka}, \citenamefont {Ren}, \citenamefont {Yacoby},\ and\
  \citenamefont {Hankiewicz}}]{scharf2019tuning}%
  \BibitemOpen
  \bibfield  {author} {\bibinfo {author} {\bibfnamefont {B.}~\bibnamefont
  {Scharf}}, \bibinfo {author} {\bibfnamefont {F.}~\bibnamefont {Pientka}},
  \bibinfo {author} {\bibfnamefont {H.}~\bibnamefont {Ren}}, \bibinfo {author}
  {\bibfnamefont {A.}~\bibnamefont {Yacoby}},\ and\ \bibinfo {author}
  {\bibfnamefont {E.~M.}\ \bibnamefont {Hankiewicz}},\ }\bibfield  {title}
  {\bibinfo {title} {{Tuning topological superconductivity in phase-controlled
  Josephson junctions with Rashba and Dresselhaus spin-orbit coupling}},\
  }\href {https://doi.org/10.1103/PhysRevB.99.214503} {\bibfield  {journal}
  {\bibinfo  {journal} {Phys. Rev. B}\ }\textbf {\bibinfo {volume} {99}},\
  \bibinfo {pages} {214503} (\bibinfo {year} {2019})}\BibitemShut {NoStop}%
\bibitem [{\citenamefont {Melo}\ \emph {et~al.}(2019)\citenamefont {Melo},
  \citenamefont {Rubbert},\ and\ \citenamefont
  {Akhmerov}}]{melo2019supercurrent}%
  \BibitemOpen
  \bibfield  {author} {\bibinfo {author} {\bibfnamefont {A.}~\bibnamefont
  {Melo}}, \bibinfo {author} {\bibfnamefont {S.}~\bibnamefont {Rubbert}},\ and\
  \bibinfo {author} {\bibfnamefont {A.~R.}\ \bibnamefont {Akhmerov}},\
  }\bibfield  {title} {\bibinfo {title} {{Supercurrent-induced Majorana bound
  states in a planar geometry}},\ }\href
  {https://scipost.org/10.21468/SciPostPhys.7.3.039} {\bibfield  {journal}
  {\bibinfo  {journal} {SciPost Phys.}\ }\textbf {\bibinfo {volume} {7}},\
  \bibinfo {pages} {39} (\bibinfo {year} {2019})}\BibitemShut {NoStop}%
\bibitem [{\citenamefont {Laeven}\ \emph {et~al.}(2020)\citenamefont {Laeven},
  \citenamefont {Nijholt}, \citenamefont {Wimmer},\ and\ \citenamefont
  {Akhmerov}}]{laeven2020enhanced}%
  \BibitemOpen
  \bibfield  {author} {\bibinfo {author} {\bibfnamefont {T.}~\bibnamefont
  {Laeven}}, \bibinfo {author} {\bibfnamefont {B.}~\bibnamefont {Nijholt}},
  \bibinfo {author} {\bibfnamefont {M.}~\bibnamefont {Wimmer}},\ and\ \bibinfo
  {author} {\bibfnamefont {A.~R.}\ \bibnamefont {Akhmerov}},\ }\bibfield
  {title} {\bibinfo {title} {{Enhanced Proximity Effect in Zigzag-Shaped
  Majorana Josephson Junctions}},\ }\href
  {https://doi.org/10.1103/PhysRevLett.125.086802} {\bibfield  {journal}
  {\bibinfo  {journal} {Phys. Rev. Lett.}\ }\textbf {\bibinfo {volume} {125}},\
  \bibinfo {pages} {086802} (\bibinfo {year} {2020})}\BibitemShut {NoStop}%
\bibitem [{\citenamefont {Volpez}\ \emph {et~al.}(2020)\citenamefont {Volpez},
  \citenamefont {Loss},\ and\ \citenamefont {Klinovaja}}]{volpez2020time}%
  \BibitemOpen
  \bibfield  {author} {\bibinfo {author} {\bibfnamefont {Y.}~\bibnamefont
  {Volpez}}, \bibinfo {author} {\bibfnamefont {D.}~\bibnamefont {Loss}},\ and\
  \bibinfo {author} {\bibfnamefont {J.}~\bibnamefont {Klinovaja}},\ }\bibfield
  {title} {\bibinfo {title} {{Time-reversal invariant topological
  superconductivity in planar Josephson bijunction}},\ }\href
  {https://doi.org/10.1103/PhysRevResearch.2.023415} {\bibfield  {journal}
  {\bibinfo  {journal} {Phys. Rev. Research}\ }\textbf {\bibinfo {volume}
  {2}},\ \bibinfo {pages} {023415} (\bibinfo {year} {2020})}\BibitemShut
  {NoStop}%
\bibitem [{\citenamefont {Paudel}\ \emph {et~al.}(2021)\citenamefont {Paudel},
  \citenamefont {Cole}, \citenamefont {Woods},\ and\ \citenamefont
  {Stanescu}}]{paudel2021enhanced}%
  \BibitemOpen
  \bibfield  {author} {\bibinfo {author} {\bibfnamefont {P.~P.}\ \bibnamefont
  {Paudel}}, \bibinfo {author} {\bibfnamefont {T.}~\bibnamefont {Cole}},
  \bibinfo {author} {\bibfnamefont {B.~D.}\ \bibnamefont {Woods}},\ and\
  \bibinfo {author} {\bibfnamefont {T.~D.}\ \bibnamefont {Stanescu}},\
  }\bibfield  {title} {\bibinfo {title} {{Enhanced topological
  superconductivity in spatially modulated planar Josephson junctions}},\
  }\href {https://doi.org/10.1103/PhysRevB.104.155428} {\bibfield  {journal}
  {\bibinfo  {journal} {Phys. Rev. B}\ }\textbf {\bibinfo {volume} {104}},\
  \bibinfo {pages} {155428} (\bibinfo {year} {2021})}\BibitemShut {NoStop}%
\bibitem [{\citenamefont {Pakizer}\ \emph {et~al.}(2021)\citenamefont
  {Pakizer}, \citenamefont {Scharf},\ and\ \citenamefont
  {Matos-Abiague}}]{pakizer2021crystalline}%
  \BibitemOpen
  \bibfield  {author} {\bibinfo {author} {\bibfnamefont {J.~D.}\ \bibnamefont
  {Pakizer}}, \bibinfo {author} {\bibfnamefont {B.}~\bibnamefont {Scharf}},\
  and\ \bibinfo {author} {\bibfnamefont {A.}~\bibnamefont {Matos-Abiague}},\
  }\bibfield  {title} {\bibinfo {title} {{Crystalline anisotropic topological
  superconductivity in planar Josephson junctions}},\ }\href
  {https://doi.org/10.1103/PhysRevResearch.3.013198} {\bibfield  {journal}
  {\bibinfo  {journal} {Phys. Rev. Research}\ }\textbf {\bibinfo {volume}
  {3}},\ \bibinfo {pages} {013198} (\bibinfo {year} {2021})}\BibitemShut
  {NoStop}%
\bibitem [{\citenamefont {Pekerten}\ \emph {et~al.}(2022)\citenamefont
  {Pekerten}, \citenamefont {Pakizer}, \citenamefont {Hawn},\ and\
  \citenamefont {Matos-Abiague}}]{pekerten2022anisotropic}%
  \BibitemOpen
  \bibfield  {author} {\bibinfo {author} {\bibfnamefont {B.}~\bibnamefont
  {Pekerten}}, \bibinfo {author} {\bibfnamefont {J.~D.}\ \bibnamefont
  {Pakizer}}, \bibinfo {author} {\bibfnamefont {B.}~\bibnamefont {Hawn}},\ and\
  \bibinfo {author} {\bibfnamefont {A.}~\bibnamefont {Matos-Abiague}},\
  }\bibfield  {title} {\bibinfo {title} {{Anisotropic topological
  superconductivity in Josephson junctions}},\ }\href
  {https://doi.org/10.1103/PhysRevB.105.054504} {\bibfield  {journal} {\bibinfo
   {journal} {Phys. Rev. B}\ }\textbf {\bibinfo {volume} {105}},\ \bibinfo
  {pages} {054504} (\bibinfo {year} {2022})}\BibitemShut {NoStop}%
\bibitem [{\citenamefont {Melo}\ \emph {et~al.}(2022)\citenamefont {Melo},
  \citenamefont {Tanev},\ and\ \citenamefont {Akhmerov}}]{melo2022greedy}%
  \BibitemOpen
  \bibfield  {author} {\bibinfo {author} {\bibfnamefont {A.}~\bibnamefont
  {Melo}}, \bibinfo {author} {\bibfnamefont {T.}~\bibnamefont {Tanev}},\ and\
  \bibinfo {author} {\bibfnamefont {A.~R.}\ \bibnamefont {Akhmerov}},\
  }\bibfield  {title} {\bibinfo {title} {{Greedy optimization of the geometry
  of Majorana Josephson junctions}},\ }\href {https://arxiv.org/abs/2205.05689}
  {\bibfield  {journal} {\bibinfo  {journal} {arXiv:2205.05689}\ } (\bibinfo
  {year} {2022})}\BibitemShut {NoStop}%
\bibitem [{\citenamefont {Ren}\ \emph {et~al.}(2019)\citenamefont {Ren},
  \citenamefont {Pientka}, \citenamefont {Hart}, \citenamefont {Pierce},
  \citenamefont {Kosowsky}, \citenamefont {Lunczer}, \citenamefont {Schlereth},
  \citenamefont {Scharf}, \citenamefont {Hankiewicz}, \citenamefont {Molenkamp}
  \emph {et~al.}}]{ren2019topological}%
  \BibitemOpen
  \bibfield  {author} {\bibinfo {author} {\bibfnamefont {H.}~\bibnamefont
  {Ren}}, \bibinfo {author} {\bibfnamefont {F.}~\bibnamefont {Pientka}},
  \bibinfo {author} {\bibfnamefont {S.}~\bibnamefont {Hart}}, \bibinfo {author}
  {\bibfnamefont {A.~T.}\ \bibnamefont {Pierce}}, \bibinfo {author}
  {\bibfnamefont {M.}~\bibnamefont {Kosowsky}}, \bibinfo {author}
  {\bibfnamefont {L.}~\bibnamefont {Lunczer}}, \bibinfo {author} {\bibfnamefont
  {R.}~\bibnamefont {Schlereth}}, \bibinfo {author} {\bibfnamefont
  {B.}~\bibnamefont {Scharf}}, \bibinfo {author} {\bibfnamefont {E.~M.}\
  \bibnamefont {Hankiewicz}}, \bibinfo {author} {\bibfnamefont {L.~W.}\
  \bibnamefont {Molenkamp}}, \emph {et~al.},\ }\bibfield  {title} {\bibinfo
  {title} {{Topological superconductivity in a phase-controlled Josephson
  junction}},\ }\href {https://doi.org/10.1038/s41586-019-1148-9} {\bibfield
  {journal} {\bibinfo  {journal} {Nature}\ }\textbf {\bibinfo {volume} {569}},\
  \bibinfo {pages} {93} (\bibinfo {year} {2019})}\BibitemShut {NoStop}%
\bibitem [{\citenamefont {Fornieri}\ \emph {et~al.}(2019)\citenamefont
  {Fornieri}, \citenamefont {Whiticar}, \citenamefont {Setiawan}, \citenamefont
  {Portol{\'e}s}, \citenamefont {Drachmann}, \citenamefont {Keselman},
  \citenamefont {Gronin}, \citenamefont {Thomas}, \citenamefont {Wang},
  \citenamefont {Kallaher} \emph {et~al.}}]{fornieri2019evidence}%
  \BibitemOpen
  \bibfield  {author} {\bibinfo {author} {\bibfnamefont {A.}~\bibnamefont
  {Fornieri}}, \bibinfo {author} {\bibfnamefont {A.~M.}\ \bibnamefont
  {Whiticar}}, \bibinfo {author} {\bibfnamefont {F.}~\bibnamefont {Setiawan}},
  \bibinfo {author} {\bibfnamefont {E.}~\bibnamefont {Portol{\'e}s}}, \bibinfo
  {author} {\bibfnamefont {A.~C.}\ \bibnamefont {Drachmann}}, \bibinfo {author}
  {\bibfnamefont {A.}~\bibnamefont {Keselman}}, \bibinfo {author}
  {\bibfnamefont {S.}~\bibnamefont {Gronin}}, \bibinfo {author} {\bibfnamefont
  {C.}~\bibnamefont {Thomas}}, \bibinfo {author} {\bibfnamefont
  {T.}~\bibnamefont {Wang}}, \bibinfo {author} {\bibfnamefont {R.}~\bibnamefont
  {Kallaher}}, \emph {et~al.},\ }\bibfield  {title} {\bibinfo {title}
  {{Evidence of topological superconductivity in planar Josephson junctions}},\
  }\href {https://doi.org/10.1038/s41586-019-1068-8} {\bibfield  {journal}
  {\bibinfo  {journal} {Nature}\ }\textbf {\bibinfo {volume} {569}},\ \bibinfo
  {pages} {89} (\bibinfo {year} {2019})}\BibitemShut {NoStop}%
\bibitem [{\citenamefont {Banerjee}\ \emph {et~al.}(2022)\citenamefont
  {Banerjee}, \citenamefont {Lesser}, \citenamefont {Rahman}, \citenamefont
  {Wang}, \citenamefont {Li}, \citenamefont {Kringhøj}, \citenamefont
  {Whiticar}, \citenamefont {Drachmann}, \citenamefont {Thomas}, \citenamefont
  {Wang}, \citenamefont {Manfra}, \citenamefont {Berg}, \citenamefont {Oreg},
  \citenamefont {Stern},\ and\ \citenamefont
  {Marcus}}]{banerjee2022signatures}%
  \BibitemOpen
  \bibfield  {author} {\bibinfo {author} {\bibfnamefont {A.}~\bibnamefont
  {Banerjee}}, \bibinfo {author} {\bibfnamefont {O.}~\bibnamefont {Lesser}},
  \bibinfo {author} {\bibfnamefont {M.~A.}\ \bibnamefont {Rahman}}, \bibinfo
  {author} {\bibfnamefont {H.~R.}\ \bibnamefont {Wang}}, \bibinfo {author}
  {\bibfnamefont {M.~R.}\ \bibnamefont {Li}}, \bibinfo {author} {\bibfnamefont
  {A.}~\bibnamefont {Kringhøj}}, \bibinfo {author} {\bibfnamefont {A.~M.}\
  \bibnamefont {Whiticar}}, \bibinfo {author} {\bibfnamefont {A.~C.~C.}\
  \bibnamefont {Drachmann}}, \bibinfo {author} {\bibfnamefont {C.}~\bibnamefont
  {Thomas}}, \bibinfo {author} {\bibfnamefont {T.}~\bibnamefont {Wang}},
  \bibinfo {author} {\bibfnamefont {M.~J.}\ \bibnamefont {Manfra}}, \bibinfo
  {author} {\bibfnamefont {E.}~\bibnamefont {Berg}}, \bibinfo {author}
  {\bibfnamefont {Y.}~\bibnamefont {Oreg}}, \bibinfo {author} {\bibfnamefont
  {A.}~\bibnamefont {Stern}},\ and\ \bibinfo {author} {\bibfnamefont {C.~M.}\
  \bibnamefont {Marcus}},\ }\bibfield  {title} {\bibinfo {title} {{Signatures
  of a topological phase transition in a planar Josephson junction}},\ }\href
  {https://arxiv.org/abs/2201.03453} {\bibfield  {journal} {\bibinfo  {journal}
  {arXiv:2201.03453}\ } (\bibinfo {year} {2022})}\BibitemShut {NoStop}%
\bibitem [{\citenamefont {Sau}\ \emph {et~al.}(2010{\natexlab{c}})\citenamefont
  {Sau}, \citenamefont {Lutchyn}, \citenamefont {Tewari},\ and\ \citenamefont
  {Das~Sarma}}]{sau2010robustness}%
  \BibitemOpen
  \bibfield  {author} {\bibinfo {author} {\bibfnamefont {J.~D.}\ \bibnamefont
  {Sau}}, \bibinfo {author} {\bibfnamefont {R.~M.}\ \bibnamefont {Lutchyn}},
  \bibinfo {author} {\bibfnamefont {S.}~\bibnamefont {Tewari}},\ and\ \bibinfo
  {author} {\bibfnamefont {S.}~\bibnamefont {Das~Sarma}},\ }\bibfield  {title}
  {\bibinfo {title} {{Robustness of Majorana fermions in proximity-induced
  superconductors}},\ }\href {https://doi.org/10.1103/PhysRevB.82.094522}
  {\bibfield  {journal} {\bibinfo  {journal} {Phys. Rev. B}\ }\textbf {\bibinfo
  {volume} {82}},\ \bibinfo {pages} {094522} (\bibinfo {year}
  {2010}{\natexlab{c}})}\BibitemShut {NoStop}%
\bibitem [{\citenamefont {Hell}\ \emph
  {et~al.}(2017{\natexlab{b}})\citenamefont {Hell}, \citenamefont {Flensberg},\
  and\ \citenamefont {Leijnse}}]{hell2017coupling}%
  \BibitemOpen
  \bibfield  {author} {\bibinfo {author} {\bibfnamefont {M.}~\bibnamefont
  {Hell}}, \bibinfo {author} {\bibfnamefont {K.}~\bibnamefont {Flensberg}},\
  and\ \bibinfo {author} {\bibfnamefont {M.}~\bibnamefont {Leijnse}},\
  }\bibfield  {title} {\bibinfo {title} {{Coupling and braiding Majorana bound
  states in networks defined in two-dimensional electron gases with
  proximity-induced superconductivity}},\ }\href
  {https://doi.org/10.1103/PhysRevB.96.035444} {\bibfield  {journal} {\bibinfo
  {journal} {Phys. Rev. B}\ }\textbf {\bibinfo {volume} {96}},\ \bibinfo
  {pages} {035444} (\bibinfo {year} {2017}{\natexlab{b}})}\BibitemShut
  {NoStop}%
\bibitem [{\citenamefont {Stern}\ and\ \citenamefont
  {Berg}(2019)}]{stern2019fractional}%
  \BibitemOpen
  \bibfield  {author} {\bibinfo {author} {\bibfnamefont {A.}~\bibnamefont
  {Stern}}\ and\ \bibinfo {author} {\bibfnamefont {E.}~\bibnamefont {Berg}},\
  }\bibfield  {title} {\bibinfo {title} {{Fractional Josephson Vortices and
  Braiding of Majorana Zero Modes in Planar Superconductor-Semiconductor
  Heterostructures}},\ }\href {https://doi.org/10.1103/PhysRevLett.122.107701}
  {\bibfield  {journal} {\bibinfo  {journal} {Phys. Rev. Lett.}\ }\textbf
  {\bibinfo {volume} {122}},\ \bibinfo {pages} {107701} (\bibinfo {year}
  {2019})}\BibitemShut {NoStop}%
\bibitem [{\citenamefont {Lesser}\ and\ \citenamefont
  {Oreg}(2022)}]{lesser2022majorana}%
  \BibitemOpen
  \bibfield  {author} {\bibinfo {author} {\bibfnamefont {O.}~\bibnamefont
  {Lesser}}\ and\ \bibinfo {author} {\bibfnamefont {Y.}~\bibnamefont {Oreg}},\
  }\bibfield  {title} {\bibinfo {title} {{Majorana zero modes induced by
  superconducting phase bias}},\ }\href
  {https://iopscience.iop.org/article/10.1088/1361-6463/ac4a37} {\bibfield
  {journal} {\bibinfo  {journal} {Journal of Physics D: Applied Physics}\
  }\textbf {\bibinfo {volume} {55}},\ \bibinfo {pages} {164001} (\bibinfo
  {year} {2022})}\BibitemShut {NoStop}%
\bibitem [{\citenamefont {Schrade}\ and\ \citenamefont
  {Fu}(2018)}]{schrade2018majorana}%
  \BibitemOpen
  \bibfield  {author} {\bibinfo {author} {\bibfnamefont {C.}~\bibnamefont
  {Schrade}}\ and\ \bibinfo {author} {\bibfnamefont {L.}~\bibnamefont {Fu}},\
  }\bibfield  {title} {\bibinfo {title} {{Majorana Superconducting Qubit}},\
  }\href {https://doi.org/10.1103/PhysRevLett.121.267002} {\bibfield  {journal}
  {\bibinfo  {journal} {Phys. Rev. Lett.}\ }\textbf {\bibinfo {volume} {121}},\
  \bibinfo {pages} {267002} (\bibinfo {year} {2018})}\BibitemShut {NoStop}%
\bibitem [{\citenamefont {Zhou}\ \emph {et~al.}(2020)\citenamefont {Zhou},
  \citenamefont {Dartiailh}, \citenamefont {Mayer}, \citenamefont {Han},
  \citenamefont {Matos-Abiague}, \citenamefont {Shabani},\ and\ \citenamefont
  {\ifmmode \check{Z}\else \v{Z}\fi{}uti\ifmmode~\acute{c}\else
  \'{c}\fi{}}}]{zhou2020phase}%
  \BibitemOpen
  \bibfield  {author} {\bibinfo {author} {\bibfnamefont {T.}~\bibnamefont
  {Zhou}}, \bibinfo {author} {\bibfnamefont {M.~C.}\ \bibnamefont {Dartiailh}},
  \bibinfo {author} {\bibfnamefont {W.}~\bibnamefont {Mayer}}, \bibinfo
  {author} {\bibfnamefont {J.~E.}\ \bibnamefont {Han}}, \bibinfo {author}
  {\bibfnamefont {A.}~\bibnamefont {Matos-Abiague}}, \bibinfo {author}
  {\bibfnamefont {J.}~\bibnamefont {Shabani}},\ and\ \bibinfo {author}
  {\bibfnamefont {I.}~\bibnamefont {\ifmmode \check{Z}\else
  \v{Z}\fi{}uti\ifmmode~\acute{c}\else \'{c}\fi{}}},\ }\bibfield  {title}
  {\bibinfo {title} {{Phase Control of Majorana Bound States in a Topological
  $\mathsf{X}$ Junction}},\ }\href
  {https://doi.org/10.1103/PhysRevLett.124.137001} {\bibfield  {journal}
  {\bibinfo  {journal} {Phys. Rev. Lett.}\ }\textbf {\bibinfo {volume} {124}},\
  \bibinfo {pages} {137001} (\bibinfo {year} {2020})}\BibitemShut {NoStop}%
\bibitem [{\citenamefont {Scappucci}\ \emph {et~al.}(2021)\citenamefont
  {Scappucci}, \citenamefont {Kloeffel}, \citenamefont {Zwanenburg},
  \citenamefont {Loss}, \citenamefont {Myronov}, \citenamefont {Zhang},
  \citenamefont {De~Franceschi}, \citenamefont {Katsaros},\ and\ \citenamefont
  {Veldhorst}}]{scappucci2020germanium}%
  \BibitemOpen
  \bibfield  {author} {\bibinfo {author} {\bibfnamefont {G.}~\bibnamefont
  {Scappucci}}, \bibinfo {author} {\bibfnamefont {C.}~\bibnamefont {Kloeffel}},
  \bibinfo {author} {\bibfnamefont {F.~A.}\ \bibnamefont {Zwanenburg}},
  \bibinfo {author} {\bibfnamefont {D.}~\bibnamefont {Loss}}, \bibinfo {author}
  {\bibfnamefont {M.}~\bibnamefont {Myronov}}, \bibinfo {author} {\bibfnamefont
  {J.-J.}\ \bibnamefont {Zhang}}, \bibinfo {author} {\bibfnamefont
  {S.}~\bibnamefont {De~Franceschi}}, \bibinfo {author} {\bibfnamefont
  {G.}~\bibnamefont {Katsaros}},\ and\ \bibinfo {author} {\bibfnamefont
  {M.}~\bibnamefont {Veldhorst}},\ }\bibfield  {title} {\bibinfo {title} {{The
  germanium quantum information route}},\ }\href
  {https://doi.org/10.1038/s41578-020-00262-z} {\bibfield  {journal} {\bibinfo
  {journal} {Nature Reviews Materials}\ }\textbf {\bibinfo {volume} {6}},\
  \bibinfo {pages} {926} (\bibinfo {year} {2021})}\BibitemShut {NoStop}%
\bibitem [{\citenamefont {Hao}\ \emph {et~al.}(2010)\citenamefont {Hao},
  \citenamefont {Tu}, \citenamefont {Cao}, \citenamefont {Zhou}, \citenamefont
  {Li}, \citenamefont {Guo}, \citenamefont {Fung}, \citenamefont {Ji},
  \citenamefont {Guo},\ and\ \citenamefont {Lu}}]{hao2010strong}%
  \BibitemOpen
  \bibfield  {author} {\bibinfo {author} {\bibfnamefont {X.-J.}\ \bibnamefont
  {Hao}}, \bibinfo {author} {\bibfnamefont {T.}~\bibnamefont {Tu}}, \bibinfo
  {author} {\bibfnamefont {G.}~\bibnamefont {Cao}}, \bibinfo {author}
  {\bibfnamefont {C.}~\bibnamefont {Zhou}}, \bibinfo {author} {\bibfnamefont
  {H.-O.}\ \bibnamefont {Li}}, \bibinfo {author} {\bibfnamefont {G.-C.}\
  \bibnamefont {Guo}}, \bibinfo {author} {\bibfnamefont {W.~Y.}\ \bibnamefont
  {Fung}}, \bibinfo {author} {\bibfnamefont {Z.}~\bibnamefont {Ji}}, \bibinfo
  {author} {\bibfnamefont {G.-P.}\ \bibnamefont {Guo}},\ and\ \bibinfo {author}
  {\bibfnamefont {W.}~\bibnamefont {Lu}},\ }\bibfield  {title} {\bibinfo
  {title} {{Strong and Tunable Spin- Orbit Coupling of One-Dimensional Holes in
  Ge/Si Core/Shell Nanowires}},\ }\href
  {https://pubs.acs.org/doi/full/10.1021/nl101181e} {\bibfield  {journal}
  {\bibinfo  {journal} {{Nano Letters}}\ }\textbf {\bibinfo {volume} {10}},\
  \bibinfo {pages} {2956} (\bibinfo {year} {2010})}\BibitemShut {NoStop}%
\bibitem [{\citenamefont {Maurand}\ \emph {et~al.}(2016)\citenamefont
  {Maurand}, \citenamefont {Jehl}, \citenamefont {Kotekar-Patil}, \citenamefont
  {Corna}, \citenamefont {Bohuslavskyi}, \citenamefont {Lavi{\'e}ville},
  \citenamefont {Hutin}, \citenamefont {Barraud}, \citenamefont {Vinet},
  \citenamefont {Sanquer} \emph {et~al.}}]{maurand2016cmos}%
  \BibitemOpen
  \bibfield  {author} {\bibinfo {author} {\bibfnamefont {R.}~\bibnamefont
  {Maurand}}, \bibinfo {author} {\bibfnamefont {X.}~\bibnamefont {Jehl}},
  \bibinfo {author} {\bibfnamefont {D.}~\bibnamefont {Kotekar-Patil}}, \bibinfo
  {author} {\bibfnamefont {A.}~\bibnamefont {Corna}}, \bibinfo {author}
  {\bibfnamefont {H.}~\bibnamefont {Bohuslavskyi}}, \bibinfo {author}
  {\bibfnamefont {R.}~\bibnamefont {Lavi{\'e}ville}}, \bibinfo {author}
  {\bibfnamefont {L.}~\bibnamefont {Hutin}}, \bibinfo {author} {\bibfnamefont
  {S.}~\bibnamefont {Barraud}}, \bibinfo {author} {\bibfnamefont
  {M.}~\bibnamefont {Vinet}}, \bibinfo {author} {\bibfnamefont
  {M.}~\bibnamefont {Sanquer}}, \emph {et~al.},\ }\bibfield  {title} {\bibinfo
  {title} {{A CMOS silicon spin qubit}},\ }\href
  {https://doi.org/10.1038/ncomms13575} {\bibfield  {journal} {\bibinfo
  {journal} {{Nature Communications}}\ }\textbf {\bibinfo {volume} {7}},\
  \bibinfo {pages} {13575} (\bibinfo {year} {2016})}\BibitemShut {NoStop}%
\bibitem [{\citenamefont {Watzinger}\ \emph {et~al.}(2018)\citenamefont
  {Watzinger}, \citenamefont {Kuku{\v{c}}ka}, \citenamefont
  {Vuku{\v{s}}i{\'c}}, \citenamefont {Gao}, \citenamefont {Wang}, \citenamefont
  {Sch{\"a}ffler}, \citenamefont {Zhang},\ and\ \citenamefont
  {Katsaros}}]{watzinger2018germanium}%
  \BibitemOpen
  \bibfield  {author} {\bibinfo {author} {\bibfnamefont {H.}~\bibnamefont
  {Watzinger}}, \bibinfo {author} {\bibfnamefont {J.}~\bibnamefont
  {Kuku{\v{c}}ka}}, \bibinfo {author} {\bibfnamefont {L.}~\bibnamefont
  {Vuku{\v{s}}i{\'c}}}, \bibinfo {author} {\bibfnamefont {F.}~\bibnamefont
  {Gao}}, \bibinfo {author} {\bibfnamefont {T.}~\bibnamefont {Wang}}, \bibinfo
  {author} {\bibfnamefont {F.}~\bibnamefont {Sch{\"a}ffler}}, \bibinfo {author}
  {\bibfnamefont {J.-J.}\ \bibnamefont {Zhang}},\ and\ \bibinfo {author}
  {\bibfnamefont {G.}~\bibnamefont {Katsaros}},\ }\bibfield  {title} {\bibinfo
  {title} {{A germanium hole spin qubit}},\ }\href
  {https://www.nature.com/articles/s41467-018-06418-4} {\bibfield  {journal}
  {\bibinfo  {journal} {{Nature Communications}}\ }\textbf {\bibinfo {volume}
  {9}},\ \bibinfo {pages} {3902} (\bibinfo {year} {2018})}\BibitemShut
  {NoStop}%
\bibitem [{\citenamefont {Hendrickx}\ \emph
  {et~al.}(2020{\natexlab{a}})\citenamefont {Hendrickx}, \citenamefont
  {Lawrie}, \citenamefont {Petit}, \citenamefont {Sammak}, \citenamefont
  {Scappucci},\ and\ \citenamefont {Veldhorst}}]{hendrickx2020single}%
  \BibitemOpen
  \bibfield  {author} {\bibinfo {author} {\bibfnamefont {N.}~\bibnamefont
  {Hendrickx}}, \bibinfo {author} {\bibfnamefont {W.}~\bibnamefont {Lawrie}},
  \bibinfo {author} {\bibfnamefont {L.}~\bibnamefont {Petit}}, \bibinfo
  {author} {\bibfnamefont {A.}~\bibnamefont {Sammak}}, \bibinfo {author}
  {\bibfnamefont {G.}~\bibnamefont {Scappucci}},\ and\ \bibinfo {author}
  {\bibfnamefont {M.}~\bibnamefont {Veldhorst}},\ }\bibfield  {title} {\bibinfo
  {title} {{A single-hole spin qubit}},\ }\href
  {https://www.nature.com/articles/s41467-020-17211-7} {\bibfield  {journal}
  {\bibinfo  {journal} {{Nature Communications}}\ }\textbf {\bibinfo {volume}
  {11}},\ \bibinfo {pages} {3478} (\bibinfo {year}
  {2020}{\natexlab{a}})}\BibitemShut {NoStop}%
\bibitem [{\citenamefont {Hendrickx}\ \emph
  {et~al.}(2020{\natexlab{b}})\citenamefont {Hendrickx}, \citenamefont
  {Franke}, \citenamefont {Sammak}, \citenamefont {Scappucci},\ and\
  \citenamefont {Veldhorst}}]{hendrickx2020fast}%
  \BibitemOpen
  \bibfield  {author} {\bibinfo {author} {\bibfnamefont {N.}~\bibnamefont
  {Hendrickx}}, \bibinfo {author} {\bibfnamefont {D.}~\bibnamefont {Franke}},
  \bibinfo {author} {\bibfnamefont {A.}~\bibnamefont {Sammak}}, \bibinfo
  {author} {\bibfnamefont {G.}~\bibnamefont {Scappucci}},\ and\ \bibinfo
  {author} {\bibfnamefont {M.}~\bibnamefont {Veldhorst}},\ }\bibfield  {title}
  {\bibinfo {title} {{Fast two-qubit logic with holes in germanium}},\ }\href
  {https://doi.org/10.1038/s41586-019-1919-3} {\bibfield  {journal} {\bibinfo
  {journal} {Nature}\ }\textbf {\bibinfo {volume} {577}},\ \bibinfo {pages}
  {487} (\bibinfo {year} {2020}{\natexlab{b}})}\BibitemShut {NoStop}%
\bibitem [{\citenamefont {Froning}\ \emph {et~al.}(2021)\citenamefont
  {Froning}, \citenamefont {Camenzind}, \citenamefont {van~der Molen},
  \citenamefont {Li}, \citenamefont {Bakkers}, \citenamefont {Zumb{\"u}hl},\
  and\ \citenamefont {Braakman}}]{froning2021ultrafast}%
  \BibitemOpen
  \bibfield  {author} {\bibinfo {author} {\bibfnamefont {F.~N.}\ \bibnamefont
  {Froning}}, \bibinfo {author} {\bibfnamefont {L.~C.}\ \bibnamefont
  {Camenzind}}, \bibinfo {author} {\bibfnamefont {O.~A.}\ \bibnamefont {van~der
  Molen}}, \bibinfo {author} {\bibfnamefont {A.}~\bibnamefont {Li}}, \bibinfo
  {author} {\bibfnamefont {E.~P.}\ \bibnamefont {Bakkers}}, \bibinfo {author}
  {\bibfnamefont {D.~M.}\ \bibnamefont {Zumb{\"u}hl}},\ and\ \bibinfo {author}
  {\bibfnamefont {F.~R.}\ \bibnamefont {Braakman}},\ }\bibfield  {title}
  {\bibinfo {title} {{Ultrafast hole spin qubit with gate-tunable spin--orbit
  switch functionality}},\ }\href
  {https://www.nature.com/articles/s41565-020-00828-6} {\bibfield  {journal}
  {\bibinfo  {journal} {Nature Nanotechnology}\ }\textbf {\bibinfo {volume}
  {16}},\ \bibinfo {pages} {308} (\bibinfo {year} {2021})}\BibitemShut
  {NoStop}%
\bibitem [{\citenamefont {Bosco}\ \emph
  {et~al.}(2021{\natexlab{a}})\citenamefont {Bosco}, \citenamefont {Benito},
  \citenamefont {Adelsberger},\ and\ \citenamefont {Loss}}]{bosco2021squeezed}%
  \BibitemOpen
  \bibfield  {author} {\bibinfo {author} {\bibfnamefont {S.}~\bibnamefont
  {Bosco}}, \bibinfo {author} {\bibfnamefont {M.}~\bibnamefont {Benito}},
  \bibinfo {author} {\bibfnamefont {C.}~\bibnamefont {Adelsberger}},\ and\
  \bibinfo {author} {\bibfnamefont {D.}~\bibnamefont {Loss}},\ }\bibfield
  {title} {\bibinfo {title} {{Squeezed hole spin qubits in Ge quantum dots with
  ultrafast gates at low power}},\ }\href
  {https://doi.org/10.1103/PhysRevB.104.115425} {\bibfield  {journal} {\bibinfo
   {journal} {Phys. Rev. B}\ }\textbf {\bibinfo {volume} {104}},\ \bibinfo
  {pages} {115425} (\bibinfo {year} {2021}{\natexlab{a}})}\BibitemShut
  {NoStop}%
\bibitem [{\citenamefont {Wang}\ \emph {et~al.}(2022)\citenamefont {Wang},
  \citenamefont {Xu}, \citenamefont {Gao}, \citenamefont {Liu}, \citenamefont
  {Ma}, \citenamefont {Zhang}, \citenamefont {Wang}, \citenamefont {Cao},
  \citenamefont {Wang}, \citenamefont {Zhang}, \citenamefont {Culcer},
  \citenamefont {Hu}, \citenamefont {Jiang}, \citenamefont {Li}, \citenamefont
  {Guo},\ and\ \citenamefont {Guo}}]{wang2022ultrafast}%
  \BibitemOpen
  \bibfield  {author} {\bibinfo {author} {\bibfnamefont {K.}~\bibnamefont
  {Wang}}, \bibinfo {author} {\bibfnamefont {G.}~\bibnamefont {Xu}}, \bibinfo
  {author} {\bibfnamefont {F.}~\bibnamefont {Gao}}, \bibinfo {author}
  {\bibfnamefont {H.}~\bibnamefont {Liu}}, \bibinfo {author} {\bibfnamefont
  {R.-L.}\ \bibnamefont {Ma}}, \bibinfo {author} {\bibfnamefont
  {X.}~\bibnamefont {Zhang}}, \bibinfo {author} {\bibfnamefont
  {Z.}~\bibnamefont {Wang}}, \bibinfo {author} {\bibfnamefont {G.}~\bibnamefont
  {Cao}}, \bibinfo {author} {\bibfnamefont {T.}~\bibnamefont {Wang}}, \bibinfo
  {author} {\bibfnamefont {J.-J.}\ \bibnamefont {Zhang}}, \bibinfo {author}
  {\bibfnamefont {D.}~\bibnamefont {Culcer}}, \bibinfo {author} {\bibfnamefont
  {X.}~\bibnamefont {Hu}}, \bibinfo {author} {\bibfnamefont {H.-W.}\
  \bibnamefont {Jiang}}, \bibinfo {author} {\bibfnamefont {H.-O.}\ \bibnamefont
  {Li}}, \bibinfo {author} {\bibfnamefont {G.-C.}\ \bibnamefont {Guo}},\ and\
  \bibinfo {author} {\bibfnamefont {G.-P.}\ \bibnamefont {Guo}},\ }\bibfield
  {title} {\bibinfo {title} {{Ultrafast coherent control of a hole spin qubit
  in a germanium quantum dot}},\ }\href
  {https://www.nature.com/articles/s41467-021-27880-7} {\bibfield  {journal}
  {\bibinfo  {journal} {{Nature Communications}}\ }\textbf {\bibinfo {volume}
  {13}},\ \bibinfo {pages} {206} (\bibinfo {year} {2022})}\BibitemShut
  {NoStop}%
\bibitem [{\citenamefont {Itoh}\ \emph {et~al.}(1993)\citenamefont {Itoh},
  \citenamefont {Hansen}, \citenamefont {Haller}, \citenamefont {Farmer},
  \citenamefont {Ozhogin}, \citenamefont {Rudnev},\ and\ \citenamefont
  {Tikhomirov}}]{itoh1993high}%
  \BibitemOpen
  \bibfield  {author} {\bibinfo {author} {\bibfnamefont {K.}~\bibnamefont
  {Itoh}}, \bibinfo {author} {\bibfnamefont {W.}~\bibnamefont {Hansen}},
  \bibinfo {author} {\bibfnamefont {E.}~\bibnamefont {Haller}}, \bibinfo
  {author} {\bibfnamefont {J.}~\bibnamefont {Farmer}}, \bibinfo {author}
  {\bibfnamefont {V.}~\bibnamefont {Ozhogin}}, \bibinfo {author} {\bibfnamefont
  {A.}~\bibnamefont {Rudnev}},\ and\ \bibinfo {author} {\bibfnamefont
  {A.}~\bibnamefont {Tikhomirov}},\ }\bibfield  {title} {\bibinfo {title}
  {{High purity isotopically enriched 70Ge and 74Ge single crystals: Isotope
  separation, growth, and properties}},\ }\href
  {https://www.cambridge.org/core/journals/journal-of-materials-research/article/high-purity-isotopically-enriched-70ge-and-74ge-single-crystals-isotope-separation-growth-and-properties/21C111A4A5629E6CA3E172C4860C33BE}
  {\bibfield  {journal} {\bibinfo  {journal} {Journal of Materials Research}\
  }\textbf {\bibinfo {volume} {8}},\ \bibinfo {pages} {1341} (\bibinfo {year}
  {1993})}\BibitemShut {NoStop}%
\bibitem [{\citenamefont {Bosco}\ and\ \citenamefont
  {Loss}(2021)}]{bosco2021fully}%
  \BibitemOpen
  \bibfield  {author} {\bibinfo {author} {\bibfnamefont {S.}~\bibnamefont
  {Bosco}}\ and\ \bibinfo {author} {\bibfnamefont {D.}~\bibnamefont {Loss}},\
  }\bibfield  {title} {\bibinfo {title} {{Fully Tunable Hyperfine Interactions
  of Hole Spin Qubits in Si and Ge Quantum Dots}},\ }\href
  {https://doi.org/10.1103/PhysRevLett.127.190501} {\bibfield  {journal}
  {\bibinfo  {journal} {Phys. Rev. Lett.}\ }\textbf {\bibinfo {volume} {127}},\
  \bibinfo {pages} {190501} (\bibinfo {year} {2021})}\BibitemShut {NoStop}%
\bibitem [{\citenamefont {Veldhorst}\ \emph {et~al.}(2017)\citenamefont
  {Veldhorst}, \citenamefont {Eenink}, \citenamefont {Yang},\ and\
  \citenamefont {Dzurak}}]{veldhorst2017silicon}%
  \BibitemOpen
  \bibfield  {author} {\bibinfo {author} {\bibfnamefont {M.}~\bibnamefont
  {Veldhorst}}, \bibinfo {author} {\bibfnamefont {H.}~\bibnamefont {Eenink}},
  \bibinfo {author} {\bibfnamefont {C.-H.}\ \bibnamefont {Yang}},\ and\
  \bibinfo {author} {\bibfnamefont {A.~S.}\ \bibnamefont {Dzurak}},\ }\bibfield
   {title} {\bibinfo {title} {{Silicon CMOS architecture for a spin-based
  quantum computer}},\ }\href
  {https://www.nature.com/articles/s41467-017-01905-6} {\bibfield  {journal}
  {\bibinfo  {journal} {{Nature Communications}}\ }\textbf {\bibinfo {volume}
  {8}},\ \bibinfo {pages} {1766} (\bibinfo {year} {2017})}\BibitemShut
  {NoStop}%
\bibitem [{\citenamefont {Hendrickx}\ \emph {et~al.}(2018)\citenamefont
  {Hendrickx}, \citenamefont {Franke}, \citenamefont {Sammak}, \citenamefont
  {Kouwenhoven}, \citenamefont {Sabbagh}, \citenamefont {Yeoh}, \citenamefont
  {Li}, \citenamefont {Tagliaferri}, \citenamefont {Virgilio}, \citenamefont
  {Capellini} \emph {et~al.}}]{hendrickx2018gate}%
  \BibitemOpen
  \bibfield  {author} {\bibinfo {author} {\bibfnamefont {N.}~\bibnamefont
  {Hendrickx}}, \bibinfo {author} {\bibfnamefont {D.}~\bibnamefont {Franke}},
  \bibinfo {author} {\bibfnamefont {A.}~\bibnamefont {Sammak}}, \bibinfo
  {author} {\bibfnamefont {M.}~\bibnamefont {Kouwenhoven}}, \bibinfo {author}
  {\bibfnamefont {D.}~\bibnamefont {Sabbagh}}, \bibinfo {author} {\bibfnamefont
  {L.}~\bibnamefont {Yeoh}}, \bibinfo {author} {\bibfnamefont {R.}~\bibnamefont
  {Li}}, \bibinfo {author} {\bibfnamefont {M.}~\bibnamefont {Tagliaferri}},
  \bibinfo {author} {\bibfnamefont {M.}~\bibnamefont {Virgilio}}, \bibinfo
  {author} {\bibfnamefont {G.}~\bibnamefont {Capellini}}, \emph {et~al.},\
  }\bibfield  {title} {\bibinfo {title} {{Gate-controlled quantum dots and
  superconductivity in planar germanium}},\ }\href
  {https://doi.org/10.1038/s41467-018-05299-x} {\bibfield  {journal} {\bibinfo
  {journal} {{Nature Communications}}\ }\textbf {\bibinfo {volume} {9}},\
  \bibinfo {pages} {2835} (\bibinfo {year} {2018})}\BibitemShut {NoStop}%
\bibitem [{\citenamefont {Hendrickx}\ \emph {et~al.}(2019)\citenamefont
  {Hendrickx}, \citenamefont {Tagliaferri}, \citenamefont {Kouwenhoven},
  \citenamefont {Li}, \citenamefont {Franke}, \citenamefont {Sammak},
  \citenamefont {Brinkman}, \citenamefont {Scappucci},\ and\ \citenamefont
  {Veldhorst}}]{hendrickx2019ballistic}%
  \BibitemOpen
  \bibfield  {author} {\bibinfo {author} {\bibfnamefont {N.~W.}\ \bibnamefont
  {Hendrickx}}, \bibinfo {author} {\bibfnamefont {M.~L.~V.}\ \bibnamefont
  {Tagliaferri}}, \bibinfo {author} {\bibfnamefont {M.}~\bibnamefont
  {Kouwenhoven}}, \bibinfo {author} {\bibfnamefont {R.}~\bibnamefont {Li}},
  \bibinfo {author} {\bibfnamefont {D.~P.}\ \bibnamefont {Franke}}, \bibinfo
  {author} {\bibfnamefont {A.}~\bibnamefont {Sammak}}, \bibinfo {author}
  {\bibfnamefont {A.}~\bibnamefont {Brinkman}}, \bibinfo {author}
  {\bibfnamefont {G.}~\bibnamefont {Scappucci}},\ and\ \bibinfo {author}
  {\bibfnamefont {M.}~\bibnamefont {Veldhorst}},\ }\bibfield  {title} {\bibinfo
  {title} {{Ballistic supercurrent discretization and micrometer-long Josephson
  coupling in germanium}},\ }\href {https://doi.org/10.1103/PhysRevB.99.075435}
  {\bibfield  {journal} {\bibinfo  {journal} {Phys. Rev. B}\ }\textbf {\bibinfo
  {volume} {99}},\ \bibinfo {pages} {075435} (\bibinfo {year}
  {2019})}\BibitemShut {NoStop}%
\bibitem [{\citenamefont {Vigneau}\ \emph {et~al.}(2019)\citenamefont
  {Vigneau}, \citenamefont {Mizokuchi}, \citenamefont {Zanuz}, \citenamefont
  {Huang}, \citenamefont {Tan}, \citenamefont {Maurand}, \citenamefont
  {Frolov}, \citenamefont {Sammak}, \citenamefont {Scappucci}, \citenamefont
  {Lefloch},\ and\ \citenamefont {De~Franceschi}}]{vigneau2019germanium}%
  \BibitemOpen
  \bibfield  {author} {\bibinfo {author} {\bibfnamefont {F.}~\bibnamefont
  {Vigneau}}, \bibinfo {author} {\bibfnamefont {R.}~\bibnamefont {Mizokuchi}},
  \bibinfo {author} {\bibfnamefont {D.~C.}\ \bibnamefont {Zanuz}}, \bibinfo
  {author} {\bibfnamefont {X.}~\bibnamefont {Huang}}, \bibinfo {author}
  {\bibfnamefont {S.}~\bibnamefont {Tan}}, \bibinfo {author} {\bibfnamefont
  {R.}~\bibnamefont {Maurand}}, \bibinfo {author} {\bibfnamefont
  {S.}~\bibnamefont {Frolov}}, \bibinfo {author} {\bibfnamefont
  {A.}~\bibnamefont {Sammak}}, \bibinfo {author} {\bibfnamefont
  {G.}~\bibnamefont {Scappucci}}, \bibinfo {author} {\bibfnamefont
  {F.}~\bibnamefont {Lefloch}},\ and\ \bibinfo {author} {\bibfnamefont
  {S.}~\bibnamefont {De~Franceschi}},\ }\bibfield  {title} {\bibinfo {title}
  {{Germanium Quantum-Well Josephson Field-Effect Transistors and
  Interferometers}},\ }\href {https://doi.org/10.1021/acs.nanolett.8b04275}
  {\bibfield  {journal} {\bibinfo  {journal} {{Nano Letters}}\ }\textbf
  {\bibinfo {volume} {19}},\ \bibinfo {pages} {1023} (\bibinfo {year}
  {2019})}\BibitemShut {NoStop}%
\bibitem [{\citenamefont {Aggarwal}\ \emph {et~al.}(2021)\citenamefont
  {Aggarwal}, \citenamefont {Hofmann}, \citenamefont {Jirovec}, \citenamefont
  {Prieto}, \citenamefont {Sammak}, \citenamefont {Botifoll}, \citenamefont
  {Mart\'{\i}-S\'anchez}, \citenamefont {Veldhorst}, \citenamefont {Arbiol},
  \citenamefont {Scappucci}, \citenamefont {Danon},\ and\ \citenamefont
  {Katsaros}}]{aggarwal2021enhancement}%
  \BibitemOpen
  \bibfield  {author} {\bibinfo {author} {\bibfnamefont {K.}~\bibnamefont
  {Aggarwal}}, \bibinfo {author} {\bibfnamefont {A.}~\bibnamefont {Hofmann}},
  \bibinfo {author} {\bibfnamefont {D.}~\bibnamefont {Jirovec}}, \bibinfo
  {author} {\bibfnamefont {I.}~\bibnamefont {Prieto}}, \bibinfo {author}
  {\bibfnamefont {A.}~\bibnamefont {Sammak}}, \bibinfo {author} {\bibfnamefont
  {M.}~\bibnamefont {Botifoll}}, \bibinfo {author} {\bibfnamefont
  {S.}~\bibnamefont {Mart\'{\i}-S\'anchez}}, \bibinfo {author} {\bibfnamefont
  {M.}~\bibnamefont {Veldhorst}}, \bibinfo {author} {\bibfnamefont
  {J.}~\bibnamefont {Arbiol}}, \bibinfo {author} {\bibfnamefont
  {G.}~\bibnamefont {Scappucci}}, \bibinfo {author} {\bibfnamefont
  {J.}~\bibnamefont {Danon}},\ and\ \bibinfo {author} {\bibfnamefont
  {G.}~\bibnamefont {Katsaros}},\ }\bibfield  {title} {\bibinfo {title}
  {{Enhancement of proximity-induced superconductivity in a planar Ge hole
  gas}},\ }\href {https://doi.org/10.1103/PhysRevResearch.3.L022005} {\bibfield
   {journal} {\bibinfo  {journal} {Phys. Rev. Research}\ }\textbf {\bibinfo
  {volume} {3}},\ \bibinfo {pages} {L022005} (\bibinfo {year}
  {2021})}\BibitemShut {NoStop}%
\bibitem [{\citenamefont {Tosato}\ \emph {et~al.}(2022)\citenamefont {Tosato},
  \citenamefont {Levajac}, \citenamefont {Wang}, \citenamefont {Boor},
  \citenamefont {Borsoi}, \citenamefont {Botifoll}, \citenamefont {Borja},
  \citenamefont {Martí-Sánchez}, \citenamefont {Arbiol}, \citenamefont
  {Sammak}, \citenamefont {Veldhorst},\ and\ \citenamefont
  {Scappucci}}]{tosato2022hard}%
  \BibitemOpen
  \bibfield  {author} {\bibinfo {author} {\bibfnamefont {A.}~\bibnamefont
  {Tosato}}, \bibinfo {author} {\bibfnamefont {V.}~\bibnamefont {Levajac}},
  \bibinfo {author} {\bibfnamefont {J.-Y.}\ \bibnamefont {Wang}}, \bibinfo
  {author} {\bibfnamefont {C.~J.}\ \bibnamefont {Boor}}, \bibinfo {author}
  {\bibfnamefont {F.}~\bibnamefont {Borsoi}}, \bibinfo {author} {\bibfnamefont
  {M.}~\bibnamefont {Botifoll}}, \bibinfo {author} {\bibfnamefont {C.~N.}\
  \bibnamefont {Borja}}, \bibinfo {author} {\bibfnamefont {S.}~\bibnamefont
  {Martí-Sánchez}}, \bibinfo {author} {\bibfnamefont {J.}~\bibnamefont
  {Arbiol}}, \bibinfo {author} {\bibfnamefont {A.}~\bibnamefont {Sammak}},
  \bibinfo {author} {\bibfnamefont {M.}~\bibnamefont {Veldhorst}},\ and\
  \bibinfo {author} {\bibfnamefont {G.}~\bibnamefont {Scappucci}},\ }\bibfield
  {title} {\bibinfo {title} {{Hard superconducting gap in a high-mobility
  semiconductor}},\ }\href {https://arxiv.org/abs/2206.00569} {\bibfield
  {journal} {\bibinfo  {journal} {arXiv:2206.00569}\ } (\bibinfo {year}
  {2022})}\BibitemShut {NoStop}%
\bibitem [{\citenamefont {Maier}\ \emph {et~al.}(2014)\citenamefont {Maier},
  \citenamefont {Klinovaja},\ and\ \citenamefont {Loss}}]{maier2014majorana}%
  \BibitemOpen
  \bibfield  {author} {\bibinfo {author} {\bibfnamefont {F.}~\bibnamefont
  {Maier}}, \bibinfo {author} {\bibfnamefont {J.}~\bibnamefont {Klinovaja}},\
  and\ \bibinfo {author} {\bibfnamefont {D.}~\bibnamefont {Loss}},\ }\bibfield
  {title} {\bibinfo {title} {{Majorana fermions in Ge/Si hole nanowires}},\
  }\href {https://doi.org/10.1103/PhysRevB.90.195421} {\bibfield  {journal}
  {\bibinfo  {journal} {Phys. Rev. B}\ }\textbf {\bibinfo {volume} {90}},\
  \bibinfo {pages} {195421} (\bibinfo {year} {2014})}\BibitemShut {NoStop}%
\bibitem [{\citenamefont {Het\'enyi}\ \emph {et~al.}(2022)\citenamefont
  {Het\'enyi}, \citenamefont {Bosco},\ and\ \citenamefont
  {Loss}}]{hetenyi2022anomalous}%
  \BibitemOpen
  \bibfield  {author} {\bibinfo {author} {\bibfnamefont {B.}~\bibnamefont
  {Het\'enyi}}, \bibinfo {author} {\bibfnamefont {S.}~\bibnamefont {Bosco}},\
  and\ \bibinfo {author} {\bibfnamefont {D.}~\bibnamefont {Loss}},\ }\bibfield
  {title} {\bibinfo {title} {{Anomalous Zero-Field Splitting for Hole Spin
  Qubits in Si and Ge Quantum Dots}},\ }\href
  {https://doi.org/10.1103/PhysRevLett.129.116805} {\bibfield  {journal}
  {\bibinfo  {journal} {Phys. Rev. Lett.}\ }\textbf {\bibinfo {volume} {129}},\
  \bibinfo {pages} {116805} (\bibinfo {year} {2022})}\BibitemShut {NoStop}%
\bibitem [{\citenamefont {Kroutvar}\ \emph {et~al.}(2004)\citenamefont
  {Kroutvar}, \citenamefont {Ducommun}, \citenamefont {Heiss}, \citenamefont
  {Bichler}, \citenamefont {Schuh}, \citenamefont {Abstreiter},\ and\
  \citenamefont {Finley}}]{kroutvar2004optically}%
  \BibitemOpen
  \bibfield  {author} {\bibinfo {author} {\bibfnamefont {M.}~\bibnamefont
  {Kroutvar}}, \bibinfo {author} {\bibfnamefont {Y.}~\bibnamefont {Ducommun}},
  \bibinfo {author} {\bibfnamefont {D.}~\bibnamefont {Heiss}}, \bibinfo
  {author} {\bibfnamefont {M.}~\bibnamefont {Bichler}}, \bibinfo {author}
  {\bibfnamefont {D.}~\bibnamefont {Schuh}}, \bibinfo {author} {\bibfnamefont
  {G.}~\bibnamefont {Abstreiter}},\ and\ \bibinfo {author} {\bibfnamefont
  {J.~J.}\ \bibnamefont {Finley}},\ }\bibfield  {title} {\bibinfo {title}
  {{Optically programmable electron spin memory using semiconductor quantum
  dots}},\ }\href {https://www.nature.com/articles/nature03008} {\bibfield
  {journal} {\bibinfo  {journal} {Nature}\ }\textbf {\bibinfo {volume} {432}},\
  \bibinfo {pages} {81} (\bibinfo {year} {2004})}\BibitemShut {NoStop}%
\bibitem [{\citenamefont {Schliemann}\ and\ \citenamefont
  {Loss}(2005)}]{schliemann2005spin}%
  \BibitemOpen
  \bibfield  {author} {\bibinfo {author} {\bibfnamefont {J.}~\bibnamefont
  {Schliemann}}\ and\ \bibinfo {author} {\bibfnamefont {D.}~\bibnamefont
  {Loss}},\ }\bibfield  {title} {\bibinfo {title} {{Spin-Hall transport of
  heavy holes in III-V semiconductor quantum wells}},\ }\href
  {https://doi.org/10.1103/PhysRevB.71.085308} {\bibfield  {journal} {\bibinfo
  {journal} {Phys. Rev. B}\ }\textbf {\bibinfo {volume} {71}},\ \bibinfo
  {pages} {085308} (\bibinfo {year} {2005})}\BibitemShut {NoStop}%
\bibitem [{\citenamefont {Moriya}\ \emph {et~al.}(2014)\citenamefont {Moriya},
  \citenamefont {Sawano}, \citenamefont {Hoshi}, \citenamefont {Masubuchi},
  \citenamefont {Shiraki}, \citenamefont {Wild}, \citenamefont {Neumann},
  \citenamefont {Abstreiter}, \citenamefont {Bougeard}, \citenamefont {Koga},\
  and\ \citenamefont {Machida}}]{moriya2014cubic}%
  \BibitemOpen
  \bibfield  {author} {\bibinfo {author} {\bibfnamefont {R.}~\bibnamefont
  {Moriya}}, \bibinfo {author} {\bibfnamefont {K.}~\bibnamefont {Sawano}},
  \bibinfo {author} {\bibfnamefont {Y.}~\bibnamefont {Hoshi}}, \bibinfo
  {author} {\bibfnamefont {S.}~\bibnamefont {Masubuchi}}, \bibinfo {author}
  {\bibfnamefont {Y.}~\bibnamefont {Shiraki}}, \bibinfo {author} {\bibfnamefont
  {A.}~\bibnamefont {Wild}}, \bibinfo {author} {\bibfnamefont {C.}~\bibnamefont
  {Neumann}}, \bibinfo {author} {\bibfnamefont {G.}~\bibnamefont {Abstreiter}},
  \bibinfo {author} {\bibfnamefont {D.}~\bibnamefont {Bougeard}}, \bibinfo
  {author} {\bibfnamefont {T.}~\bibnamefont {Koga}},\ and\ \bibinfo {author}
  {\bibfnamefont {T.}~\bibnamefont {Machida}},\ }\bibfield  {title} {\bibinfo
  {title} {{Cubic Rashba Spin-Orbit Interaction of a Two-Dimensional Hole Gas
  in a Strained-$\mathrm{Ge}/\mathrm{SiGe}$ Quantum Well}},\ }\href
  {https://doi.org/10.1103/PhysRevLett.113.086601} {\bibfield  {journal}
  {\bibinfo  {journal} {Phys. Rev. Lett.}\ }\textbf {\bibinfo {volume} {113}},\
  \bibinfo {pages} {086601} (\bibinfo {year} {2014})}\BibitemShut {NoStop}%
\bibitem [{\citenamefont {Bleibaum}\ and\ \citenamefont
  {Wachsmuth}(2006)}]{bleibaum2006spin}%
  \BibitemOpen
  \bibfield  {author} {\bibinfo {author} {\bibfnamefont {O.}~\bibnamefont
  {Bleibaum}}\ and\ \bibinfo {author} {\bibfnamefont {S.}~\bibnamefont
  {Wachsmuth}},\ }\bibfield  {title} {\bibinfo {title} {{Spin Hall effect in
  semiconductor heterostructures with cubic Rashba spin-orbit interaction}},\
  }\href {https://doi.org/10.1103/PhysRevB.74.195330} {\bibfield  {journal}
  {\bibinfo  {journal} {Phys. Rev. B}\ }\textbf {\bibinfo {volume} {74}},\
  \bibinfo {pages} {195330} (\bibinfo {year} {2006})}\BibitemShut {NoStop}%
\bibitem [{\citenamefont {Bosco}\ \emph {et~al.}(2022)\citenamefont {Bosco},
  \citenamefont {Scarlino}, \citenamefont {Klinovaja},\ and\ \citenamefont
  {Loss}}]{bosco2022fully}%
  \BibitemOpen
  \bibfield  {author} {\bibinfo {author} {\bibfnamefont {S.}~\bibnamefont
  {Bosco}}, \bibinfo {author} {\bibfnamefont {P.}~\bibnamefont {Scarlino}},
  \bibinfo {author} {\bibfnamefont {J.}~\bibnamefont {Klinovaja}},\ and\
  \bibinfo {author} {\bibfnamefont {D.}~\bibnamefont {Loss}},\ }\bibfield
  {title} {\bibinfo {title} {{Fully Tunable Longitudinal Spin-Photon
  Interactions in Si and Ge Quantum Dots}},\ }\href
  {https://doi.org/10.1103/PhysRevLett.129.066801} {\bibfield  {journal}
  {\bibinfo  {journal} {Phys. Rev. Lett.}\ }\textbf {\bibinfo {volume} {129}},\
  \bibinfo {pages} {066801} (\bibinfo {year} {2022})}\BibitemShut {NoStop}%
\bibitem [{\citenamefont {Dantas}\ \emph {et~al.}(2022)\citenamefont {Dantas},
  \citenamefont {Legg}, \citenamefont {Bosco}, \citenamefont {Loss},\ and\
  \citenamefont {Klinovaja}}]{dantas2022}%
  \BibitemOpen
  \bibfield  {author} {\bibinfo {author} {\bibfnamefont {R.~M.~A.}\
  \bibnamefont {Dantas}}, \bibinfo {author} {\bibfnamefont {H.~F.}\
  \bibnamefont {Legg}}, \bibinfo {author} {\bibfnamefont {S.}~\bibnamefont
  {Bosco}}, \bibinfo {author} {\bibfnamefont {D.}~\bibnamefont {Loss}},\ and\
  \bibinfo {author} {\bibfnamefont {J.}~\bibnamefont {Klinovaja}},\ }\bibfield
  {title} {\bibinfo {title} {Determination of spin-orbit interaction in
  semiconductor nanostructures via non-linear transport},\ }\href
  {https://arxiv.org/abs/2210.05429} {\bibfield  {journal} {\bibinfo  {journal}
  {arXiv:2210.05429}\ } (\bibinfo {year} {2022})}\BibitemShut {NoStop}%
\bibitem [{\citenamefont {Kells}\ \emph {et~al.}(2012)\citenamefont {Kells},
  \citenamefont {Meidan},\ and\ \citenamefont {Brouwer}}]{kells2012near}%
  \BibitemOpen
  \bibfield  {author} {\bibinfo {author} {\bibfnamefont {G.}~\bibnamefont
  {Kells}}, \bibinfo {author} {\bibfnamefont {D.}~\bibnamefont {Meidan}},\ and\
  \bibinfo {author} {\bibfnamefont {P.~W.}\ \bibnamefont {Brouwer}},\
  }\bibfield  {title} {\bibinfo {title} {{Near-zero-energy end states in
  topologically trivial spin-orbit coupled superconducting nanowires with a
  smooth confinement}},\ }\href {https://doi.org/10.1103/PhysRevB.86.100503}
  {\bibfield  {journal} {\bibinfo  {journal} {Phys. Rev. B}\ }\textbf {\bibinfo
  {volume} {86}},\ \bibinfo {pages} {100503} (\bibinfo {year}
  {2012})}\BibitemShut {NoStop}%
\bibitem [{\citenamefont {Moore}\ \emph
  {et~al.}(2018{\natexlab{a}})\citenamefont {Moore}, \citenamefont {Stanescu},\
  and\ \citenamefont {Tewari}}]{moore2018two}%
  \BibitemOpen
  \bibfield  {author} {\bibinfo {author} {\bibfnamefont {C.}~\bibnamefont
  {Moore}}, \bibinfo {author} {\bibfnamefont {T.~D.}\ \bibnamefont
  {Stanescu}},\ and\ \bibinfo {author} {\bibfnamefont {S.}~\bibnamefont
  {Tewari}},\ }\bibfield  {title} {\bibinfo {title} {{Two-terminal charge
  tunneling: Disentangling Majorana zero modes from partially separated Andreev
  bound states in semiconductor-superconductor heterostructures}},\ }\href
  {https://doi.org/10.1103/PhysRevB.97.165302} {\bibfield  {journal} {\bibinfo
  {journal} {Phys. Rev. B}\ }\textbf {\bibinfo {volume} {97}},\ \bibinfo
  {pages} {165302} (\bibinfo {year} {2018}{\natexlab{a}})}\BibitemShut
  {NoStop}%
\bibitem [{\citenamefont {Moore}\ \emph
  {et~al.}(2018{\natexlab{b}})\citenamefont {Moore}, \citenamefont {Zeng},
  \citenamefont {Stanescu},\ and\ \citenamefont {Tewari}}]{moore2018quantized}%
  \BibitemOpen
  \bibfield  {author} {\bibinfo {author} {\bibfnamefont {C.}~\bibnamefont
  {Moore}}, \bibinfo {author} {\bibfnamefont {C.}~\bibnamefont {Zeng}},
  \bibinfo {author} {\bibfnamefont {T.~D.}\ \bibnamefont {Stanescu}},\ and\
  \bibinfo {author} {\bibfnamefont {S.}~\bibnamefont {Tewari}},\ }\bibfield
  {title} {\bibinfo {title} {{Quantized zero-bias conductance plateau in
  semiconductor-superconductor heterostructures without topological Majorana
  zero modes}},\ }\href {https://doi.org/10.1103/PhysRevB.98.155314} {\bibfield
   {journal} {\bibinfo  {journal} {Phys. Rev. B}\ }\textbf {\bibinfo {volume}
  {98}},\ \bibinfo {pages} {155314} (\bibinfo {year}
  {2018}{\natexlab{b}})}\BibitemShut {NoStop}%
\bibitem [{\citenamefont {Huang}\ \emph {et~al.}(2018)\citenamefont {Huang},
  \citenamefont {Pan}, \citenamefont {Liu}, \citenamefont {Sau}, \citenamefont
  {Stanescu},\ and\ \citenamefont {Das~Sarma}}]{huang2018metamorphosis}%
  \BibitemOpen
  \bibfield  {author} {\bibinfo {author} {\bibfnamefont {Y.}~\bibnamefont
  {Huang}}, \bibinfo {author} {\bibfnamefont {H.}~\bibnamefont {Pan}}, \bibinfo
  {author} {\bibfnamefont {C.-X.}\ \bibnamefont {Liu}}, \bibinfo {author}
  {\bibfnamefont {J.~D.}\ \bibnamefont {Sau}}, \bibinfo {author} {\bibfnamefont
  {T.~D.}\ \bibnamefont {Stanescu}},\ and\ \bibinfo {author} {\bibfnamefont
  {S.}~\bibnamefont {Das~Sarma}},\ }\bibfield  {title} {\bibinfo {title}
  {{Metamorphosis of Andreev bound states into Majorana bound states in
  pristine nanowires}},\ }\href {https://doi.org/10.1103/PhysRevB.98.144511}
  {\bibfield  {journal} {\bibinfo  {journal} {Phys. Rev. B}\ }\textbf {\bibinfo
  {volume} {98}},\ \bibinfo {pages} {144511} (\bibinfo {year}
  {2018})}\BibitemShut {NoStop}%
\bibitem [{\citenamefont {Pe\~naranda}\ \emph {et~al.}(2018)\citenamefont
  {Pe\~naranda}, \citenamefont {Aguado}, \citenamefont {San-Jose},\ and\
  \citenamefont {Prada}}]{penaranda2018quantifying}%
  \BibitemOpen
  \bibfield  {author} {\bibinfo {author} {\bibfnamefont {F.}~\bibnamefont
  {Pe\~naranda}}, \bibinfo {author} {\bibfnamefont {R.}~\bibnamefont {Aguado}},
  \bibinfo {author} {\bibfnamefont {P.}~\bibnamefont {San-Jose}},\ and\
  \bibinfo {author} {\bibfnamefont {E.}~\bibnamefont {Prada}},\ }\bibfield
  {title} {\bibinfo {title} {{Quantifying wave-function overlaps in
  inhomogeneous Majorana nanowires}},\ }\href
  {https://doi.org/10.1103/PhysRevB.98.235406} {\bibfield  {journal} {\bibinfo
  {journal} {Phys. Rev. B}\ }\textbf {\bibinfo {volume} {98}},\ \bibinfo
  {pages} {235406} (\bibinfo {year} {2018})}\BibitemShut {NoStop}%
\bibitem [{\citenamefont {Reeg}\ \emph
  {et~al.}(2018{\natexlab{a}})\citenamefont {Reeg}, \citenamefont {Dmytruk},
  \citenamefont {Chevallier}, \citenamefont {Loss},\ and\ \citenamefont
  {Klinovaja}}]{reeg2018zero}%
  \BibitemOpen
  \bibfield  {author} {\bibinfo {author} {\bibfnamefont {C.}~\bibnamefont
  {Reeg}}, \bibinfo {author} {\bibfnamefont {O.}~\bibnamefont {Dmytruk}},
  \bibinfo {author} {\bibfnamefont {D.}~\bibnamefont {Chevallier}}, \bibinfo
  {author} {\bibfnamefont {D.}~\bibnamefont {Loss}},\ and\ \bibinfo {author}
  {\bibfnamefont {J.}~\bibnamefont {Klinovaja}},\ }\bibfield  {title} {\bibinfo
  {title} {{Zero-energy Andreev bound states from quantum dots in proximitized
  Rashba nanowires}},\ }\href {https://doi.org/10.1103/PhysRevB.98.245407}
  {\bibfield  {journal} {\bibinfo  {journal} {Phys. Rev. B}\ }\textbf {\bibinfo
  {volume} {98}},\ \bibinfo {pages} {245407} (\bibinfo {year}
  {2018}{\natexlab{a}})}\BibitemShut {NoStop}%
\bibitem [{\citenamefont {Vuik}\ \emph {et~al.}(2019)\citenamefont {Vuik},
  \citenamefont {Nijholt}, \citenamefont {Akhmerov},\ and\ \citenamefont
  {Wimmer}}]{vuik2019reproducing}%
  \BibitemOpen
  \bibfield  {author} {\bibinfo {author} {\bibfnamefont {A.}~\bibnamefont
  {Vuik}}, \bibinfo {author} {\bibfnamefont {B.}~\bibnamefont {Nijholt}},
  \bibinfo {author} {\bibfnamefont {A.~R.}\ \bibnamefont {Akhmerov}},\ and\
  \bibinfo {author} {\bibfnamefont {M.}~\bibnamefont {Wimmer}},\ }\bibfield
  {title} {\bibinfo {title} {{Reproducing topological properties with
  quasi-Majorana states}},\ }\href
  {https://scipost.org/10.21468/SciPostPhys.7.5.061} {\bibfield  {journal}
  {\bibinfo  {journal} {SciPost Phys.}\ }\textbf {\bibinfo {volume} {7}},\
  \bibinfo {pages} {61} (\bibinfo {year} {2019})}\BibitemShut {NoStop}%
\bibitem [{\citenamefont {Das~Sarma}\ and\ \citenamefont
  {Pan}(2021)}]{sarma2021disorder}%
  \BibitemOpen
  \bibfield  {author} {\bibinfo {author} {\bibfnamefont {S.}~\bibnamefont
  {Das~Sarma}}\ and\ \bibinfo {author} {\bibfnamefont {H.}~\bibnamefont
  {Pan}},\ }\bibfield  {title} {\bibinfo {title} {{Disorder-induced zero-bias
  peaks in Majorana nanowires}},\ }\href
  {https://doi.org/10.1103/PhysRevB.103.195158} {\bibfield  {journal} {\bibinfo
   {journal} {Phys. Rev. B}\ }\textbf {\bibinfo {volume} {103}},\ \bibinfo
  {pages} {195158} (\bibinfo {year} {2021})}\BibitemShut {NoStop}%
\bibitem [{\citenamefont {Hess}\ \emph {et~al.}(2021)\citenamefont {Hess},
  \citenamefont {Legg}, \citenamefont {Loss},\ and\ \citenamefont
  {Klinovaja}}]{hess2021local}%
  \BibitemOpen
  \bibfield  {author} {\bibinfo {author} {\bibfnamefont {R.}~\bibnamefont
  {Hess}}, \bibinfo {author} {\bibfnamefont {H.~F.}\ \bibnamefont {Legg}},
  \bibinfo {author} {\bibfnamefont {D.}~\bibnamefont {Loss}},\ and\ \bibinfo
  {author} {\bibfnamefont {J.}~\bibnamefont {Klinovaja}},\ }\bibfield  {title}
  {\bibinfo {title} {{Local and nonlocal quantum transport due to Andreev bound
  states in finite Rashba nanowires with superconducting and normal
  sections}},\ }\href {https://doi.org/10.1103/PhysRevB.104.075405} {\bibfield
  {journal} {\bibinfo  {journal} {Phys. Rev. B}\ }\textbf {\bibinfo {volume}
  {104}},\ \bibinfo {pages} {075405} (\bibinfo {year} {2021})}\BibitemShut
  {NoStop}%
\bibitem [{\citenamefont {Watzinger}\ \emph {et~al.}(2016)\citenamefont
  {Watzinger}, \citenamefont {Kloeffel}, \citenamefont {Vukušić},
  \citenamefont {Rossell}, \citenamefont {Sessi}, \citenamefont {Kukučka},
  \citenamefont {Kirchschlager}, \citenamefont {Lausecker}, \citenamefont
  {Truhlar}, \citenamefont {Glaser}, \citenamefont {Rastelli}, \citenamefont
  {Fuhrer}, \citenamefont {Loss},\ and\ \citenamefont
  {Katsaros}}]{watzinger2016heavy}%
  \BibitemOpen
  \bibfield  {author} {\bibinfo {author} {\bibfnamefont {H.}~\bibnamefont
  {Watzinger}}, \bibinfo {author} {\bibfnamefont {C.}~\bibnamefont {Kloeffel}},
  \bibinfo {author} {\bibfnamefont {L.}~\bibnamefont {Vukušić}}, \bibinfo
  {author} {\bibfnamefont {M.~D.}\ \bibnamefont {Rossell}}, \bibinfo {author}
  {\bibfnamefont {V.}~\bibnamefont {Sessi}}, \bibinfo {author} {\bibfnamefont
  {J.}~\bibnamefont {Kukučka}}, \bibinfo {author} {\bibfnamefont
  {R.}~\bibnamefont {Kirchschlager}}, \bibinfo {author} {\bibfnamefont
  {E.}~\bibnamefont {Lausecker}}, \bibinfo {author} {\bibfnamefont
  {A.}~\bibnamefont {Truhlar}}, \bibinfo {author} {\bibfnamefont
  {M.}~\bibnamefont {Glaser}}, \bibinfo {author} {\bibfnamefont
  {A.}~\bibnamefont {Rastelli}}, \bibinfo {author} {\bibfnamefont
  {A.}~\bibnamefont {Fuhrer}}, \bibinfo {author} {\bibfnamefont
  {D.}~\bibnamefont {Loss}},\ and\ \bibinfo {author} {\bibfnamefont
  {G.}~\bibnamefont {Katsaros}},\ }\bibfield  {title} {\bibinfo {title}
  {{Heavy-Hole States in Germanium Hut Wires}},\ }\href
  {https://doi.org/10.1021/acs.nanolett.6b02715} {\bibfield  {journal}
  {\bibinfo  {journal} {{Nano Letters}}\ }\textbf {\bibinfo {volume} {16}},\
  \bibinfo {pages} {6879} (\bibinfo {year} {2016})}\BibitemShut {NoStop}%
\bibitem [{\citenamefont {Lu}\ \emph {et~al.}(2017)\citenamefont {Lu},
  \citenamefont {Harris}, \citenamefont {Huang}, \citenamefont {Chuang},
  \citenamefont {Li},\ and\ \citenamefont {Liu}}]{lu2017effective}%
  \BibitemOpen
  \bibfield  {author} {\bibinfo {author} {\bibfnamefont {T.}~\bibnamefont
  {Lu}}, \bibinfo {author} {\bibfnamefont {C.}~\bibnamefont {Harris}}, \bibinfo
  {author} {\bibfnamefont {S.-H.}\ \bibnamefont {Huang}}, \bibinfo {author}
  {\bibfnamefont {Y.}~\bibnamefont {Chuang}}, \bibinfo {author} {\bibfnamefont
  {J.-Y.}\ \bibnamefont {Li}},\ and\ \bibinfo {author} {\bibfnamefont
  {C.}~\bibnamefont {Liu}},\ }\bibfield  {title} {\bibinfo {title} {{Effective
  g factor of low-density two-dimensional holes in a Ge quantum well}},\ }\href
  {https://aip.scitation.org/doi/full/10.1063/1.4990569} {\bibfield  {journal}
  {\bibinfo  {journal} {{Applied Physics Letters}}\ }\textbf {\bibinfo {volume}
  {111}},\ \bibinfo {pages} {102108} (\bibinfo {year} {2017})}\BibitemShut
  {NoStop}%
\bibitem [{\citenamefont {Hofmann}\ \emph {et~al.}(2019)\citenamefont
  {Hofmann}, \citenamefont {Jirovec}, \citenamefont {Borovkov}, \citenamefont
  {Prieto}, \citenamefont {Ballabio}, \citenamefont {Frigerio}, \citenamefont
  {Chrastina}, \citenamefont {Isella},\ and\ \citenamefont
  {Katsaros}}]{hofmann2019assessing}%
  \BibitemOpen
  \bibfield  {author} {\bibinfo {author} {\bibfnamefont {A.}~\bibnamefont
  {Hofmann}}, \bibinfo {author} {\bibfnamefont {D.}~\bibnamefont {Jirovec}},
  \bibinfo {author} {\bibfnamefont {M.}~\bibnamefont {Borovkov}}, \bibinfo
  {author} {\bibfnamefont {I.}~\bibnamefont {Prieto}}, \bibinfo {author}
  {\bibfnamefont {A.}~\bibnamefont {Ballabio}}, \bibinfo {author}
  {\bibfnamefont {J.}~\bibnamefont {Frigerio}}, \bibinfo {author}
  {\bibfnamefont {D.}~\bibnamefont {Chrastina}}, \bibinfo {author}
  {\bibfnamefont {G.}~\bibnamefont {Isella}},\ and\ \bibinfo {author}
  {\bibfnamefont {G.}~\bibnamefont {Katsaros}},\ }\bibfield  {title} {\bibinfo
  {title} {{Assessing the potential of Ge/SiGe quantum dots as hosts for
  singlet-triplet qubits}},\ }\href {https://arxiv.org/abs/1910.05841}
  {\bibfield  {journal} {\bibinfo  {journal} {arXiv:1910.05841}\ } (\bibinfo
  {year} {2019})}\BibitemShut {NoStop}%
\bibitem [{\citenamefont {Gao}\ \emph {et~al.}(2020)\citenamefont {Gao},
  \citenamefont {Wang}, \citenamefont {Watzinger}, \citenamefont {Hu},
  \citenamefont {Ran{\v{c}}i{\'c}}, \citenamefont {Zhang}, \citenamefont
  {Wang}, \citenamefont {Yao}, \citenamefont {Wang}, \citenamefont
  {Kuku{\v{c}}ka} \emph {et~al.}}]{gao2020site}%
  \BibitemOpen
  \bibfield  {author} {\bibinfo {author} {\bibfnamefont {F.}~\bibnamefont
  {Gao}}, \bibinfo {author} {\bibfnamefont {J.-H.}\ \bibnamefont {Wang}},
  \bibinfo {author} {\bibfnamefont {H.}~\bibnamefont {Watzinger}}, \bibinfo
  {author} {\bibfnamefont {H.}~\bibnamefont {Hu}}, \bibinfo {author}
  {\bibfnamefont {M.~J.}\ \bibnamefont {Ran{\v{c}}i{\'c}}}, \bibinfo {author}
  {\bibfnamefont {J.-Y.}\ \bibnamefont {Zhang}}, \bibinfo {author}
  {\bibfnamefont {T.}~\bibnamefont {Wang}}, \bibinfo {author} {\bibfnamefont
  {Y.}~\bibnamefont {Yao}}, \bibinfo {author} {\bibfnamefont {G.-L.}\
  \bibnamefont {Wang}}, \bibinfo {author} {\bibfnamefont {J.}~\bibnamefont
  {Kuku{\v{c}}ka}}, \emph {et~al.},\ }\bibfield  {title} {\bibinfo {title}
  {{Site-controlled uniform Ge/Si hut wires with electrically tunable
  spin--orbit coupling}},\ }\href {https://doi.org/10.1002/adma.201906523}
  {\bibfield  {journal} {\bibinfo  {journal} {Advanced Materials}\ }\textbf
  {\bibinfo {volume} {32}},\ \bibinfo {pages} {1906523} (\bibinfo {year}
  {2020})}\BibitemShut {NoStop}%
\bibitem [{\citenamefont {Alidoust}\ \emph {et~al.}(2021)\citenamefont
  {Alidoust}, \citenamefont {Shen},\ and\ \citenamefont {\ifmmode
  \check{Z}\else \v{Z}\fi{}uti\ifmmode~\acute{c}\else
  \'{c}\fi{}}}]{alidoust2021cubic}%
  \BibitemOpen
  \bibfield  {author} {\bibinfo {author} {\bibfnamefont {M.}~\bibnamefont
  {Alidoust}}, \bibinfo {author} {\bibfnamefont {C.}~\bibnamefont {Shen}},\
  and\ \bibinfo {author} {\bibfnamefont {I.}~\bibnamefont {\ifmmode
  \check{Z}\else \v{Z}\fi{}uti\ifmmode~\acute{c}\else \'{c}\fi{}}},\ }\bibfield
   {title} {\bibinfo {title} {{Cubic spin-orbit coupling and anomalous
  Josephson effect in planar junctions}},\ }\href
  {https://doi.org/10.1103/PhysRevB.103.L060503} {\bibfield  {journal}
  {\bibinfo  {journal} {Phys. Rev. B}\ }\textbf {\bibinfo {volume} {103}},\
  \bibinfo {pages} {L060503} (\bibinfo {year} {2021})}\BibitemShut {NoStop}%
\bibitem [{\citenamefont {Mayer}\ \emph {et~al.}(2022)\citenamefont {Mayer},
  \citenamefont {Sierra},\ and\ \citenamefont
  {Hankiewicz}}]{mayer2022intrinsic}%
  \BibitemOpen
  \bibfield  {author} {\bibinfo {author} {\bibfnamefont {J.~B.}\ \bibnamefont
  {Mayer}}, \bibinfo {author} {\bibfnamefont {M.~A.}\ \bibnamefont {Sierra}},\
  and\ \bibinfo {author} {\bibfnamefont {E.~M.}\ \bibnamefont {Hankiewicz}},\
  }\bibfield  {title} {\bibinfo {title} {{Intrinsic emergence of Majorana modes
  in Luttinger $j=\frac{3}{2}$ systems}},\ }\href
  {https://doi.org/10.1103/PhysRevB.105.224513} {\bibfield  {journal} {\bibinfo
   {journal} {Phys. Rev. B}\ }\textbf {\bibinfo {volume} {105}},\ \bibinfo
  {pages} {224513} (\bibinfo {year} {2022})}\BibitemShut {NoStop}%
\bibitem [{\citenamefont {Terrazos}\ \emph {et~al.}(2021)\citenamefont
  {Terrazos}, \citenamefont {Marcellina}, \citenamefont {Wang}, \citenamefont
  {Coppersmith}, \citenamefont {Friesen}, \citenamefont {Hamilton},
  \citenamefont {Hu}, \citenamefont {Koiller}, \citenamefont {Saraiva},
  \citenamefont {Culcer},\ and\ \citenamefont {Capaz}}]{terrazos2021theory}%
  \BibitemOpen
  \bibfield  {author} {\bibinfo {author} {\bibfnamefont {L.~A.}\ \bibnamefont
  {Terrazos}}, \bibinfo {author} {\bibfnamefont {E.}~\bibnamefont
  {Marcellina}}, \bibinfo {author} {\bibfnamefont {Z.}~\bibnamefont {Wang}},
  \bibinfo {author} {\bibfnamefont {S.~N.}\ \bibnamefont {Coppersmith}},
  \bibinfo {author} {\bibfnamefont {M.}~\bibnamefont {Friesen}}, \bibinfo
  {author} {\bibfnamefont {A.~R.}\ \bibnamefont {Hamilton}}, \bibinfo {author}
  {\bibfnamefont {X.}~\bibnamefont {Hu}}, \bibinfo {author} {\bibfnamefont
  {B.}~\bibnamefont {Koiller}}, \bibinfo {author} {\bibfnamefont {A.~L.}\
  \bibnamefont {Saraiva}}, \bibinfo {author} {\bibfnamefont {D.}~\bibnamefont
  {Culcer}},\ and\ \bibinfo {author} {\bibfnamefont {R.~B.}\ \bibnamefont
  {Capaz}},\ }\bibfield  {title} {\bibinfo {title} {{Theory of hole-spin qubits
  in strained germanium quantum dots}},\ }\href
  {https://doi.org/10.1103/PhysRevB.103.125201} {\bibfield  {journal} {\bibinfo
   {journal} {Phys. Rev. B}\ }\textbf {\bibinfo {volume} {103}},\ \bibinfo
  {pages} {125201} (\bibinfo {year} {2021})}\BibitemShut {NoStop}%
\bibitem [{\citenamefont {Luttinger}(1956)}]{luttinger1956quantum}%
  \BibitemOpen
  \bibfield  {author} {\bibinfo {author} {\bibfnamefont {J.~M.}\ \bibnamefont
  {Luttinger}},\ }\bibfield  {title} {\bibinfo {title} {{Quantum Theory of
  Cyclotron Resonance in Semiconductors: General Theory}},\ }\href
  {https://doi.org/10.1103/PhysRev.102.1030} {\bibfield  {journal} {\bibinfo
  {journal} {Phys. Rev.}\ }\textbf {\bibinfo {volume} {102}},\ \bibinfo {pages}
  {1030} (\bibinfo {year} {1956})}\BibitemShut {NoStop}%
\bibitem [{\citenamefont {Winkler}(2003)}]{winkler2003spin}%
  \BibitemOpen
  \bibfield  {author} {\bibinfo {author} {\bibfnamefont {R.}~\bibnamefont
  {Winkler}},\ }\href@noop {} {\emph {\bibinfo {title} {{Spin–Orbit Coupling
  Effects in Two-Dimensional Electron and Hole Systems}}}}\ (\bibinfo
  {publisher} {Springer-Verlag},\ \bibinfo {address} {Berlin, Heidelberg, New
  York},\ \bibinfo {year} {2003})\BibitemShut {NoStop}%
\bibitem [{\citenamefont {Xiong}\ \emph {et~al.}(2021)\citenamefont {Xiong},
  \citenamefont {Guan}, \citenamefont {Luo},\ and\ \citenamefont
  {Li}}]{xiong2021emergence}%
  \BibitemOpen
  \bibfield  {author} {\bibinfo {author} {\bibfnamefont {J.-X.}\ \bibnamefont
  {Xiong}}, \bibinfo {author} {\bibfnamefont {S.}~\bibnamefont {Guan}},
  \bibinfo {author} {\bibfnamefont {J.-W.}\ \bibnamefont {Luo}},\ and\ \bibinfo
  {author} {\bibfnamefont {S.-S.}\ \bibnamefont {Li}},\ }\bibfield  {title}
  {\bibinfo {title} {{Emergence of strong tunable linear Rashba spin-orbit
  coupling in two-dimensional hole gases in semiconductor quantum wells}},\
  }\href {https://doi.org/10.1103/PhysRevB.103.085309} {\bibfield  {journal}
  {\bibinfo  {journal} {Phys. Rev. B}\ }\textbf {\bibinfo {volume} {103}},\
  \bibinfo {pages} {085309} (\bibinfo {year} {2021})}\BibitemShut {NoStop}%
\bibitem [{Note1()}]{Note1}%
  \BibitemOpen
  \bibinfo {note} {Another way to break inversion symmetry is by an inversion
  asymmetric confinement potential~\cite {bosco2021hole}.}\BibitemShut {Stop}%
\bibitem [{\citenamefont {Lodari}\ \emph {et~al.}(2022)\citenamefont {Lodari},
  \citenamefont {Kong}, \citenamefont {Rendell}, \citenamefont {Tosato},
  \citenamefont {Sammak}, \citenamefont {Veldhorst}, \citenamefont {Hamilton},\
  and\ \citenamefont {Scappucci}}]{lodari2022lightly}%
  \BibitemOpen
  \bibfield  {author} {\bibinfo {author} {\bibfnamefont {M.}~\bibnamefont
  {Lodari}}, \bibinfo {author} {\bibfnamefont {O.}~\bibnamefont {Kong}},
  \bibinfo {author} {\bibfnamefont {M.}~\bibnamefont {Rendell}}, \bibinfo
  {author} {\bibfnamefont {A.}~\bibnamefont {Tosato}}, \bibinfo {author}
  {\bibfnamefont {A.}~\bibnamefont {Sammak}}, \bibinfo {author} {\bibfnamefont
  {M.}~\bibnamefont {Veldhorst}}, \bibinfo {author} {\bibfnamefont
  {A.}~\bibnamefont {Hamilton}},\ and\ \bibinfo {author} {\bibfnamefont
  {G.}~\bibnamefont {Scappucci}},\ }\bibfield  {title} {\bibinfo {title}
  {{Lightly strained germanium quantum wells with hole mobility exceeding one
  million}},\ }\href {https://doi.org/10.1063/5.0083161} {\bibfield  {journal}
  {\bibinfo  {journal} {Applied Physics Letters}\ }\textbf {\bibinfo {volume}
  {120}},\ \bibinfo {pages} {122104} (\bibinfo {year} {2022})}\BibitemShut
  {NoStop}%
\bibitem [{\citenamefont {Wang}\ \emph {et~al.}(2021)\citenamefont {Wang},
  \citenamefont {Marcellina}, \citenamefont {Hamilton}, \citenamefont {Cullen},
  \citenamefont {Rogge}, \citenamefont {Salfi},\ and\ \citenamefont
  {Culcer}}]{wang2021optimal}%
  \BibitemOpen
  \bibfield  {author} {\bibinfo {author} {\bibfnamefont {Z.}~\bibnamefont
  {Wang}}, \bibinfo {author} {\bibfnamefont {E.}~\bibnamefont {Marcellina}},
  \bibinfo {author} {\bibfnamefont {A.~R.}\ \bibnamefont {Hamilton}}, \bibinfo
  {author} {\bibfnamefont {J.~H.}\ \bibnamefont {Cullen}}, \bibinfo {author}
  {\bibfnamefont {S.}~\bibnamefont {Rogge}}, \bibinfo {author} {\bibfnamefont
  {J.}~\bibnamefont {Salfi}},\ and\ \bibinfo {author} {\bibfnamefont
  {D.}~\bibnamefont {Culcer}},\ }\bibfield  {title} {\bibinfo {title} {{Optimal
  operation points for ultrafast, highly coherent Ge hole spin-orbit qubits}},\
  }\href {https://doi.org/10.1038/s41534-021-00386-2} {\bibfield  {journal}
  {\bibinfo  {journal} {npj Quantum Information}\ }\textbf {\bibinfo {volume}
  {7}},\ \bibinfo {pages} {54} (\bibinfo {year} {2021})}\BibitemShut {NoStop}%
\bibitem [{\citenamefont {Bir}\ \emph {et~al.}(1974)\citenamefont {Bir},
  \citenamefont {Pikus} \emph {et~al.}}]{bir1974symmetry}%
  \BibitemOpen
  \bibfield  {author} {\bibinfo {author} {\bibfnamefont {G.~L.}\ \bibnamefont
  {Bir}}, \bibinfo {author} {\bibfnamefont {G.~E.}\ \bibnamefont {Pikus}},
  \emph {et~al.},\ }\href@noop {} {\emph {\bibinfo {title} {{Symmetry and
  strain-induced effects in semiconductors}}}},\ Vol.\ \bibinfo {volume} {484}\
  (\bibinfo  {publisher} {Wiley New York},\ \bibinfo {year} {1974})\BibitemShut
  {NoStop}%
\bibitem [{\citenamefont {Miserev}\ and\ \citenamefont
  {Sushkov}(2017)}]{miserev2017dimensional}%
  \BibitemOpen
  \bibfield  {author} {\bibinfo {author} {\bibfnamefont {D.~S.}\ \bibnamefont
  {Miserev}}\ and\ \bibinfo {author} {\bibfnamefont {O.~P.}\ \bibnamefont
  {Sushkov}},\ }\bibfield  {title} {\bibinfo {title} {{Dimensional reduction of
  the Luttinger Hamiltonian and $g$-factors of holes in symmetric
  two-dimensional semiconductor heterostructures}},\ }\href
  {https://doi.org/10.1103/PhysRevB.95.085431} {\bibfield  {journal} {\bibinfo
  {journal} {Phys. Rev. B}\ }\textbf {\bibinfo {volume} {95}},\ \bibinfo
  {pages} {085431} (\bibinfo {year} {2017})}\BibitemShut {NoStop}%
\bibitem [{\citenamefont {Marcellina}\ \emph {et~al.}(2017)\citenamefont
  {Marcellina}, \citenamefont {Hamilton}, \citenamefont {Winkler},\ and\
  \citenamefont {Culcer}}]{marcellina2017spin}%
  \BibitemOpen
  \bibfield  {author} {\bibinfo {author} {\bibfnamefont {E.}~\bibnamefont
  {Marcellina}}, \bibinfo {author} {\bibfnamefont {A.~R.}\ \bibnamefont
  {Hamilton}}, \bibinfo {author} {\bibfnamefont {R.}~\bibnamefont {Winkler}},\
  and\ \bibinfo {author} {\bibfnamefont {D.}~\bibnamefont {Culcer}},\
  }\bibfield  {title} {\bibinfo {title} {{Spin-orbit interactions in
  inversion-asymmetric two-dimensional hole systems: A variational analysis}},\
  }\href {https://doi.org/10.1103/PhysRevB.95.075305} {\bibfield  {journal}
  {\bibinfo  {journal} {Phys. Rev. B}\ }\textbf {\bibinfo {volume} {95}},\
  \bibinfo {pages} {075305} (\bibinfo {year} {2017})}\BibitemShut {NoStop}%
\bibitem [{\citenamefont {Michal}\ \emph {et~al.}(2021)\citenamefont {Michal},
  \citenamefont {Venitucci},\ and\ \citenamefont
  {Niquet}}]{michal2021longitudinal}%
  \BibitemOpen
  \bibfield  {author} {\bibinfo {author} {\bibfnamefont {V.~P.}\ \bibnamefont
  {Michal}}, \bibinfo {author} {\bibfnamefont {B.}~\bibnamefont {Venitucci}},\
  and\ \bibinfo {author} {\bibfnamefont {Y.-M.}\ \bibnamefont {Niquet}},\
  }\bibfield  {title} {\bibinfo {title} {{Longitudinal and transverse electric
  field manipulation of hole spin-orbit qubits in one-dimensional channels}},\
  }\href {https://doi.org/10.1103/PhysRevB.103.045305} {\bibfield  {journal}
  {\bibinfo  {journal} {Phys. Rev. B}\ }\textbf {\bibinfo {volume} {103}},\
  \bibinfo {pages} {045305} (\bibinfo {year} {2021})}\BibitemShut {NoStop}%
\bibitem [{\citenamefont {Mao}\ \emph {et~al.}(2012)\citenamefont {Mao},
  \citenamefont {Gong}, \citenamefont {Dumitrescu}, \citenamefont {Tewari},\
  and\ \citenamefont {Zhang}}]{mao2012hole}%
  \BibitemOpen
  \bibfield  {author} {\bibinfo {author} {\bibfnamefont {L.}~\bibnamefont
  {Mao}}, \bibinfo {author} {\bibfnamefont {M.}~\bibnamefont {Gong}}, \bibinfo
  {author} {\bibfnamefont {E.}~\bibnamefont {Dumitrescu}}, \bibinfo {author}
  {\bibfnamefont {S.}~\bibnamefont {Tewari}},\ and\ \bibinfo {author}
  {\bibfnamefont {C.}~\bibnamefont {Zhang}},\ }\bibfield  {title} {\bibinfo
  {title} {{Hole-Doped Semiconductor Nanowire on Top of an $s$-Wave
  Superconductor: A New and Experimentally Accessible System for Majorana
  Fermions}},\ }\href {https://doi.org/10.1103/PhysRevLett.108.177001}
  {\bibfield  {journal} {\bibinfo  {journal} {Phys. Rev. Lett.}\ }\textbf
  {\bibinfo {volume} {108}},\ \bibinfo {pages} {177001} (\bibinfo {year}
  {2012})}\BibitemShut {NoStop}%
\bibitem [{\citenamefont {Chiu}\ \emph {et~al.}(2016)\citenamefont {Chiu},
  \citenamefont {Teo}, \citenamefont {Schnyder},\ and\ \citenamefont
  {Ryu}}]{chiu2016classification}%
  \BibitemOpen
  \bibfield  {author} {\bibinfo {author} {\bibfnamefont {C.-K.}\ \bibnamefont
  {Chiu}}, \bibinfo {author} {\bibfnamefont {J.~C.~Y.}\ \bibnamefont {Teo}},
  \bibinfo {author} {\bibfnamefont {A.~P.}\ \bibnamefont {Schnyder}},\ and\
  \bibinfo {author} {\bibfnamefont {S.}~\bibnamefont {Ryu}},\ }\bibfield
  {title} {\bibinfo {title} {{Classification of topological quantum matter with
  symmetries}},\ }\href {https://doi.org/10.1103/RevModPhys.88.035005}
  {\bibfield  {journal} {\bibinfo  {journal} {Rev. Mod. Phys.}\ }\textbf
  {\bibinfo {volume} {88}},\ \bibinfo {pages} {035005} (\bibinfo {year}
  {2016})}\BibitemShut {NoStop}%
\bibitem [{\citenamefont {Tewari}\ and\ \citenamefont
  {Sau}(2012)}]{tewari2012topological}%
  \BibitemOpen
  \bibfield  {author} {\bibinfo {author} {\bibfnamefont {S.}~\bibnamefont
  {Tewari}}\ and\ \bibinfo {author} {\bibfnamefont {J.~D.}\ \bibnamefont
  {Sau}},\ }\bibfield  {title} {\bibinfo {title} {{Topological Invariants for
  Spin-Orbit Coupled Superconductor Nanowires}},\ }\href
  {https://doi.org/10.1103/PhysRevLett.109.150408} {\bibfield  {journal}
  {\bibinfo  {journal} {Phys. Rev. Lett.}\ }\textbf {\bibinfo {volume} {109}},\
  \bibinfo {pages} {150408} (\bibinfo {year} {2012})}\BibitemShut {NoStop}%
\bibitem [{\citenamefont {Kittel}\ and\ \citenamefont
  {McEuen}(2005)}]{kittel2005kittel}%
  \BibitemOpen
  \bibfield  {author} {\bibinfo {author} {\bibfnamefont {C.}~\bibnamefont
  {Kittel}}\ and\ \bibinfo {author} {\bibfnamefont {P.}~\bibnamefont
  {McEuen}},\ }\href@noop {} {\emph {\bibinfo {title} {{Introduction to Solid
  State Physics}}}},\ \bibinfo {edition} {8th}\ ed.\ (\bibinfo  {publisher}
  {John Wiley \& Sons},\ \bibinfo {year} {2005})\BibitemShut {NoStop}%
\bibitem [{\citenamefont {Reeg}\ \emph
  {et~al.}(2018{\natexlab{b}})\citenamefont {Reeg}, \citenamefont {Loss},\ and\
  \citenamefont {Klinovaja}}]{reeg2018proximity}%
  \BibitemOpen
  \bibfield  {author} {\bibinfo {author} {\bibfnamefont {C.}~\bibnamefont
  {Reeg}}, \bibinfo {author} {\bibfnamefont {D.}~\bibnamefont {Loss}},\ and\
  \bibinfo {author} {\bibfnamefont {J.}~\bibnamefont {Klinovaja}},\ }\bibfield
  {title} {\bibinfo {title} {{Proximity effect in a two-dimensional electron
  gas coupled to a thin superconducting layer}},\ }\href
  {https://doi.org/10.3762/bjnano.9.118} {\bibfield  {journal} {\bibinfo
  {journal} {Beilstein Journal of Nanotechnology}\ }\textbf {\bibinfo {volume}
  {9}},\ \bibinfo {pages} {1263} (\bibinfo {year}
  {2018}{\natexlab{b}})}\BibitemShut {NoStop}%
\bibitem [{\citenamefont {Reeg}\ \emph
  {et~al.}(2018{\natexlab{c}})\citenamefont {Reeg}, \citenamefont {Loss},\ and\
  \citenamefont {Klinovaja}}]{reeg2018metallization}%
  \BibitemOpen
  \bibfield  {author} {\bibinfo {author} {\bibfnamefont {C.}~\bibnamefont
  {Reeg}}, \bibinfo {author} {\bibfnamefont {D.}~\bibnamefont {Loss}},\ and\
  \bibinfo {author} {\bibfnamefont {J.}~\bibnamefont {Klinovaja}},\ }\bibfield
  {title} {\bibinfo {title} {{Metallization of a Rashba wire by a
  superconducting layer in the strong-proximity regime}},\ }\href
  {https://doi.org/10.1103/PhysRevB.97.165425} {\bibfield  {journal} {\bibinfo
  {journal} {Phys. Rev. B}\ }\textbf {\bibinfo {volume} {97}},\ \bibinfo
  {pages} {165425} (\bibinfo {year} {2018}{\natexlab{c}})}\BibitemShut
  {NoStop}%
\bibitem [{\citenamefont {Prada}\ \emph {et~al.}(2020)\citenamefont {Prada},
  \citenamefont {San-Jose}, \citenamefont {de~Moor}, \citenamefont {Geresdi},
  \citenamefont {Lee}, \citenamefont {Klinovaja}, \citenamefont {Loss},
  \citenamefont {Nyg{\aa}rd}, \citenamefont {Aguado},\ and\ \citenamefont
  {Kouwenhoven}}]{prada2020andreev}%
  \BibitemOpen
  \bibfield  {author} {\bibinfo {author} {\bibfnamefont {E.}~\bibnamefont
  {Prada}}, \bibinfo {author} {\bibfnamefont {P.}~\bibnamefont {San-Jose}},
  \bibinfo {author} {\bibfnamefont {M.~W.}\ \bibnamefont {de~Moor}}, \bibinfo
  {author} {\bibfnamefont {A.}~\bibnamefont {Geresdi}}, \bibinfo {author}
  {\bibfnamefont {E.~J.}\ \bibnamefont {Lee}}, \bibinfo {author} {\bibfnamefont
  {J.}~\bibnamefont {Klinovaja}}, \bibinfo {author} {\bibfnamefont
  {D.}~\bibnamefont {Loss}}, \bibinfo {author} {\bibfnamefont {J.}~\bibnamefont
  {Nyg{\aa}rd}}, \bibinfo {author} {\bibfnamefont {R.}~\bibnamefont {Aguado}},\
  and\ \bibinfo {author} {\bibfnamefont {L.~P.}\ \bibnamefont {Kouwenhoven}},\
  }\bibfield  {title} {\bibinfo {title} {{From Andreev to Majorana bound states
  in hybrid superconductor--semiconductor nanowires}},\ }\href
  {https://www.nature.com/articles/s42254-020-0228-y} {\bibfield  {journal}
  {\bibinfo  {journal} {Nature Reviews Physics}\ }\textbf {\bibinfo {volume}
  {2}},\ \bibinfo {pages} {575} (\bibinfo {year} {2020})}\BibitemShut {NoStop}%
\bibitem [{\citenamefont {Sauls}(2018)}]{sauls2018andreev}%
  \BibitemOpen
  \bibfield  {author} {\bibinfo {author} {\bibfnamefont {J.}~\bibnamefont
  {Sauls}},\ }\bibfield  {title} {\bibinfo {title} {{Andreev bound states and
  their signatures}},\ }\href {https://doi.org/10.1098/rsta.2018.0140}
  {\bibfield  {journal} {\bibinfo  {journal} {Philosophical Transactions of the
  Royal Society A: Mathematical, Physical and Engineering Sciences}\ }\textbf
  {\bibinfo {volume} {376}},\ \bibinfo {pages} {20180140} (\bibinfo {year}
  {2018})}\BibitemShut {NoStop}%
\bibitem [{\citenamefont {Prada}\ \emph {et~al.}(2012)\citenamefont {Prada},
  \citenamefont {San-Jose},\ and\ \citenamefont {Aguado}}]{prada2012transport}%
  \BibitemOpen
  \bibfield  {author} {\bibinfo {author} {\bibfnamefont {E.}~\bibnamefont
  {Prada}}, \bibinfo {author} {\bibfnamefont {P.}~\bibnamefont {San-Jose}},\
  and\ \bibinfo {author} {\bibfnamefont {R.}~\bibnamefont {Aguado}},\
  }\bibfield  {title} {\bibinfo {title} {{Transport spectroscopy of $NS$
  nanowire junctions with Majorana fermions}},\ }\href
  {https://doi.org/10.1103/PhysRevB.86.180503} {\bibfield  {journal} {\bibinfo
  {journal} {Phys. Rev. B}\ }\textbf {\bibinfo {volume} {86}},\ \bibinfo
  {pages} {180503} (\bibinfo {year} {2012})}\BibitemShut {NoStop}%
\bibitem [{\citenamefont {Rainis}\ \emph {et~al.}(2013)\citenamefont {Rainis},
  \citenamefont {Trifunovic}, \citenamefont {Klinovaja},\ and\ \citenamefont
  {Loss}}]{rainis2013towards}%
  \BibitemOpen
  \bibfield  {author} {\bibinfo {author} {\bibfnamefont {D.}~\bibnamefont
  {Rainis}}, \bibinfo {author} {\bibfnamefont {L.}~\bibnamefont {Trifunovic}},
  \bibinfo {author} {\bibfnamefont {J.}~\bibnamefont {Klinovaja}},\ and\
  \bibinfo {author} {\bibfnamefont {D.}~\bibnamefont {Loss}},\ }\bibfield
  {title} {\bibinfo {title} {{Towards a realistic transport modeling in a
  superconducting nanowire with Majorana fermions}},\ }\href
  {https://doi.org/10.1103/PhysRevB.87.024515} {\bibfield  {journal} {\bibinfo
  {journal} {Phys. Rev. B}\ }\textbf {\bibinfo {volume} {87}},\ \bibinfo
  {pages} {024515} (\bibinfo {year} {2013})}\BibitemShut {NoStop}%
\bibitem [{Note2()}]{Note2}%
  \BibitemOpen
  \bibinfo {note} {This problem is solved by considering not only the lowest
  energy level $E_0$ but the four lowest energy levels and counting how many
  are close to zero. The number of pairs of MBSs corresponds to $|Q_\protect
  \mathbb {Z}|$.}\BibitemShut {Stop}%
\bibitem [{Note3()}]{Note3}%
  \BibitemOpen
  \bibinfo {note} {The technical details of how to fit the slopes of the phase
  transition curves are as follows. The Zeeman energies $\Delta _{Z,i}$ at
  which $E(k_x=0)=0$ are determined for 100 evenly spaced values of the
  superconducting phase difference $\phi _i$, with $i=1, \protect \dots , 100$,
  $\phi _1=0.8\pi $ and $\phi _{100}=1.2\pi $. The slopes are determined by
  fitting a linear function to the pairs $(\phi _i, \Delta _{Z,i})$. The
  minimum Zeeman field is reached at index $M$, i.e., $\phi ^{(m)} \approx \phi
  _M$, which is not exactly at $\pi $. Around this minimum the phase diagram is
  parabolic. To not distort the linear fit, points around the minimum must be
  excluded. The left slope is determined using the points $(\phi _i, \Delta
  _{Z,i})$ with $i=1, \protect \dots , M-26$. The right slope is calculated
  using the points $(\phi _i, \Delta _{Z,i})$ with $i=M+26, \protect \dots ,
  100$.}\BibitemShut {Stop}%
\bibitem [{\citenamefont {Braunecker}\ \emph {et~al.}(2010)\citenamefont
  {Braunecker}, \citenamefont {Japaridze}, \citenamefont {Klinovaja},\ and\
  \citenamefont {Loss}}]{braunecker2010spin}%
  \BibitemOpen
  \bibfield  {author} {\bibinfo {author} {\bibfnamefont {B.}~\bibnamefont
  {Braunecker}}, \bibinfo {author} {\bibfnamefont {G.~I.}\ \bibnamefont
  {Japaridze}}, \bibinfo {author} {\bibfnamefont {J.}~\bibnamefont
  {Klinovaja}},\ and\ \bibinfo {author} {\bibfnamefont {D.}~\bibnamefont
  {Loss}},\ }\bibfield  {title} {\bibinfo {title} {{Spin-selective Peierls
  transition in interacting one-dimensional conductors with spin-orbit
  interaction}},\ }\href {https://doi.org/10.1103/PhysRevB.82.045127}
  {\bibfield  {journal} {\bibinfo  {journal} {Phys. Rev. B}\ }\textbf {\bibinfo
  {volume} {82}},\ \bibinfo {pages} {045127} (\bibinfo {year}
  {2010})}\BibitemShut {NoStop}%
\bibitem [{\citenamefont {Vaitiek{\.e}nas}\ \emph {et~al.}(2021)\citenamefont
  {Vaitiek{\.e}nas}, \citenamefont {Liu}, \citenamefont {Krogstrup},\ and\
  \citenamefont {Marcus}}]{vaitiekenas2021zero}%
  \BibitemOpen
  \bibfield  {author} {\bibinfo {author} {\bibfnamefont {S.}~\bibnamefont
  {Vaitiek{\.e}nas}}, \bibinfo {author} {\bibfnamefont {Y.}~\bibnamefont
  {Liu}}, \bibinfo {author} {\bibfnamefont {P.}~\bibnamefont {Krogstrup}},\
  and\ \bibinfo {author} {\bibfnamefont {C.}~\bibnamefont {Marcus}},\
  }\bibfield  {title} {\bibinfo {title} {{Zero-bias peaks at zero magnetic
  field in ferromagnetic hybrid nanowires}},\ }\href
  {https://www.nature.com/articles/s41567-020-1017-3} {\bibfield  {journal}
  {\bibinfo  {journal} {Nature Physics}\ }\textbf {\bibinfo {volume} {17}},\
  \bibinfo {pages} {43} (\bibinfo {year} {2021})}\BibitemShut {NoStop}%
\bibitem [{\citenamefont {Bosco}\ \emph
  {et~al.}(2021{\natexlab{b}})\citenamefont {Bosco}, \citenamefont
  {Het\'enyi},\ and\ \citenamefont {Loss}}]{bosco2021hole}%
  \BibitemOpen
  \bibfield  {author} {\bibinfo {author} {\bibfnamefont {S.}~\bibnamefont
  {Bosco}}, \bibinfo {author} {\bibfnamefont {B.}~\bibnamefont {Het\'enyi}},\
  and\ \bibinfo {author} {\bibfnamefont {D.}~\bibnamefont {Loss}},\ }\bibfield
  {title} {\bibinfo {title} {{Hole Spin Qubits in $\mathrm{Si}$ FinFETs With
  Fully Tunable Spin-Orbit Coupling and Sweet Spots for Charge Noise}},\ }\href
  {https://doi.org/10.1103/PRXQuantum.2.010348} {\bibfield  {journal} {\bibinfo
   {journal} {PRX Quantum}\ }\textbf {\bibinfo {volume} {2}},\ \bibinfo {pages}
  {010348} (\bibinfo {year} {2021}{\natexlab{b}})}\BibitemShut {NoStop}%
\bibitem [{\citenamefont {Drichko}\ \emph {et~al.}(2018)\citenamefont
  {Drichko}, \citenamefont {Dmitriev}, \citenamefont {Malysh}, \citenamefont
  {Smirnov}, \citenamefont {von K{\"a}nel}, \citenamefont {Kummer},
  \citenamefont {Chrastina},\ and\ \citenamefont
  {Isella}}]{drichko2018effective}%
  \BibitemOpen
  \bibfield  {author} {\bibinfo {author} {\bibfnamefont {I.~L.}\ \bibnamefont
  {Drichko}}, \bibinfo {author} {\bibfnamefont {A.~A.}\ \bibnamefont
  {Dmitriev}}, \bibinfo {author} {\bibfnamefont {V.}~\bibnamefont {Malysh}},
  \bibinfo {author} {\bibfnamefont {I.~Y.}\ \bibnamefont {Smirnov}}, \bibinfo
  {author} {\bibfnamefont {H.}~\bibnamefont {von K{\"a}nel}}, \bibinfo {author}
  {\bibfnamefont {M.}~\bibnamefont {Kummer}}, \bibinfo {author} {\bibfnamefont
  {D.}~\bibnamefont {Chrastina}},\ and\ \bibinfo {author} {\bibfnamefont
  {G.}~\bibnamefont {Isella}},\ }\bibfield  {title} {\bibinfo {title}
  {{Effective g factor of 2D holes in strained Ge quantum wells}},\ }\href
  {https://doi.org/10.1063/1.5025413} {\bibfield  {journal} {\bibinfo
  {journal} {Journal of Applied Physics}\ }\textbf {\bibinfo {volume} {123}},\
  \bibinfo {pages} {165703} (\bibinfo {year} {2018})}\BibitemShut {NoStop}%
\bibitem [{\citenamefont {Lesser}\ \emph {et~al.}(2021)\citenamefont {Lesser},
  \citenamefont {Flensberg}, \citenamefont {von Oppen},\ and\ \citenamefont
  {Oreg}}]{lesser2021three}%
  \BibitemOpen
  \bibfield  {author} {\bibinfo {author} {\bibfnamefont {O.}~\bibnamefont
  {Lesser}}, \bibinfo {author} {\bibfnamefont {K.}~\bibnamefont {Flensberg}},
  \bibinfo {author} {\bibfnamefont {F.}~\bibnamefont {von Oppen}},\ and\
  \bibinfo {author} {\bibfnamefont {Y.}~\bibnamefont {Oreg}},\ }\bibfield
  {title} {\bibinfo {title} {{Three-phase Majorana zero modes at tiny magnetic
  fields}},\ }\href {https://doi.org/10.1103/PhysRevB.103.L121116} {\bibfield
  {journal} {\bibinfo  {journal} {Phys. Rev. B}\ }\textbf {\bibinfo {volume}
  {103}},\ \bibinfo {pages} {L121116} (\bibinfo {year} {2021})}\BibitemShut
  {NoStop}%
\bibitem [{\citenamefont {Lesser}\ \emph {et~al.}(2022)\citenamefont {Lesser},
  \citenamefont {Oreg},\ and\ \citenamefont {Stern}}]{lesser2022one}%
  \BibitemOpen
  \bibfield  {author} {\bibinfo {author} {\bibfnamefont {O.}~\bibnamefont
  {Lesser}}, \bibinfo {author} {\bibfnamefont {Y.}~\bibnamefont {Oreg}},\ and\
  \bibinfo {author} {\bibfnamefont {A.}~\bibnamefont {Stern}},\ }\bibfield
  {title} {\bibinfo {title} {{One-dimensional topological superconductivity
  based entirely on phase control}},\ }\href {https://arxiv.org/abs/2206.13537}
  {\bibfield  {journal} {\bibinfo  {journal} {arXiv:2206.13537}\ } (\bibinfo
  {year} {2022})}\BibitemShut {NoStop}%
\end{thebibliography}%

\end{document}